\documentclass[10pt,journal,compsoc]{IEEEtran}

\usepackage{tikz}
\usetikzlibrary{shapes,arrows.meta,chains,calc}
\usepackage{graphicx}
\graphicspath{{figures/}}
\usepackage[labelformat=simple]{subcaption}

\usepackage{amsmath}
\usepackage{amssymb}
\usepackage{amsthm}
\usepackage{empheq}

\DeclareMathOperator*{\argmin}{argmin}

\usepackage{mathtools} % for "\DeclarePairedDelimiter" macro
\DeclarePairedDelimiter{\floor}{\lfloor}{\rfloor}

\usepackage{enumerate}
\usepackage{enumitem}
\setlist[itemize]{noitemsep, topsep=0pt}
\setlist[enumerate]{noitemsep, topsep=0pt}
\usepackage{color}

\usepackage{xcolor}
\usepackage[listings,skins,breakable]{tcolorbox}

\usepackage{algorithm}
\usepackage[noend]{algpseudocode}

\usepackage{forloop}
\newcounter{ct}

\usepackage{booktabs}
\usepackage{multicol}
\usepackage{multirow}
\newcommand{\tabincell}[2]{\begin{tabular}{@{}#1@{}}#2\end{tabular}}
\usepackage{xargs}

\usepackage[backend=biber, citestyle=numeric-comp, bibstyle=ieee]{biblatex}
\bibliography{refs}
\usepackage{array}
\newcolumntype{L}[1]{>{\raggedright\let\newline\\\arraybackslash\hspace{0pt}}m{#1}}
\newcolumntype{C}[1]{>{\centering\let\newline\\\arraybackslash\hspace{0pt}}m{#1}}
\newcolumntype{R}[1]{>{\raggedleft\let\newline\\\arraybackslash\hspace{0pt}}m{#1}}

\usepackage[normalem]{ulem}

\usepackage{balance}

\usepackage[bottom]{footmisc}

\usepackage[hidelinks]{hyperref}

\DeclareMathAlphabet{\zcal}{\encodingdefault}{cmr}{m}{n}

\begin{document}

\title{Joint Planning of Network Slicing and Mobile Edge Computing: Models and Algorithms}

\author{Bin~Xiang, %~\IEEEmembership{Member,~IEEE,}
        Jocelyne~Elias, %~\IEEEmembership{Fellow,~IEEE,}
        Fabio~Martignon, %~\IEEEmembership{Fellow,~IEEE,}
        and~Elisabetta~Di~Nitto%,~\IEEEmembership{Fellow,~IEEE}% <-this % stops a space
\IEEEcompsocitemizethanks{
\IEEEcompsocthanksitem B. Xiang and E. Di Nitto are with the
Dipartimento di Elettronica, Informazione e Bioingegneria,
Politecnico di Milano, Milan, Italy, 20133.\protect\\
E-mail: \{bin.xiang, elisabetta.dinitto\}@polimi.it.
\IEEEcompsocthanksitem J. Elias is with the Department of Computer Science and Engineering (DISI), University of Bologna, Bologna, Italy, 40126.\protect\\
E-mail: jocelyne.elias@unibo.it.
\IEEEcompsocthanksitem F. Martignon is with the
Department of Management, Information and Production Engineering, University of Bergamo, Bergamo, Italy, 24044.\protect\\
E-mail: fabio.martignon@unibg.it.
}% <-this % stops an unwanted space
%\thanks{Manuscript received XXX XX, 20XX; revised XXX XX, 20XX.}
}

%\markboth{IEEE TRANSACTIONS ON CLOUD COMPUTING, VOL. XXX, NO. XXX, XXX 20XX}%
%{}

\IEEEtitleabstractindextext{%
\begin{abstract}

Multi-access Edge Computing (MEC) facilitates the deployment of critical applications with stringent QoS requirements, latency in particular.
This paper considers the problem of jointly planning the availability of computational resources at the edge, the slicing of mobile network and edge computation resources, and the routing of heterogeneous traffic types to the various slices. These aspects are intertwined and must be addressed together to provide the desired QoS to all mobile users and traffic types still keeping costs under control.
We formulate our problem as a mixed-integer nonlinear program (MINLP) and we define a heuristic, named Neighbor Exploration and Sequential Fixing (NESF), to facilitate the solution of the problem. The approach allows network operators to fine tune the network operation cost and the total latency experienced by users. We evaluate the performance of the proposed model and heuristic against two natural greedy approaches. We show the impact of the variation of all the considered parameters (viz., different types of traffic, tolerable latency, network topology and bandwidth, computation and link capacity) on the defined model. Numerical results demonstrate that NESF is very effective, achieving near-optimal planning and resource allocation solutions in a very short computing time even for large-scale network scenarios.

\end{abstract}

}

\maketitle

\IEEEdisplaynontitleabstractindextext

\IEEEpeerreviewmaketitle

\IEEEraisesectionheading{\section{Introduction}\label{sec:introduction}}
\IEEEPARstart{N}{ext} generation mobile networks aim to meet different users' Quality of Service (QoS) requirements in several demanding application scenarios and use cases. Among the others, controlling \textit{latency} is certainly one of the key QoS requirements that mobile operators have to deal with. In fact, the classification devised by the International Telecommunications Union-Radio communication Sector (ITU-R), shows that mission-critical services depend on strong latency constraints. For example, in some use cases (e.g., autonomous driving), the tolerable latency is expected to reach less than 1~ms~\cite{xiang2017}.

To address such constraints various ingredients are emerging. First of all, through \textit{Network Slicing}, the physical network infrastructure can be split into several isolated logical networks, each dedicated to applications with specific latency requirements, thus enabling an efficient and dynamic use of network resources \cite{zhang2017network}.

\textcolor{black}{Second, \textit{Multi-access Edge Computing (MEC)} provides an IT service environment and cloud-computing capabilities at the edge of the mobile network, within the Radio Access Network and in close proximity to mobile subscribers~\cite{hu2015mobile}. Through this approach, the latency experienced by mobile users can be consistently reduced. However, the computation power that can be offered by an edge cloud is quite limited in comparison with a remote cloud. Fortunately, this problem can be addressed by enabling cooperation among multiple edge clouds, scenario that can be realized in next-generation mobile networks (5G and beyond) as they will be likely built in an ultra-dense manner, where the edge clouds attached to base stations will also be massively deployed and connected to each other in a specific topology. }

In this line, we study the case of a complex network organized in multiple \emph{edge clouds}, each of which may be connected to the Radio Access Network of a certain location. All such edge clouds are connected through an arbitrary topology. This way, each edge cloud can serve end user traffic by relying not only on its own resources, but also offloading some traffic to its neighbors when needed.
We specifically consider multiple classes of traffic and corresponding requirements, including voice, video, web, among others.
For every class of traffic incoming from the corresponding Radio Access Network, the edge cloud decides whether to serve it or offload it to some other edge cloud. This decision depends on the QoS requirements associated to the specific class of traffic and on the current status of the edge cloud. 

Our main objective is to ensure that the infrastructure is able to serve all possible types of traffic within the boundaries of their QoS requirements and of the available resources.

In this work, therefore, we propose a complete approach, named \textit{Joint Planning and Slicing of mobile Network and edge Computation resources} (JPSNC), which solves the problem of \textit{operating} cost-efficient edge networks. The approach jointly takes into account the overall budget that the operator uses in order to allocate and operate computing capabilities in its edge network, and \textit{allocates resources}, aiming at minimizing the network operation cost and the total traffic latency of transmitting, outsourcing and processing user traffic, under the constraint of user tolerable latency for each class of traffic.

This turns out to be a mixed-integer nonlinear programming (MINLP) optimization problem, which is an $\mathcal{NP}$-hard problem \cite{kannan1978computational}.
To tackle this challenge, we transform it into an equivalent mixed-integer quadratically constrained programming (MIQCP) problem, which can be solved more efficiently through the Branch and Bound method.
Based on this reformulation, we further propose an effective heuristic, named \textit{Neighbor Exploration and Sequential Fixing (NESF)}, that permits to obtain near-optimal solutions in a very short computing time, even for the large-scale scenarios we considered in our numerical analysis. Furthermore, we propose two simple heuristics, based on a greedy approach. They provide benchmarks for our algorithms, obtain (slightly) sub-optimal solutions with respect to NESF, and are still very fast.
Finally, we systematically analyze and discuss with a thorough numerical evaluation the impact of all considered parameters (viz. the overall planning budget of the operator, different types of traffic, tolerable latency, network topology and bandwidth, computation and link capacity) on the optimal and approximate solutions obtained from our proposed model and heuristics. Numerical results demonstrate that our proposed model and heuristics can provide very efficient resource allocation and network planning solution for multiple edge networks.

This work takes the root from a previous paper~\cite{xiang2019joint} where we focused exclusively on minimizing the latency of traffic in a hierarchical network, keeping the network and computation capacity fixed. In this paper, we have completely revised our optimization model to cope with a joint network planning, slicing and edge computing problem, aimed at minimizing both the total latency and operation cost for arbitrary network topologies.

The remainder of this paper is organized as follows. 
Section~\ref{section-systemarchitecture} introduces the network system architecture we consider.
Section~\ref{section-systemoverview} provides an intuitive overview of the proposed approach by using a simple example.
Section~\ref{section-problemformulation} illustrates the proposed mathematical model and Section~\ref{section-heuristics} the heuristics. Section~\ref{section-numericalresults} discusses numerical results in a set of typical network topologies and scenarios.
Section~\ref{section-relatedwork} discusses related work.
Finally, Section~\ref{section-conclusion} concludes the paper.

\section{System Architecture}
\label{section-systemarchitecture}

\begin{figure}[ht]
    \centering
    \includegraphics[width=0.40\textwidth]{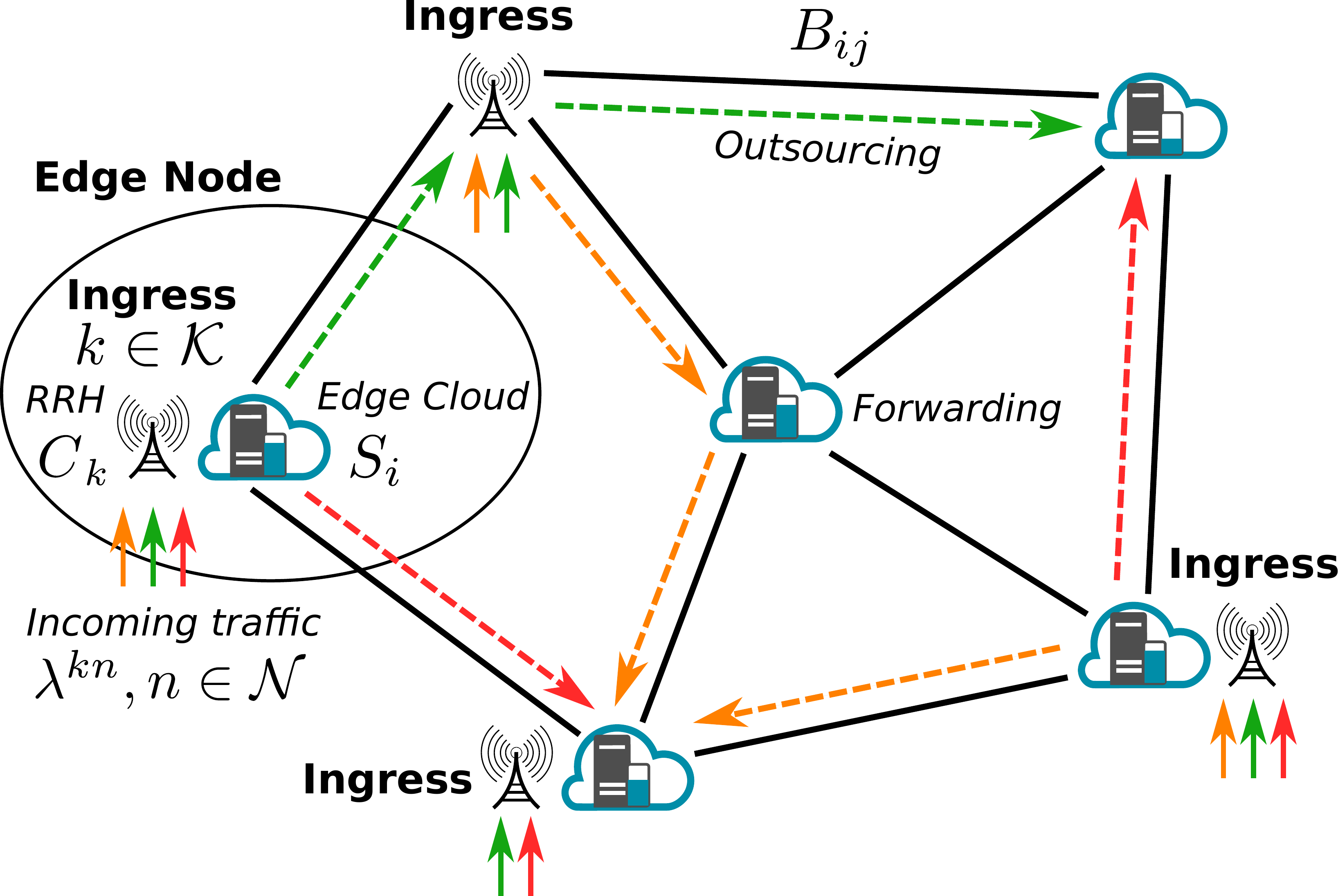}%
    \caption{Network system architecture.}
    \label{fig_arch}
\end{figure}

Figure \ref{fig_arch} illustrates our reference network architecture.
We consider an edge network composed of \emph{Edge Nodes}. Each of such nodes can be equipped with any of the following three capabilities: 
\begin{itemize}
    \item the ability of acquiring traffic from mobile devices through the Remote Radio Head (RRH), such nodes are those we call \emph{Ingress Nodes};
    \item the ability of executing network or application level services requiring computational power, this is done thanks to the availability of an \emph{Edge Cloud} on the node;
    \item the ability to route traffic to other nodes. 
\end{itemize}
Not all nodes must have all the three capabilities, so, in this respect, the edge network can be constituted of heterogeneous nodes.

Each link $(i,j)$ between any two edge nodes, $i$ and~$j$, has a fixed bandwidth, denoted by $B_{ij}$. Each \emph{Ingress Node}~$k$ has a specific ingress network capacity $C_k$, which is a measure of its ability to accept traffic incoming from mobile devices.
Nodes able to perform some computation have a computation capacity $S_i$. One of the objectives of the planning model presented in this paper is to determine the optimal value of the computation capacity that must be made available at each node.

We assume that users' incoming data in each Ingress Node is aggregated according to the corresponding \emph{traffic type} $n \in \mathcal{N}$. Examples of traffic types can be video, game, data from sensors, and the like. \textcolor{black}{
They group demands or services having the same requirements.
We assume that the network has a set of slices of different types and that each slice aggregates traffic of the same type. 
Therefore, all demands or services in the same slice could be treated in the same manner and could share network resources in a soft way like the concept of \emph{soft slicing} introduced in \cite{IETFJuly2017}. 
Our slicing model is also similar in part to the one introduced in \cite{MobicomFiore}, where the authors assume that mobile subscribers consume a variety of heterogeneous services and the operator owning the infrastructure implements a set of slices where \emph{each slice} is dedicated to a \emph{different subset of services}.
}

In Figure~\ref{fig_arch} traffic of different types is shown as arrows of different colors. From each Ingress Node, traffic can be split and processed on all edge clouds in the network; the dashed arrows shown in the figure represent possible outsourcing paths of the traffic pieces from different Ingress Nodes. Different slices of the ingress network capacity $C_k$ and the edge cloud computation capacity $S_i$ are allocated to serve the different types of traffic based on the corresponding Service Level Agreements (SLAs), which, in this paper are focused on keeping latency under control. Thus, another objective of our model is to find the allocation of traffic to the edge clouds that allows us to minimize the total latency, which is expressed in terms of the latency at the ingress node, due to the limitations of the wireless network, plus the latency due to the traffic processing computation, plus the latency occurring in the communication links internal to the network system architecture.

We assume that the edge network is controlled by a \emph{management component} which is in charge of achieving the optimal utilization of its resources, in terms of network and computation, still guaranteeing the SLA associated to each traffic type accepted by the network. This component monitors the network by periodically computing the network capacity of each ingress node (through broadcast messages exchanged in the network) and the bandwidth of each link in the network topology. Moreover, it knows the maximum available computation capacity of all computation nodes.
With these pieces of information as input, and knowing the SLA associated to each traffic type, the management component periodically solves an optimization problem that provides as output the identification of a proper network configuration and traffic allocation. In particular, it will identify:
i) the amount of computational capacity to be assigned to each node so that, with the foreseen traffic, the node usage remains below a certain level of its capacity;
ii) which node is taking care of which traffic type; and
iii) the nodes through which each traffic type must be routed toward its destination.

For simplicity, the optimization problem is based on the assumption that the system is time-slotted, where time is divided into equal-length short slots (short periods where network parameters can be considered as fixed and traffic shows only small variations).
We observe that our proposed heuristic (NESF) exhibits a short computing time so that it is feasible to run the problem periodically and to adjust the configuration of the system network based on the actual evolution of the traffic.

\textcolor{black}{In the next section, we give an intuition of the solution applied by the management component in the case of a simple network, while in Section~\ref{section-problemformulation} we formalize the optimization problem and in Section~\ref{section-heuristics} we present some heuristics that make the problem tractable in realistic cases.}

\section{Overview of Planning and Allocation}
\label{section-systemoverview}

\begin{figure}[!t]
    \centering
    %\captionsetup{skip=0pt}
    %\captionsetup[subfigure]{skip=0pt}
    \begin{subfigure}[t]{0.47\linewidth}\centering
        \includegraphics[width=\textwidth]{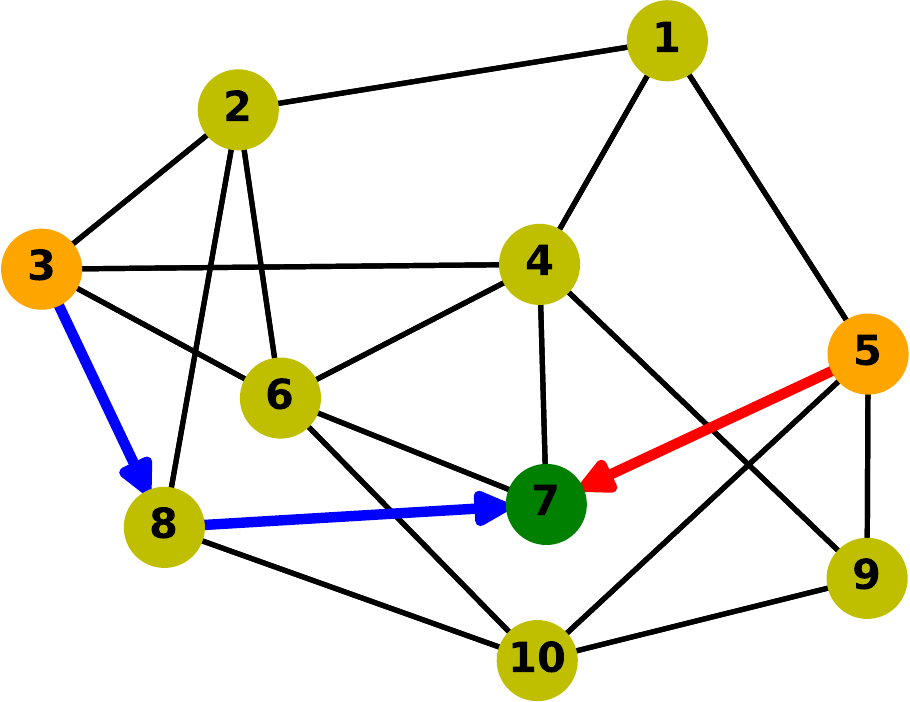}
        \caption{Minimizing both latency and computation costs}
        \label{toy-b}
    \end{subfigure}
    \hfill
    \begin{subfigure}[t]{0.47\linewidth}\centering
        \includegraphics[width=\textwidth]{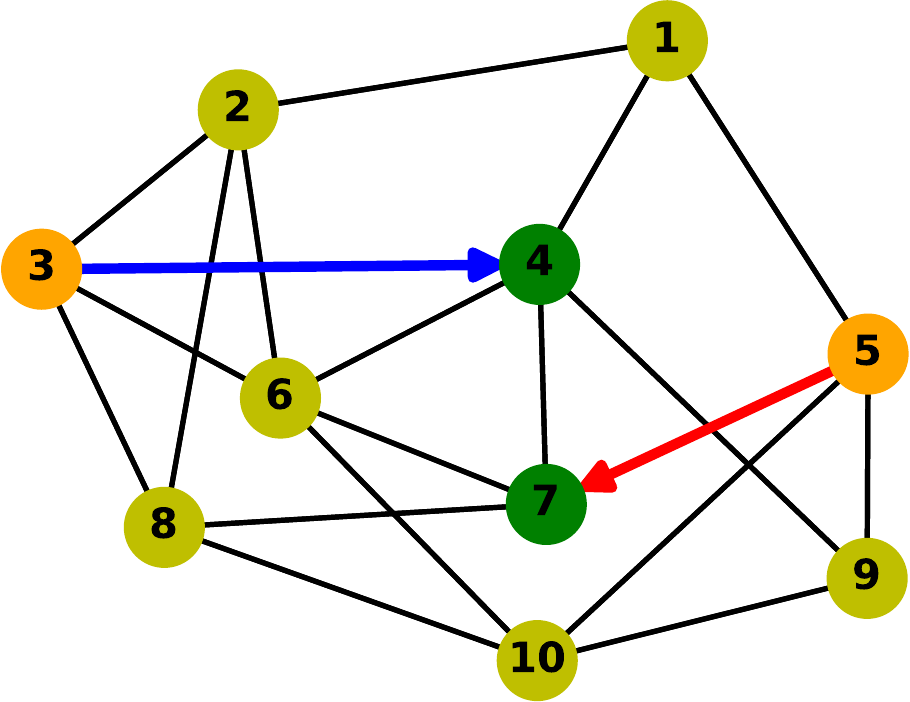}
        \caption{With the same settings, but a change $\lambda^{n5,t2}=40Gb/s$}
        \label{toy-c}
    \end{subfigure}
    \caption{Toy example for a network with 10 nodes and 20 edges (average degree: 4.0).}
    \label{fig-toyExample}
\end{figure}

In this section we refer to a simple but still meaningful edge network and we show how the management component behaves in the presence of two types of traffic. \textcolor{black}{In Section~\ref{section-problemformulation} we present in detail the optimization model that computes the allocation of computational and network resources as well as the optimal routing paths and we show how all values are computed. Here the goal is to provide the intuition beyond the proposed optimization approach.} 

The example we consider is shown in Figure~\ref{fig-toyExample} and consists of 10 nodes, two of which are ingress nodes (labeled as $n3$ and $n5$ in the figure and colored in orange), connected together with an average degree of 4. For simplicity, we assume that the bandwidth of all links is $B_{l} = 100 Gb/s$, and the wireless network capacity of the two ingress nodes is, respectively, $C_{n3} = 50 Gb/s$ and $C_{n5} = 60 Gb/s$.
Every node in the network has a computation capacity that can take one of the following values: $D_0 = 0 Gb/s$ (i.e., no computation capacity is made available at the current time), $D_1 = 30 Gb/s$, $D_2 = 40 Gb/s$, and $D_3 = 50 Gb/s$\footnote{Note that computation capacity is often expressed in cycles/s. As discussed in Section~\ref{section-numericalresults}, for homogeneity with the other values, we have transformed it into $Gb/s$.}. 
Given the above edge network, let us assume the management component estimates that node $n3$ will receive traffic of type $t1$ at rate $\lambda^{n3, t1} = 25 Gb/s$ and type $t2$ at rate $\lambda^{n3, t2} = 20 Gb/s$, while node $n5$ will receive the two types of traffic with rates $\lambda^{n5, t1} = 15 Gb/s$ and $\lambda^{n5, t2} = 35 Gb/s$, respectively. Finally, let us assume that the network operator has set an upper bound on the \emph{power budget} to be used (i.e., the total amount of computational power) $P = 300 Gb/s$ and has defined in its SLA a tolerable latency for the two types of traffic, respectively, to the following values: $\tau_{t1} = 1 ms$ and $\tau_{t2} = 2 ms$.

\textcolor{black}{Under the above assumptions and constraints, the management component will solve the optimization problem, and will decide to offload part of the traffic from the two ingress nodes to an intermediate node as shown in Figure~\ref{toy-b}. More specifically, the management component will assign at ingress node $n3$ a wireless network capacity slice of $27 Gb/s$ (out of the total $50Gb/s$) to $t1$ and of $23 Gb/s$ to $t2$, while at ingress node $n5$ it will assign $22 Gb/s$ (out of the total $60Gb/s$) to $t1$ and $38 Gb/s$ to $t2$.} Moreover, it will assign a computation capacity $D_2$ to nodes $n3$ and $n5$ and $D_3$ to $n7$, while it will switch off the computation capacity of the other nodes. This will lead to a total computation capacity of $130 Gb/s$, which is well below the available computation capacity budget $P$. Given that $t1$ is the traffic type with the most demanding constraint in terms of latency, the management component decides to use the full $D_2$ capacity of $n3$ to process traffic $t1$ from $n3$. Applying the same strategy within node $n5$ would result in a waste of resources because the $t1$ traffic of $n5$ will take only $15 Gb/s$ of the available computation capacity, and the remaining one will not be sufficient to handle the expected total amount of $t2$ traffic. Since moving the $t1$ traffic of one hop would still allow the system to fulfill the SLA, the decision will be then to configure the network to route such traffic to $n7$.
The reason for choosing $n7$ is mainly because it is one of the nearest neighbors of both $n3$ and $n5$ (with 2 hops to $n3$ and 1 hop to $n5$) and, with its $D_3$ capacity, will be able to handle both $t2$ traffic from $n3$ and $t1$ traffic from $n5$. Specifically, the percentage of computation capacity of $n7$ allocated for $n3,t2$ and $n5,t1$ is $64\%$ and $36\%$, respectively. $t2$ traffic from $n5$ is, instead, processed locally at $n5$ itself.

Let us now assume that the management component observes a change in the $\lambda^{n5,t2}$ traffic rate, which increases to $\lambda^{n5,t2} = 40 Gb/s$. It will then run again the optimization algorithm that will output the configuration illustrated in Figure~\ref{toy-c}.
The slicing of the wireless network capacity for ingress node $n3$ will not vary, while \textcolor{black}{the total wireless capacity at ingress node $n5$ will be redistributed as follows:} a slice of $42 Gb/s$ will be assigned to $t2$ and, as a consequence, a slice of $18 Gb/s$, smaller than before, to $t1$. Moreover, \textcolor{black}{the computation capacity to be allocated to each node will be recomputed.}
Capacity $D_2$ will be allocated to $n3$, which will process $t1$ locally, and $D_1$ will be allocated to the neighbor node $n4$, which will handle the $t2$ traffic from $n3$. Capacity $D_3$ will be allocated to $n5$ to process $t2$ locally and, finally, $D_1$ will be allocated to $n7$ to process $t1$ incoming from $n5$. Both ingress nodes will \textcolor{black}{offload part of their traffic to} the nearest 1-hop neighbor and the total computation capacity will be equal to $150 Gb/s$.

Notice that, by manually analyzing the initial configuration of Figure~\ref{toy-b}, we may think that a better solution \textcolor{black}{to the increase of $\lambda^{n5,t2}$} would be to simply increase the computation capacity of $n5$ to $D_3$ as in this way the network \textcolor{black}{configuration} will remain almost the same as before and the total computation capacity will be $140 Gb/s$, smaller than the one of Figure~\ref{toy-c}. However, a more in-depth analysis shows that, even if this solution is certainly feasible, it is less optimal than the one of Figure~\ref{toy-c} in terms of $t1$ total latency\textcolor{black}{, which, as described in detail in Section~\ref{section-problemformulation}, depends on both the wireless network latency and the outsourcing latency. The main reason for this increase in the latency is that traffic $t1$ from node $n5$ will suffer from a larger latency in the wireless ingress network due to a smaller allocated slice, and also from a relatively high latency due to the traffic computation on $n7$. According to the model we formalize in the next section, the total latency for $t1$ in this case is $0.72 ms$, while, as it will be shown in Section~\ref{subsection-results-small}, it is $0.47 ms$ in the case of Figure~\ref{toy-c}, thanks to the fact that node $n5$ has the computation capacity of $n7$ entirely dedicated to the $t1$ traffic it introduces in the network.} %In Section~\ref{section-problemformulation} we show how such values are computed and, in general, the optimization model that computes the optimal allocation of computational and network resources as well as the optimal routing paths. 

\section{Problem Formulation}
\label{section-problemformulation}

\textcolor{black}{In this section we provide the mathematical formulation of our \textit{Joint Planning and Slicing of mobile Network and edge Computation resources} (JPSNC) model.}
Table~\ref{notation} summarizes the notation used throughout this section.
For brevity, we simplify expression $\forall n\in\mathcal{N}$ as $\forall n$, and apply the same rule to other set symbols like $\mathcal{E},\mathcal{K},\mathcal{L}$, etc. throughout the rest of this paper unless otherwise specified.

\setlength{\textfloatsep}{5pt}
\begin{table}[!h]
    \caption{Summary of used notations.}
    \label{notation}
    \centering
    \begin{tabular}{@{\hspace{2pt}}l@{\hspace{4pt}}l@{\hspace{2pt}}}
        \toprule
        Parameters & Definition \\
        \midrule
        $\mathcal{N}$ & Set of traffic types \\
        $\mathcal{E}$ & Set of edge nodes in the edge networks \\
        $\mathcal{K}$ & Set of ingress nodes, where $\mathcal{K} \subseteq \mathcal{E}$ \\
        $\mathcal{L}$ & Set of directed links in the networks \\
        $B_{ij}$ & Bandwidth of the link from node $i$ to $j$, where $(i,j)\in\mathcal{L}$\\
        $C_k$ & Network capacity of ingress edge node $k \in \mathcal{K}$ \\
        $D_a$ & Levels of computation capacities ($a\in\mathcal{A}=\{1,2,3 \ldots\}$) \\
        $P$ & Planning budget of computation capacity \\
        $\lambda^{kn}$ & User traffic rate of type $n$ in ingress node $k$\\
        $\tau_n$ & Tolerable delay for serving the total traffic of type $n$ \\
        $\kappa_i$ & Cost of using one unit of computation capacity on node $i$ \\
        $w$ & Weight to balance among total latency and operation cost \\
        \midrule
        Variables & Definition \\
        \midrule
        $c^{kn}$ & Slice of the network capacity for traffic $kn$ \\
        $b^{kn}_i$ & \textcolor{black}{Whether traffic $kn$ is processed on node $i$ or not}\\
        $\alpha^{kn}_i$ & Percentage of traffic $kn$ processed on node $i$ \\
        $\beta^{kn}_i$ & Percentage of $i$'s computation capacity sliced to traffic $kn$ \\
        $\delta^a_i$ & Decision for planning computation capacity on node $i$ \\
        $\mathcal{R}^{kn}_i$ & Set of links for routing the traffic piece $\alpha^{kn}_i$ from $k$ to $i$ \\
        \bottomrule
    \end{tabular}
\end{table}
The goal of our formulation is to minimize a weighted sum of the total latency and network operation cost for serving several types of user traffic under the constraints of users' maximum tolerable latency and network planning budget. This allows the network operator to fine tune its needs in terms of quality of service provided to its users and cost of the planned network. Different types of traffic, with heterogeneous requirements, need to be accommodated, and may enter the network from different ingress nodes. 

In the following, we first focus on the network planning issue and its related cost, as well as on the traffic routing issue, and then detail all components that contribute to the overall latency experienced by users, which we capture in our model.

\subsection{Network Planning and Routing}

{\it\textbf{Network Planning}}:
We assume that, in each edge node, some processing capacity can be made available, thus enabling MEC capabilities.
This action will result in an operation cost that will increase at the increase of the amount of processing capacity.
To model more closely real network scenarios, we assume that only a discrete set of capacity values can be chosen by the network operator and made available. Therefore, we adopt a piecewise-constant function $S_i$ for the processing capacity of an edge node, in line with~\cite{santoyo2018latency}. This is defined as:
\begin{equation}
    S_i = \sum\nolimits_{a\in\mathcal{A}} \delta^a_i D_a,\; \forall i, \label{eq_si}
\end{equation}
where $D_a$ is a capacity level ($a\in\mathcal{A}$) and $\delta^a_i \in \{0,1\}$ is a binary decision variable for capacity planning, satisfying the following constraint (only one level of capacity can be made available on a node, including zero, i.e., no processing capability):
\begin{equation}
    \sum\nolimits_{a \in \mathcal{A}} \delta^a_i = 1 - \delta^0_i,\; \forall i, \label{con_cap}
\end{equation}
where $\delta^0_i$ is a binary variable that indicates whether node~$i$ has currently available some computation power or not.
This constraint implies that $S_i$ can be set as either 0 (no computation power) or exactly one capacity level, $D_a$.

To save on operation costs, in the case an edge node is not supposed to be exploited to process some traffic, then no processing capacity is made available on it.
We introduce binary variable $b^{kn}_i$ to indicate whether traffic $kn$ is processed on node $i$ (we will use the expression ``traffic $kn$'' in the following, for brevity, to indicate the user traffic of type $n$ from ingress point $k$). Then the following constraint should be satisfied:
\begin{equation}
    b^{kn}_i \leqslant 1-\delta^0_i \leqslant \sum_{k' \in \mathcal{K}}\sum_{n' \in \mathcal{N}} b^{k' n'}_i,\; \forall k, \forall n, \forall i, \label{con_z_b}
\end{equation}

We also consider a total planning budget, $P$, for the available computation capacity, introducing the following constraint:
\begin{equation}
    \sum\nolimits_{i \in \mathcal{E}} S_i \leqslant P. \label{con_P}
\end{equation}
Then, the total operation cost can be expressed as:
\begin{equation}
    J = \sum\nolimits_{i \in \mathcal{E}} \kappa_i S_i, \label{eq_J}
\end{equation}
where $\kappa_i$ is the cost of using one unit of computation capacity (in the example of Section~\ref{section-systemoverview} this will be $1 Gb/s$) on node $i$.

{\it\textbf{Network Routing}}:
We assume that each type of traffic can be split into multiple pieces only at its ingress node. Each piece can then be offloaded to another edge computing node independently of the other pieces, but it cannot be further split (we say that each piece is \emph{unsplittable}).

\textcolor{black}{
The reason for using \emph{unsplittable} routing in our optimization model is twofold:
first of all, network slicing in the 5G architecture should be performed in an isolated manner for security and privacy reasons, especially for specific customer services \cite{IETFNovember2020,FCC2018,IETFJuly2017}. Hence, considering unsplittable routing is, in practice, reasonable. 
Second, this choice is beneficial to reduce the complexity of our optimization problem since splitting the traffic across edge nodes could significantly increase the complexity without a strong justification, especially for the kind of user services mentioned above (with security and privacy requirements).
In general, we consider that the user traffic or the virtual operator traffic passes through a predefined set of nodes along a given (unique) path, like a given chain of nodes providing services to the user/virtual provider.}

Each link $l\in\mathcal{L}$ may carry different traffic pieces, $\alpha^{kn}_i$ %(remind that $\alpha^{kn}_i$ is the fraction of traffic $kn$ to be processed at node $i$).
(we denote by $\alpha^{kn}_i$ the percentage of traffic $kn$ processed at node $i$, and with $\beta^{kn}_i$ the percentage of computation capacity $S_i$ sliced for traffic~$kn$).
Then, the traffic flow $kn$ on $l$, $f^{kn}_l$, can be expressed as the sum of all pieces of traffic that pass through such link:
\begin{equation}
    f^{kn}_l = \sum_{i\in\mathcal{E}:\; l\in\mathcal{R}^{kn}_i}\alpha^{kn}_i,\;\forall k,\forall n,\forall l, \label{con_f}
\end{equation}
where $\mathcal{R}^{kn}_i \subset \mathcal{L}$ denotes a routing path (set of traversed links) for the traffic piece $\alpha^{kn}_i\lambda^{kn}$ from ingress $k$ to node~$i$.
The following constraint ensures that the total traffic on each link does not exceed its capacity:
\begin{equation}
    B_{ij} > \sum\limits_{k\in\mathcal{K}}\sum\limits_{n\in\mathcal{N}} f^{k n}_{ij}\lambda^{k n},\; \forall (i,j) \in \mathcal{L}.\label{con_link}
\end{equation}

The traffic flow conservation constraint is enforced by the following constraint:
\begin{equation}
    \sum_{j\in\mathcal{I}_i} f^{kn}_{ji} - \sum_{j\in\mathcal{O}_i} f^{kn}_{ij}=
    \left\{\begin{array}{@{}l@{}l@{}}
        \alpha^{kn}_i-1, &\; \text{if}\;i=k,\\
        \alpha^{kn}_i,   &\; \text{otherwise},
    \end{array}\right.\\
    \,\forall k, \forall n, \forall i,\label{con_fcc}
\end{equation}
where $\mathcal{I}_i=\{j\in\mathcal{E}\,|\,(j,i)\in\mathcal{L}\}$ and $\mathcal{O}_i=\{j\in\mathcal{E}\,|\,(i,j)\in\mathcal{L}\}$ are the set of nodes connected by the incoming and outgoing links of node $i$, respectively. The fulfillment of this constraint guarantees \emph{continuity} of the routing path. Moreover, the routing path $\mathcal{R}^{kn}_i$ should be \emph{acyclic}. 

\textcolor{black}{To sum up, we consider at each ingress node aggregates of traffic, each corresponding to a type of traffic/service; an aggregate of type $n$ at ingress node $k$ has a total rate $\lambda^{kn}$. We split such aggregate (only) at ingress node $k$ into several pieces $\{\alpha^{kn}_i\lambda^{kn}, i \in \mathcal{E}\}$, where $\alpha^{kn}_i$ represents the percentage of traffic $kn$ processed at node $i$. We then determine for each piece $\alpha^{kn}_i\lambda^{kn}$ a single path $\mathcal{R}^{kn}_i$ between ingress node $k$ and edge node $i$. Note that $\alpha^{kn}_i$ may be null for some edge nodes $i$ and the selection of processing nodes depends, among other factors, on latency constraints specified in the next section since not all nodes are used to process a given traffic.
In practice, since we deal with large aggregates, each single demand inside the aggregate follows a single path, (since it largely ``fits'' in the fraction of traffic that follows a single path).}

\subsection{Latency Components}

The latency in each ingress edge node is modeled as the sum of the \emph{wireless network latency} and the \emph{outsourcing latency} which, in turn, is composed of the \emph{processing latency} in some edge cloud and then \emph{link latency} between edge clouds.

{\it\textbf{Wireless Network Latency}}:
We model the transmission of traffic in each user ingress point as an $M|M|1$ processing queue.
The \emph{wireless network latency} for transmitting the user traffic of type $n$ from ingress point $k$, 
denoted by $t^{kn}_W$, can therefore be expressed as:
\begin{equation}
    t^{kn}_{W} = \frac{1}{c^{kn} - \lambda^{kn}}, \; \forall k,\forall n, \label{eq_tw}
\end{equation}
\noindent
where $c^{kn}$ is the capacity of the network slice allocated for traffic $kn$ in the ingress edge network (a decision variable in our model) and $\lambda^{kn}$ is the traffic rate.
The following constraints ensure that the capacity of all slices does not exceed the total capacity $C_k$ of each ingress edge node, and $c^{kn}$ is higher than the corresponding $\lambda^{kn}$ value:
\begin{alignat}{1}
    & \sum\nolimits_{n \in \mathcal{N}} c^{kn} \leqslant C_k, \; \forall k, \label{c_net} \\
    & \; \lambda^{kn} < c^{kn}, \; \forall k, \forall n. \label{con_cn_n}
\end{alignat}

{\it\textbf{Processing Latency}}:
We assume that each type of traffic can be segmented and processed on different edge clouds,
and each edge cloud can slice its computation capacity to serve different types of traffic from different ingress nodes. As introduced before, we indicate with $\alpha^{kn}_i$ the percentage of traffic $kn$ processed at node $i$, and with $\beta^{kn}_i$ the percentage of computation capacity $S_i$ sliced for traffic~$kn$.
The processing of user traffic is described by an $M|M|1$ model.
Let $t^{kn,i}_P$ denote the \emph{processing latency} of edge cloud~$i$ for traffic $kn$.
Then, based on the computational capacity $\beta^{kn}_i S_i$ sliced for traffic $kn$, with an amount $\alpha^{kn}_i \lambda^{kn}$ to be served, $\forall k,\forall n,\forall i,\; t^{kn,i}_P$ is expressed as:
\begin{equation}
    t^{kn,i}_P = \left\{\begin{array}{cl}
        \frac{1}{\beta^{kn}_i S_i - \alpha^{kn}_i \lambda^{kn}}, & \text{if } \alpha^{kn}_i > 0, \\
        0, & \text{otherwise}.
    \end{array}\right.\label{eq_tp}
\end{equation}
In the above equation, when traffic $kn$ is not processed on edge cloud $i$, the corresponding value is $0$; at the same time, no computation resource of $i$ should be sliced to traffic $kn$ (i.e., $\beta^{kn}_i=0$). The corresponding constraint is written as:
\begin{equation}
    \left\{\begin{array}{ll}
        \alpha^{kn}_i \lambda^{kn} < \beta^{kn}_i S_i, & \text{if } \alpha^{kn}_i > 0,\\
        \alpha^{kn}_i = \beta^{kn}_i = 0, & \text{otherwise}.
    \end{array}\right.\label{con_4_tp}
\end{equation}
$\alpha^{kn}_i$ and $\beta^{kn}_i$ also have to fulfill the following consistency constraints:
\begin{align}
    &\sum\nolimits_{i \in \mathcal{E}} \alpha^{kn}_i = 1, \; \forall k, \forall n, \label{c_alpha}\\
    &\sum\nolimits_{k \in \mathcal{K}}\sum\nolimits_{n \in \mathcal{N}} \beta^{kn}_i \leqslant 1, \; \forall i. \label{c_beta}
\end{align}

{\it\textbf{Link Latency}}:
Let $t^{kn,i}_L$ denote the \emph{link latency} for routing traffic $kn$ to node $i$. In each ingress node, the incoming traffic is routed in a multi-path way, i.e., different types or pieces of the traffic may be dispatched to different nodes via different paths.
$\forall k,\forall n,\forall i$, $t^{kn,i}_L$ is defined as:
\begin{equation}
    t^{kn,i}_L =
    \left\{\begin{array}{@{}c@{}l@{}}
        \sum\limits_{l\in\mathcal{R}^{kn}_i}\frac{1}{B_l - \sum\limits_{k'\in\mathcal{K}}\sum\limits_{n'\in\mathcal{N}} f^{k' n'}_l\lambda^{k' n'}}, &\; \text{if } \alpha^{kn}_i > 0\,\&\, i \neq k,\\
        0, &\; \text{otherwise}.
    \end{array}\right.\label{eq_tl}
\end{equation}
Recall that $\mathcal{R}^{kn}_i$ is a routing path for the traffic piece $\alpha^{kn}_i\lambda^{kn}$ from ingress $k$ to node $i$.
The \emph{link latency} is accounted for only if a certain traffic piece is processed on node $i$ (i.e. $\alpha^{kn}_i > 0$) and $i\neq k$.

{\it\textbf{Total Latency}}:
Now we can define the \emph{outsourcing latency} for traffic $kn$, which depends on the longest serving time among edge clouds:
\begin{equation}
    t^{kn}_{PL} = \max_{i\in\mathcal{E}}\{t^{kn,i}_P + t^{kn,i}_L\}, \; \forall k, \forall n. \label{eq_t_pl}
\end{equation}
The latency experienced by each type of traffic coming from the ingress nodes, can therefore be defined as $t^{kn}_W + t^{kn}_{PL}$, and also should respect the tolerable latency requirement:
\begin{equation}
    t^{kn}_W + t^{kn}_{PL} \leqslant \tau_n, \; \forall k,\forall n. \label{con_tau}
\end{equation}
For each traffic type $n$, we consider the maximum value among different ingress nodes with respect to the wireless network latency and outsourcing latency, i.e., $\max_{k\in\mathcal{K}}\{t^{kn}_W + t^{kn}_{PL}\}$.
Then, we define the total latency as follows:
\begin{equation}
    T = \sum_{n\in\mathcal{N}}\max_{k\in\mathcal{K}}\{t^{kn}_W + t^{kn}_{PL}\}.\label{eq_T}
\end{equation}

{\color{black}
The way we model latency and delay is aligned with other approaches in the literature. 
The work of Ma et al.~\cite{ma2019cost} presents a system delay model which has the same components adopted in our paper; the communication delay in the wireless access is modeled as in our work (using an $M|M|1$-like expression). Moreover, this work also assumes that traffic is processed across a subset of computing nodes and the service time of edge hosts and cloud instances are exponentially distributed, hence the service processes of mobile edge and cloud can be modeled as $M|M|1$ queues in each time interval. The same assumption is made in~\cite{niu2015when}.
In \cite{chen2018computation}, the authors assume that both the congestion delay and the computation delay at each small-cell Base Station (by considering a Poisson arrival of the computation tasks) can be modeled as an $M|M|1$ queuing system; the work in
\cite{luong2018joint} assumes that the baseband processing of each Virtual Machine (VM) on each User Equipment packets can be described as an $M|M|1$ processing queue, where the service time at the VM of each physical server follows an exponential distribution.
Finally, the works \cite{li20175g,liu2017multiobjective,tang2017system,zhuang2016energy} also adopt similar choices concerning the delay modeling.
}

\vspace{-10pt}
\subsection{Optimization Problem - JPSNC}
\label{subection_OptimizationProblemJPSNC}
Our goal in the \textit{Joint Planning and Slicing of mobile Network and edge Computation resources} (JPSNC) problem is to minimize the total latency and the operation cost, under the constraints of maximum tolerable delay for each traffic type coming from ingress nodes and the total planning budget for making available processing-capable nodes:
\begin{alignat}{1}
    \mathcal{P}\zcal{0} : \min_{\substack{c^{kn},b^{kn}_i,\alpha^{kn}_i,\\\beta^{kn}_i,\delta^a_i,\mathcal{R}^{kn}_i}} & \; T + w J, \notag \\
    \text{s.t.} \quad\;\; & \; \eqref{eq_si} - \eqref{eq_T}, \notag
\end{alignat}
where $w \geq 0$ is a weight that permits to set the desired balance between the total latency and operation cost.
Problem $\mathcal{P}\zcal{0}$ contains both nonlinear and indicator constraints, therefore, it is a mixed-integer nonlinear programming (MINLP) problem, which is hard to be solved directly \cite{kannan1978computational}, as discussed in Section~\ref{subsection-reformulationShort}.

{
\color{black}
We observe that we can give priority to one component of the objective function (latency $T$ or operational cost $J$) with respect to the other by setting the weight $w$. This is obtained by setting $w$ such that (if $T$ is privileged) improving latency is preferred even if this increases the operational cost of the planned network at its maximum (a similar reasoning is applied if the cost $J$ is privileged over~$T$).

To this aim, we first compute the bounds for the values of $T$ and $J$ approximately as follows:

\begin{enumerate}
    \item $\min (\kappa_i) \cdot \sum {\lambda^{kn}} \leqslant J \leqslant \max (\kappa_i) \cdot P$; %    final range: $\approx [13.8, 30]$
    \item $|\mathcal{N}| \cdot ( \frac{1}{\max(C_k)} + \frac{1}{\max(D_a)} ) < T \leqslant \sum \tau_n$. %    final range: $\approx [0.2, 11]$
\end{enumerate}

For the lower bound of $J=\sum(\kappa_i \cdot S_i) \geqslant \min(\kappa_i) \cdot \sum(S_i)$, we observe that the total computation power should cover the total traffic rate, to avoid infeasibility, hence we have: $\min(\kappa_i) \cdot \sum{\lambda^{kn}}$.
For computing the bounds of $T$, we use its definition and the tolerable latency to get the upper bound, while for the lower bound, we use the definitions (wireless latency and computation latency, for link latency, we get 0 due to the lower bound) and let the denominators reach the maximum. 
The values of $w$ that enforce the desired priority in the optimization process can therefore be computed as $w_L = \frac{T_{min}}{J_{max}}$ and $w_U = \frac{T_{max}}{J_{min}}$.
}

\subsection{JPSNC Reformulation}
\label{subsection-reformulationShort}

\textcolor{black}{
Problem $\mathcal{P}\zcal{0}$ formulated in Section~\ref{section-problemformulation} cannot be solved directly and efficiently due to the following reasons:
\begin{itemize}
    \item We aim at identifying the optimal routing (the routing path $\mathcal{R}^{kn}_i$ is a variable in our model, since many paths may exist from each ingress node $k$ to a generic node $i$ in the network); furthermore, we must ensure that such routing is acyclic and ensures continuity and unsplittability of traffic pieces.
    \item Variables $\mathcal{R}^{kn}_i$ and $\alpha^{kn}_i$ are reciprocally dependent: to find the optimal routing, the percentage of traffic processed at each node $i$ should be known, and at the same time, to solve the optimal traffic allocation, the routing path should be known.
    \item The processing latency, defined in the previous sections, depends on three decision variables in our model and the corresponding formula~\eqref{eq_tp} is (highly) nonlinear.
    \item $\mathcal{P}\zcal{0}$ contains indicator functions and constraints, e.g. \eqref{eq_tp}, \eqref{con_4_tp}, \eqref{eq_tl}, which cannot be directly and easily processed by most solvers.
\end{itemize}
}
\textcolor{black}{
To deal with the above issues, we propose an equivalent reformulation of $\mathcal{P}\zcal{0}$ (called Problem $\mathcal{P}\zcal{1}$), which can be solved very efficiently with the Branch and Bound method. Moreover, the reformulated problem can be further relaxed and, based on that, we propose in the next section an heuristic algorithm which can get near-optimal solutions in a shorter computing time.
More specifically, in $\mathcal{P}\zcal{1}$, we first reformulate the processing latency and link latency constraints (viz., constraints~(\ref{eq_tp}) and~(\ref{eq_tl})), and we deal, at the same time, with the computation planning problem.
Then, we handle the difficulties related to variables $\mathcal{R}^{kn}_i$ and the corresponding routing constraints.
Appendix \ref{section-problemreformulation} contains all details about the problem reformulation.
Since some constraints are quadratic while the others are linear, $\mathcal{P}\zcal{1}$ is a mixed-integer quadratically constrained programming (MIQCP) problem, for which commercial and freely available solvers can be used, as we will illustrate in the numerical evaluation section.
}

\section{Heuristics}
\label{section-heuristics}

Hereafter, we illustrate our proposed heuristic, named \emph{Neighbor Exploration and Sequential Fixing} - NESF, which proceeds by exploring and utilizing the neighbors of each ingress node for hosting (a part of) the traffic along an \emph{objective descent direction}, that is, by trying to minimize the objective function (which, we recall, is a weighted sum of the total latency and operation cost). During each step where we explore potential candidates for computation offloading, we partially fix the main binary decision variables in the reformulated problem $\mathcal{P}\zcal{1}$ and then solve the so-reduced problem by using the Branch and Bound method. Our exploring strategy provides excellent results, in practice, achieving near-optimal solution in many network scenarios, as we will illustrate in the Numerical Results Section.

\begin{figure}[ht]
    \captionsetup{skip=3pt}
    \centering
    \includegraphics[width=0.80\linewidth]{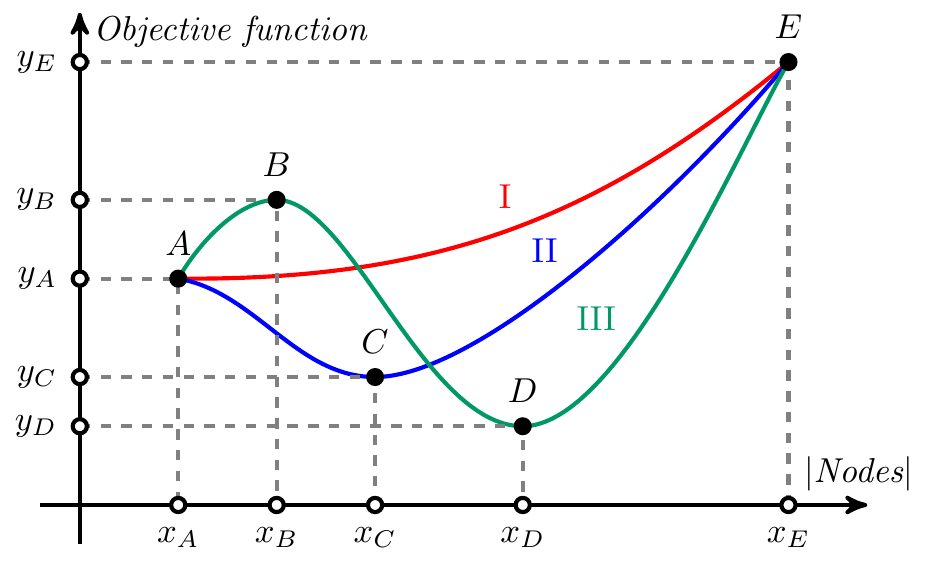}%
    \caption{Three typical variations of the objective function value versus the number of computing nodes made available.}
    \label{fig_path}
    \vspace{-6pt}
\end{figure}

The detailed exploring strategy is illustrated in Figure~\ref{fig_path}, which shows three typical variation paths of the objective function value versus the number of computing nodes made available in the network (note that these 3 trends are independent from each other, in the sense that either of them, or a combination of them, can be experienced in a given network instance). Point~$A$ represents the stage where a minimum required number of computing nodes ($x_A$) is opened to ensure the feasibility of the problem. For instance, if the ingress nodes can host all the traffic under all the constraints, $x_A=|\mathcal{K}|$. Point $E$ indicates the maximum number of computing nodes that can be made available in the network; any point above $x_E$ will violate the computation budget or tolerable latency constraints.

During the search phase of our heuristic, which is executed in Algorithms \ref{alg_attempt} and \ref{alg_planning}, detailed hereafter, we first try to obtain (or get as closer as possible to) point $A$ and the corresponding objective value $y_A$. If $A$ can not be found within the computation budget, the problem is infeasible. Otherwise, we continue to explore computation candidates from the $h$-hop neighbors of each ingress node, and allocate them to serve different types of traffic. The objective value is obtained by solving $\mathcal{P}\zcal{1}$ with new configurations of the decision variables.
The change of the objective value may hence exhibit one of the three patterns (I, II and III) illustrated in Figure~\ref{fig_path}.

The objective value increases monotonically in path I. In path II, it first decreases to point $C$ then increases to point~$E$; finally, path III shows a more complex pattern which has one local maximum point $B$ and one minimum point $D$. In case~I, the network system has just enough computation power to serve the traffic. Hence, adding more computation capacity to the system does not guarantee to decrease delay, while it will increase on the other hand computation costs. In case~II, few ingress nodes in the system may support a relatively high traffic load. Equipping some of their neighbors with more computation capabilities (with total amount less than $x_C$) can still decrease the total system costs. After point~$C$, the objective value shows a similar trend to case~I. In case~III, several ingress nodes may serve high traffic load. At the beginning, adding some computing nodes (with total amount less than $x_B$) may be not enough to decrease the delay costs to a certain degree, and this will also increase the total installation costs. After point $B$, the objective value varies like in case II and has a minimum at point $D$.

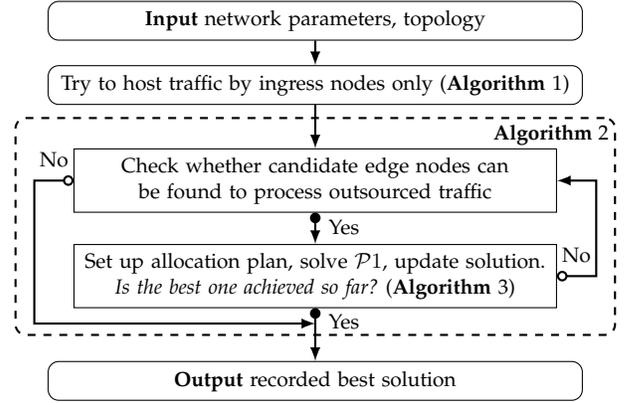
\begin{figure}[!t]
    \centering

\begin{tikzpicture}[%
    >=latex, %triangle 45,              % Nice arrows; your taste may be different
    start chain=going below,    % General flow is top-to-bottom
    node distance=4mm and 22mm, % Global setup of box spacing
    every join/.style={norm},   % Default linetype for connecting boxes
    scale = 0.82, transform shape,
    ]

\tikzset{
    % Node styles
    base/.style={draw, on chain, on grid, align=center, minimum height=4ex, minimum width=5ex, inner sep=4pt},
    proc/.style={base, rectangle, text width=22.5em},
    term/.style={proc, rounded corners, text width=25em},
    % Box style
    box/.style={draw, dashed, thick, rectangle, rounded corners, anchor=north},
    % Line styles
    norm/.style={thick, ->, draw},
    % Tip styles
    o/.tip={Circle[open,length=4pt]},
    */.tip={Circle[length=4pt]},
}

    % draw nodes
    \node [term] (n1) {\textbf{Input} network parameters, topology};
    \node [term, join] (n2) {Try to host traffic by ingress nodes only (\textbf{Algorithm} 1)};
    \node [proc, join, below=15.5mm of n2] (t1) {\textcolor{black}{Check whether candidate edge nodes can be found to process outsourced traffic}};
    \node [proc, join=by *->, below=15.5mm of t1] (t2) {Set up allocation plan, solve $\mathcal{P}\zcal{1}$, update solution.\\ \textit{\textcolor{black}{Is the best one achieved so far?}} (\textbf{Algorithm} 3)};%by exploration
    \path (t1.south) to node [pos=0.45, right=1mm] {Yes} (t2);
    \node [term, join=by *->, below=17mm of t2] (n3) {\textbf{Output} recorded best solution};
    \path (t2.south) to node [pos=0.26, right=1mm] {Yes} (n3);

    % draw lines
    \draw [thick, o->] (t2.east) -- node[above=1mm, pos=0.5] {No} ++(1.9em,0) |- (t1);
    \draw [thick, o->] (t1.west) -- node[above=1mm, pos=0.5] {No} ++(-1.9em,0) |- ([yshift=-2.5mm]t2.south);

    % draw dash box
    \node [box, minimum width=29em, minimum height=10.5em, shift={(0mm,-2.0mm)}] (b1) at (n2.south) {};
    \node [below left] (a2) at (b1.north east) {\textbf{Algorithm} 2};

\end{tikzpicture}

    \caption{Flowchart of our NESF heuristic.}
    \label{alg_flow}
    \vspace{-6pt}
\end{figure}

To summarize, our heuristic aims at reaching the minimum points $A$ (I), $C$ (II) and $D$ (III) in Figure \ref{fig_path}, and its flowchart is shown in Figure \ref{alg_flow}.
The main idea behind Algorithm~\ref{alg_attempt} is to check whether the ingress nodes can host all the traffic without activating additional MEC units, thus saving some computation cost.
Algorithm \ref{alg_searching} aims at searching the $h$-hop neighbors of each ingress node for making them process part of the traffic (the outsourced traffic), while Algorithm \ref{alg_planning} aims at setting up the allocation plan for outsourced traffic and try to solve $\mathcal{P}\zcal{1}$ to obtain the best solution.
The three proposed algorithms are described into detail in the following subsections. The definition of the new notation introduced in these algorithms is summarized for clarity in Table \ref{notation_alg}.

\begin{table}[b!] % [b!]
    \centering
    \captionsetup{skip=3pt}
    \caption{Notations used in the algorithms.}
    \setlength{\tabcolsep}{3pt}
    \begin{tabular}{ll}
        %\multicolumn{3}{c}{Parameter settings.}\\
        \toprule
        Notation & Definition\\
        \midrule
        $S^e_k$ & \emph{Estimated} available computation of ingress node $k\in\mathcal{K}$ \\
        $\mathcal{K}^u$ & Ingress nodes that cannot host all traffic ($S^e_k \leqslant 0$) \\
        $H$ & Maximum searching depth of our heuristic \\
        $\mathcal{G}^{h}_k$ & $h$-hop neighbors ($h\leqslant H$) of ingress node $k\in\mathcal{K}$ \\
        $Q_k$ & Candidates for computing traffic from ingress node $k\in\mathcal{K}$ \\
        $S^o_k$ & \emph{Overall} computation of ingress node $k\in\mathcal{K}$ \\
        $S^l_i$ & Maximum \emph{left} computation of node $i\in\mathcal{E}$ \\
        $\mathcal{K}^b_i$ & Ingress nodes who \emph{booked} computation from node $i\in\mathcal{E}$ \\
        $d_{ik}$ & Count of hops from node $i$ to ingress node $k\in\mathcal{K}$ \\
        $O_{\mathcal{P}}$ & Objective function value of problem $\mathcal{P}$ \\
       \bottomrule
    \end{tabular}
    \label{notation_alg}
\end{table}

\subsection{Attempt of serving traffic without additional MEC units}
In Algorithm \ref{alg_attempt}, the main idea is to check whether ingress nodes can host all the traffic,
without using other MEC units in order to save both computation cost and latency.
To this end, we first individuate the subset of ingress nodes (denoted as $\mathcal{K}^u$) that cannot host all the traffic that enters the network through them.
This is done by checking if $S^e_k \;(= D_m-\sum_{n\in\mathcal{N}}\lambda^{kn}) \leqslant 0$ (lines 1-2), that is, if some computing capacity is still available or not at ingress nodes (recall that $D_m$ is the maximum computation capacity that can be made available). 
Then, if $\mathcal{K}^u \neq \varnothing$, $\forall k\in\mathcal{K}^u$, we try to find the set of its neighbor ingress nodes $k'\in [(\mathcal{K}-\mathcal{K}^u) \cap (\bigcup^H_h\mathcal{G}^h_k)]$ that can cover $S^e_k$ (i.e., $S^e_{k'}+S^e_k \!>\! 0$),
where $\mathcal{G}^h_k \subset \mathcal{E}$ is the set of node $k$'s $h$-hop neighbor nodes ($h=1, \ldots, H$).
If found, they are stored as candidates in a list, $Q_k$, ordered with increasing distance (hop count) from $k$ (lines 3-7).
If $\mathcal{K}^u = \varnothing$ or sufficient nodes in $Q_k$ have been found to process the extra traffic from $\mathcal{K}^u$ (line 9), then $\forall k\in\mathcal{K}^u$, the corresponding traffic is allocated to nodes in $Q_k$ starting from the top (choosing the closest ones) and repeatedly (covering all the traffic types), beginning with less latency to more latency-tolerant traffic.

This is implemented by setting the corresponding variables $b^{kn}_i$, $\delta^a_i$ and $\gamma^{kn,i}_l$ in $\mathcal{P}\zcal{1}$ to save the costs and also accelerate the algorithm.
Finally, $\mathcal{P}\zcal{1}$ with the fixed variables is solved by using \textit{Branch and Bound} method to obtain the solution (lines 10-11). If $\mathcal{P}\zcal{1}$ is feasible with these settings,
\textcolor{black}{
the objective value $O_{\mathcal{P}\zcal{1}}$ is stored to be used in the next searching and resource allocation phases of Algorithm~\ref{alg_planning}.
}
\begin{algorithm}[!t]
    \caption{\textit{Attempt of serving traffic with ingress nodes only}}
    \label{alg_attempt}
    \begin{algorithmic}[1]
        %\Statex {\color{blue} $\triangleright$ Attempt to host traffic by only ingress themselves}
        \State $S^e_k = D_m-\sum_{n\in\mathcal{N}}\lambda^{kn},\;\forall k\in\mathcal{K}$; % {\color{gray} $\triangleright$ Computation left}
        %\State Compute estimated left computation $S^e_k,\,\forall k\in\mathcal{K}$;
        \State $\mathcal{K}^u = \{k\in\mathcal{K}\mid S^e_k\leqslant 0\}$; % {\color{gray} $\triangleright$ Unable to host traffic}
        %\State Mark the ingress nodes that cannot host all traffic by $\mathcal{K}^u$;
        %\State $\mathcal{N}_k = \downarrow sort(\mathcal{N}),\, key=(\lambda^{kn},\, \tau_n),\; \forall k\in\mathcal{K}$;
        \State Compute $k$'s $h$-hop neighbors $\mathcal{G}^{h}_k,\,h\leqslant H,\,\forall k \in \mathcal{K}$;
        %\State Initialize candidates $Q_k=\{k\},\; \forall k\in\mathcal{K}$;
        \State $Q_k=\{k\},\; \forall k\in\mathcal{K}$, $O_t=-1$;
        %\If{$\mathcal{K}^u\neq\varnothing\;\&\;\sum_{k\in\mathcal{K}}S^e_k > 0$}
            \For {$k\in\mathcal{K}^u$} % {\color{gray} $\triangleright$ \textbf{if} $\mathcal{K}^u\neq\varnothing$}
                \State $\mathcal{X}=\{k'\in [(\mathcal{K}\!-\!\mathcal{K}^u) \cap (\bigcup^H_h\mathcal{G}^h_k)] \mid S^e_{k'}+S^e_k > 0\}$;
                %\State Find ingress nodes ($\mathcal{K}^n$) in $\mathcal{G}_k$ to host $S^e_k$;
                %\If{$\mathcal{K}^n\neq\varnothing$} add $\mathcal{K}^n$ to $Q_k$ by ascending hop;
                %\State Find ingress nodes in $\bigcup^H_h\mathcal{G}^h_k$ to \textit{cover} $S^e_k$;
                %\If{$found$} add them to $Q_k$ by ascending hop;
                %\EndIf
                \State $Q_k=Q_k \cup \mathcal{X}$, rank $Q_k$ by increasing hop count to $k$;
            \EndFor
        %\EndIf
        \State Rank $\mathcal{N}$ as $\mathcal{N}_k$ by descending $(\lambda^{kn}, \tau_n),\forall k\in\mathcal{K}$;
        \If{$\mathcal{K}^u=\varnothing$ \textbf{or} $\bigwedge_{k\in\mathcal{K}^u}(|Q_k| > 1)$}
            %\markcomment{1}{\color{gray} $\triangleright$ via turning \textit{on/off} corresponding $b^{kn}_i$ in $\mathcal{P}\zcal{1}$}
            %\State Turn off node set $\mathcal{E}-\bigcup_{k\in\mathcal{K}}Q_k$ in $\mathcal{P}\zcal{1}$;
            %\State Host $\mathcal{N}_k$ by $Q_k$ in order and repeatedly, $\forall k\in\mathcal{K}$;
            %\State Solve $\mathcal{P}\zcal{1}$ via \textbf{\textit{B\&B}} to obtain optimal solution;
            \State Allocate $Q_k$ to $\mathcal{N}_k$ in order and repeatedly, $\forall k\in\mathcal{K}$;
            \State Solve $\mathcal{P}\zcal{1}$ by \textbf{B\&B} to obtain obj. fct. value $O_{\mathcal{P}\zcal{1}}$;
            \If{$O_{\mathcal{P}\zcal{1}}>0$}
                $O_t=O_{\mathcal{P}\zcal{1}}$;
            \EndIf
        \EndIf
        %\State Record objective as $O_t=O_{\mathcal{P}\zcal{1}}$ (default -1);
    \end{algorithmic}
\end{algorithm}

\vspace{-2em}
\subsection{Neighbor search for computation candidates}

\begin{algorithm}[!t]
    \caption{\textit{Priority searching of computation candidates}}
    \label{alg_searching}
    \begin{algorithmic}[1]
        %\Statex {\color{blue} $\triangleright$ Search $h$-hop neighbors for outsourcing traffic}
        %\State $\mathcal{K}^s\!=\uparrow sort(\mathcal{K}), key=(S^e_k, -\lambda^{kn}),n = \argmax_{n'}{\tau_{n'}}$;
        \State Rank ingress nodes as $\mathcal{K}^s$ by ascending $(S^e_k, -\lambda^{kn})$;
        %\State $\hat{k} = \mathcal{K}^s(0),\;h_k = 1\;\forall k\in\mathcal{K},\;\mathcal{K}^f=\{\}$;
        \State $\hat{k} = \mathcal{K}^s(0),h_k = 1,S^o_k=S^e_k\,(\forall k\in\mathcal{K}),\mathcal{K}^b_i=\varnothing\,(\forall i\in\mathcal{E})$;
        \While{$|\bigcup_{k\in\mathcal{K}}Q_k| < \floor{\frac{P}{min{(D_a})}}$ \textbf{and} $\mathcal{K}^s\neq\varnothing$} %{\color{gray} $\triangleright$ Within the budget}
            %\State Pick nodes for $k\in\mathcal{K}^s$ from its $h_k$-hop neighbors;
            %\State Initialize temporary candidates $\mathcal{B}=\varnothing$;
            \State $\mathcal{B}=\varnothing$;
            \For{\textcolor{black}{$i\in( \mathcal{G}^{h_{\hat{k}}}_{\hat{k}} - [\mathcal{K} \cup Q_{\hat{k}}] )$}}
                %\State Compute $i$'s max left computation $S^l_i$;
                \State $S^l_i = D_m + \sum_{k\in\mathcal{K}^b_i}S^e_k$
                \If{$S^l_i + S^e_{\hat{k}} > 0$}
                    $\mathcal{B}=\mathcal{B} \cup \{i\}$;
                \EndIf
            \EndFor
            %\If{$\mathcal{B}=\varnothing$} {\color{gray} $\triangleright$ search next hop or ingress}
            %    \If{$h_{\hat{k}}<H$}
            %        %{\color{gray} $\triangleright$ Increase hop searching level}
            %        $h_{\hat{k}} = h_{\hat{k}} + 1$, \textbf{continue};
            %    \EndIf
            %    \State Add $\hat{k}$ to $\mathcal{K}^f$ and find next unfinished $k$;
            %    \If{\textit{found}} $\hat{k}=k$, \textbf{continue}; \textbf{else break};
            %    \EndIf
            \If{$\mathcal{B}=\varnothing$}
                \State $h_{\hat{k}}\!+\!+$, update $\mathcal{K}^s,\hat{k}$ when $h_{\hat{k}}\!>\!H$ and \textbf{continue}; 
            \EndIf
            \State Rank $\mathcal{B}$ by descending $(S^l_i,-d_{ik}\!:\!k\in\mathcal{K}^s)$, $\hat{\imath}=\mathcal{B}(0)$;
            \State $Q_{\hat{k}}=Q_{\hat{k}}\cup\{\hat{\imath}\},\;\mathcal{K}^b_{\hat{\imath}}=\mathcal{K}^b_{\hat{\imath}} \cup \{\hat{k}\},\;S^b=D_m$;
            %\markcomment{1}{\color{blue} Try to spread $\hat{\imath}$ to $k\in\mathcal{K}^s\backslash\{\hat{k}\}$ and update $Q_k,\,\mathcal{K}^b_{\hat{\imath}}$;}
            \For{$k\!\in\!\mathcal{K}^s\backslash\{\hat{k}\}$, \textbf{if} $(\hat{\imath}\!\in\!\bigcup^H_h\mathcal{G}^h_k)\,\&\,(S^b\!>\!\overline{\lambda^k})$}
                \State $Q_k=Q_k\cup\{\hat{\imath}\},\;\mathcal{K}^b_{\hat{\imath}}=\mathcal{K}^b_{\hat{\imath}} \cup \{k\},\;S^b=S^b-\overline{\lambda^k}$;
            \EndFor
            %\markcomment{1}{\color{blue} Update $S^o_k\;\forall k\in\mathcal{K}^b_{\hat{\imath}}$ and $\hat{k} = \argmin_{k\in\mathcal{K}^s} S^o_k$;}
            \State $S^o_k = S^o_k + (D_m + \sum_{k'\in\mathcal{K}^b_{\hat{\imath}}\cap\mathcal{K}^u-\{k\}}S^e_{k'}),\;\forall k\!\in\!\mathcal{K}^b_{\hat{\imath}}$;
            \State $\hat{k} = \argmin_{k\in\mathcal{K}^s} S^o_k$;
            \If{$S^o_{\hat{k}} \leqslant 0$} \textbf{continue}; \textbf{else} $skip := (S^o_{\hat{k}}\! \leqslant\! r D_m)$;
            \EndIf
            %\State \textit{Set up allocation plan and solve $\mathcal{P}\zcal{1}$} ({\bf Algorithm} \ref{alg_planning});
            \State Run ({\bf Algorithm} \ref{alg_planning}) to obtain $O_t$;
        \EndWhile
        \State \textbf{Return} $O_t$; % {\color{gray} $\triangleright$ updated in {\bf Algorithm} \ref{alg_planning}}
    \end{algorithmic}
\end{algorithm}

This section describes Algorithm~\ref{alg_searching}, upon which Algorithm \ref{alg_planning} is based to provide the final solution.
Algorithm~\ref{alg_searching} proceeds as follows.
We first assign a rank (or a priority value) to each ingress node taking into account the amount of incoming traffic and the computation capacity.
Then, we handle the outsourced traffic offloading task (i.e., choose the best subset of computational nodes) starting from the ingress node with the highest priority.

In more detail, set $\mathcal{K}^s$ is set $\mathcal{K}$ sorted by the ascending value of the tuple $(S^e_k, -\lambda^{kn})$, i.e., the ingress node with the lowest estimated available (left) computation $S^e_k$ and the higher amount of traffic of type $n$ has the highest rank/priority in our Algorithm 2, where $n$ represents the traffic type having the maximum tolerable latency (lines 1-2).
The process of determining the best subset of computation nodes for processing the outsourced traffic of each ingress node is executed \textit{hop-by-hop}, starting with ingress node $\hat{k} = \mathcal{K}^s(0)$, until any one of the following three conditions is satisfied:\\
(1) the number of computation nodes opened for processing traffic exceeds the maximum budget $\floor{\frac{P}{min{(D_a})}}$, or\\
(2) all ingress nodes are completely scanned (line 3), or\\
(3) the algorithm could not improve further the solution (Algorithm \ref{alg_planning}, lines 8, 10).

In the searching phase, we first try to identify the set of temporary candidate computation nodes $\mathcal{B}$ for ingress~\textcolor{black}{$\hat{k}$ ($\mathcal{B} \subseteq (\mathcal{G}^{h_{\hat{k}}}_{\hat{k}} - [\mathcal{K} \cup Q_{\hat{k}}])\,)$}, by checking if the maximum available computation capacity of $i \in \mathcal{B}$, $S^l_i$ could help $\hat{k}$ to cover $S^e_{\hat{k}}$ (lines 4-7).
$S^l_i$ is computed as the difference between $i$'s maximum installable computation capacity $D_m$ and the total computation booked from $i$ by ingress nodes in $\mathcal{K}^b_i\subseteq\mathcal{K}$, i.e., $\sum_{k\in\mathcal{K}^b_i}S^e_k$, where $\mathcal{K}^b_i$ is the set of ingress nodes that booked computation from node $i$.
If $\mathcal{B}=\varnothing$, we increase the number of hops $h_{\hat{k}}$ for ingress $\hat{k}$.
If not (we are done with $\hat{k}$), we move to the next ingress node in the set $\mathcal{K}^s$ (lines 8-9).

At this point we rank $\mathcal{B}$ by descending values of tuple $(S^l_i,-d_{ik}\!:\!k\in\mathcal{K}^s)$, where $d_{ik}$ is the count of hops from node $i$ to ingress node $k\in\mathcal{K}^s$.
The first computation node $\hat{\imath}$ is selected as the one to compute the traffic of $\hat{k}$, and $\hat{k}$ is added into the corresponding set $\mathcal{K}^b_{\hat{\imath}}$.
To make full use of computation node $\hat{\imath}$, we further spread it to help other ingress nodes $\mathcal{K}^s\backslash\{\hat{k}\}$, if $\hat{\imath}$ is their neighbor within $H$ hops and has sufficient computation budget (lines~10-13).
\textcolor{black}{
Then, given such computation node $\hat{\imath}$ and for each ingress node $k$, we update the value of the overall computation, $S^o_k$, due to the full use of computation nodes $\hat{\imath}$ (line~14). Hence, ingress~$k$ with the minimum support $S^o_k$ will be chosen as the next searching target and Algorithm~\ref{alg_searching} continues as follows.
}

The next searching target $\hat{k}$ is set to $k\in\mathcal{K}^s$ with the minimum $S^o_k$ value (lines 15-16).
If $S^o_{\hat{k}}\leqslant 0$, this means that the current computation configuration could not host all the traffic; hence, the algorithm will go back to the \textit{while} loop and \textit{continue} to the next searching.
Otherwise, we set a flag $skip := (S^o_{\hat{k}}\! \leqslant\! r D_m)$ where $r$ is set to a small value (i.e., $0.1$).
\textcolor{black}{If $skip$ is \textit{true}, it indicates that $\hat{k}$ has a high traffic load, and this may cause the processing latency to increase. This flag is used in Algorithm \ref{alg_planning}. In fact, this step implements the strategy of skipping point $B$ to avoid the local minimum (point $A$) in path III shown in Figure \ref{fig_path}.}
Finally, based on $Q_k$, we run Algorithm \ref{alg_planning} to obtain the objective value $O_t$ and the corresponding solution.

\vspace{-1em}
\subsection{Resource Allocation and Final Solution}
In Algorithm \ref{alg_planning},
we first relax problem $\mathcal{P}\zcal{1}$ to $\tilde{\mathcal{P}}\zcal{1}$, replacing binary variables $b^{kn}_i$, $\delta^a_i$ and $\gamma^{kn,i}_l$ with continuous ones.
Given the set $Q_k$ (by Algorithm~\ref{alg_searching}) of candidate computation nodes for processing the outsourced traffic of ingress node~$k$, the goal is to allocate node $k$'s different traffic types to the computation nodes in $Q_k$ starting with the traffic with the most stringent constraint in terms of latency. Unused computation nodes are turned off.
These two steps (lines 1-2) provide a partial guiding information and also an acceleration for solving the relaxed problem, thus obtaining quite fast the relaxed optimal values of $\tilde{b}^{kn}_i$.

If $\tilde{\mathcal{P}}\zcal{1}$ is infeasible ($O_{\tilde{\mathcal{P}}\zcal{1}}<0$), we check whether both the previous best solution exists ($O_t>0$) and the algorithm does not $skip$. If yes, the searching process breaks and returns $O_t$ (line 10).
Otherwise, the algorithm will continue searching to avoid getting stuck in a local optimum point in path III (see Figure \ref{fig_path}), according to the following.

Hence, if $\tilde{\mathcal{P}}\zcal{1}$ is feasible (line~3), the obtained $\tilde{b}^{kn}_i$ value can be regarded as the probability of processing traffic $kn$ at node $i$. Based on this, for each ingress $k$, we rank the candidates in descending order of the probabilities $\sum_{n\in\mathcal{N}}\tilde{b}^{kn}_i$. Then we revert to the original problem $\mathcal{P}\zcal{1}$, set the upper bound for $\mathcal{P}\zcal{1}$ \textcolor{black}{if possible}, allocate the candidates to host all types of traffic in order and repeatedly for each ingress node, and also turn off the unused nodes (lines 5-7).
\textcolor{black}{
By solving~$\mathcal{P}\zcal{1}$, we obtain the current solution and compare it with the previous best one ($O_t$). If the solution gets worse, the whole searching process breaks out and returns the recorded best result (line 8). Otherwise (if the solution is improving), the current solution is updated as the best one and the searching process continues.
}
\begin{algorithm}[!t]
    \caption{\textit{Allocating resources and obtaining the solution}}
    \label{alg_planning}
    \begin{algorithmic}[1]
        %\markcomment{0}{\color{blue} $\triangleright$ Set up allocation plan and obtain the best solution}
        \State Relax $b^{kn}_i,\delta^a_i,\gamma^{kn,i}_l$ to continuous ones ($\mathcal{P}\zcal{1}\rightarrow\tilde{\mathcal{P}}\zcal{1}$);
        %\If{$O_t>0$} set $O_t$ as $\tilde{\mathcal{P}}\zcal{1}$'s \textit{upper bound};
        %\EndIf
        %\State Turn off node set $\mathcal{E}-\bigcup_{k\in\mathcal{K}}Q_k$ in $\tilde{\mathcal{P}}\zcal{1}$;
        %\For{$k\in\mathcal{K}$, \textbf{if} $|Q_k|\leqslant|\mathcal{N}|$}
        %    \State Host $\mathcal{N}_k$ by $Q_k$ in $\tilde{\mathcal{P}}\zcal{1}$ in order and repeatedly;
        %\EndFor
        %\State Solve $\tilde{\mathcal{P}}\zcal{1}$ via \textbf{\textit{B\&B}} to obtain optimal $\tilde{b}^{kn}_i$;
        \State Allocate $Q_k$ to $\mathcal{N}_k$ \textit{partially} and solve $\tilde{\mathcal{P}}\zcal{1}$ to obtain $\tilde{b}^{kn}_i$;
        \If{$O_{\tilde{\mathcal{P}}\zcal{1}}>0$} % {\color{gray} $\triangleright$ $\tilde{\mathcal{P}}\zcal{1}$ is Feasible}
            %\State $Q^s_k = \downarrow sort(Q_k),\; key=\sum_{n\in\mathcal{N}}\tilde{b}^{kn}_i,\; \forall k\in\mathcal{K}$;
            \State Rank candidates as $Q^s_k$ by descending $\sum_{n\in\mathcal{N}}\tilde{b}^{kn}_i$;
            \State Revert to the original problem $\mathcal{P}\zcal{1}$;
            \If{$O_t>0$} set $O_t$ as $\mathcal{P}\zcal{1}$'s \textit{upper bound};
            \EndIf
            %\State Turn off node set $\mathcal{E}-\bigcup_{k\in\mathcal{K}}Q_k$ in $\mathcal{P}\zcal{1}$;
            %\State Host $\mathcal{N}_k$ by $Q^s_k$ in order and repeatedly, $\forall k\in\mathcal{K}$;
            %\State Solve $\mathcal{P}\zcal{1}$ via \textbf{\textit{B\&B}} to obtain optimal solution;
            \State Allocate $Q^s_k$ to $\mathcal{N}_k$ and solve $\mathcal{P}\zcal{1}$;
            \If{$0\!<\!O_t\&(O_t\!<\!O_{\mathcal{P}\zcal{1}}||O_{\mathcal{P}\zcal{1}}\!<\!0)\&\overline{skip}$} \textbf{break};
            \EndIf
            \If{$0\!<\!O_{\mathcal{P}\zcal{1}}\&(O_{\mathcal{P}\zcal{1}}\!<\!O_t||O_t\!<\!0)$} $O_t\!=\!O_{\mathcal{P}\zcal{1}}$;
            \EndIf
        \ElsIf{$O_t>0\;\&\;\overline{skip}$} \textbf{break};
        \EndIf
    \end{algorithmic}
\end{algorithm}

\vspace{-1em}
\subsection{Summary and Acceleration Technique}
Essentially, the proposed heuristic described in the above subsections exploits the $\mathcal{P}\zcal{1}$ formulation limiting the search space only to the nodes that are within a limited number of hops $h<H$ from the ingress nodes. We expect this is a realistic assumption based on the consideration that the main purpose of edge networks is to keep the traffic as close as possible to the ingress nodes and, therefore, to the users. Thanks to this approach, we are able to make the $\mathcal{P}\zcal{1}$ problem more tractable and solvable in a short time even in the case of complex edge networks (see Section~\ref{section-numericalresults}). 

We can further improve the solution time by eliminating from the problem formulation all unneeded variables. 
In particular, we modify $\mathcal{P}\zcal{1}$ by adding a scope $k$ (where $k$ is the ingress node) to $\mathcal{E}$ and $\mathcal{L}$. $\mathcal{E}_k \subseteq \mathcal{E}$ represents the set of $h$-hop neighbor nodes ($h\leqslant H$) of $k$ and $\mathcal{L}_k \subseteq \mathcal{L}$ the set of links inside this neighborhood. This way, the solver will be able to skip all variables outside the considered $k$ scope, thus reducing the time needed to load, store, analyze and prune the problem. Such modification does not change the result produced by the heuristic but it results in a consistent improvement (up to 1 order of magnitude) in the computing time needed to obtain the solution in our numerical analysis.

\section{Evaluation}
\label{section-numericalresults}
The goal of this evaluation is to show that: i) our $\mathcal{P}\zcal{1}$ model offers an appropriate solution to the edge network optimization problem we have discussed in this paper, ii) our \emph{NESF} heuristic computes a solution which is aligned with the optimal one, and iii) when compared with two benchmark heuristics, \emph{Greedy} and \emph{Greedy-Fair}, \emph{NESF} offers better results within similar ranges of computing time.

Consistently, the rest of this section is organized as follows:
Section~\ref{subsection-benchmark} describes the heuristics we have compared with; \textcolor{black}{Section~\ref{subsec_network_topo} presents the network topologies we have considered in the experiments;} Section~\ref{subsection-setup} describes the setup for our experiments; Section~\ref{subsection-results-small} discusses about optimal solution and the results obtained by the heuristics in the small network scenario presented in Section~\ref{section-systemoverview}; Section~\ref{subsection-results-large} analyzes the results achieved by the heuristics when the network parameters vary; Finally, Section~\ref{subsection-time} discusses about the computing time needed to find a solution.

\subsection{Benchmark Heuristics}
\label{subsection-benchmark}

We propose two benchmark heuristics, based on a greedy approach, which can be naturally devised in our context:

\emph{Greedy}:
With this approach, each ingress node uses its neighbor nodes computation facilities to guarantee a low overall latency for its incoming traffic.
Hence, each ingress node first tries to locally process all incoming traffic. %one type by one type,
If its computation capacity is sufficient, a feasible solution is obtained; otherwise, the extra traffic is split and outsourced to its 1-hop neighbors, and so on, until it is completely processed (if possible).

\emph{Greedy-Fair}: It is a variant of \textit{Greedy} which performs a sort of \textit{``fair" traffic offloading} on neighbor nodes. More specifically, it proceeds as follows:
1) compute the maximum number of available computing nodes, based on the power budget and the average computation capacity of a node;
2) divide such maximum number (budget) into $|\mathcal{K}|$ parts according to the ratio of the total traffic rate among ingress nodes, and choose for each ingress node the corresponding number of computing nodes from its nearest $h$-hop neighbors. Each ingress node spreads its load on its neighbors proportionally to the corresponding distance ($\frac{1}{hop+1}$), for example, if the load is outsourced to two \textit{1-hop} neighbors, the ratio is $(1 : \frac{1}{2}: \frac{1}{2})$ = $(0.5 : 0.25: 0.25)$.

{\color{black}
\subsection{Network Topologies}\label{subsec_network_topo}

We experimented with our optimization approach using multiple network topologies. 

\subsubsection{Random graphs}
We exploited Erd\"os-R\'enyi random graphs~\cite{Erd1959On} by specifying the number of nodes and edges.
As the original Erd\"os-R\'enyi algorithm may produce disconnected random graphs with isolated nodes and components, to generate a connected network graph, we patched it with a simple strategy that connects isolated nodes to randomly sampled nodes (up to 10 nodes) in the graph.
We generated several kinds of topologies with different numbers of nodes and edges, shown in Figure~\ref{net_topo}, that span from a quasi-tree shape topology (Figure~\ref{topo_tree_shape}) to a more general, highly connected one with 100 nodes and 150 edges (Figure~\ref{topo_5f}).
\textcolor{black}{The structural information for all topologies is shown in Table~\ref{tab_topo}. All topology datasets are publicly available in our repository\footnote{\url{https://github.com/bnxng/Topo4EdgePlanning}}}.
These topologies can be considered representative of various edge network configurations where multiple edge nodes are distributed in various ways over the territory. Due to space constraints, in the following we present and discuss the results obtained for a representative topology, i.e., the one in Figure~\ref{topo_5e}, as well as those for the small topology of Figure \ref{fig-toyExample}, used to compare our proposed heuristics to the optimal solution. The full set of results is available online\footnote{\label{note_sup}\url{http://xiang.faculty.polimi.it/files/SupplementaryResults.pdf}}.

\begin{table}[!h]
    \captionsetup{skip=3pt}
    \caption{Structural information of the topologies used in the experiments.}
    \label{tab_topo}
    \setlength{\tabcolsep}{1.5pt}
    \centering
    \begin{tabular}{c|c|c|c|c|c}
        \toprule
        Topology & \#Node & \#Edge & \#Ingress & Degree (Min, Max, Avg) & Diameter\\
        \midrule
        \emph{10N20E}          & $10 $ & $20 $ & $2$ & $(3.0,\; 5.0,\; 4.0)$ & $3 $ \\
        \emph{20N30E}          & $20 $ & $30 $ & $3$ & $(1.0,\; 5.0,\; 3.0)$ & $6 $ \\
        \emph{40N60E}          & $40 $ & $60 $ & $3$ & $(1.0,\; 7.0,\; 3.0)$ & $8 $ \\
        \emph{50N50E}          & $50 $ & $50 $ & $3$ & $(1.0,\; 4.0,\; 2.0)$ & $15$ \\
        \emph{60N90E}          & $60 $ & $90 $ & $3$ & $(1.0,\; 6.0,\; 3.0)$ & $7 $ \\
        \emph{80N120E}         & $80 $ & $120$ & $3$ & $(1.0,\; 6.0,\; 3.0)$ & $9 $ \\
        \emph{100N150E}        & $100$ & $150$ & $3$ & $(1.0,\; 7.0,\; 3.0)$ & $9 $ \\
        \emph{Citt\`{a} Studi} & $30 $ & $35 $ & $6$ & $(1.0,\; 6.0,\; 2.3)$ & $10$ \\
        \bottomrule
    \end{tabular}
\end{table}

\begin{figure*}[!t]
    \centering
    \captionsetup{skip=3pt}
    \captionsetup[subfigure]{skip=3pt}
    \begin{subfigure}[b]{0.30\linewidth}\centering
        \includegraphics[width=\textwidth]{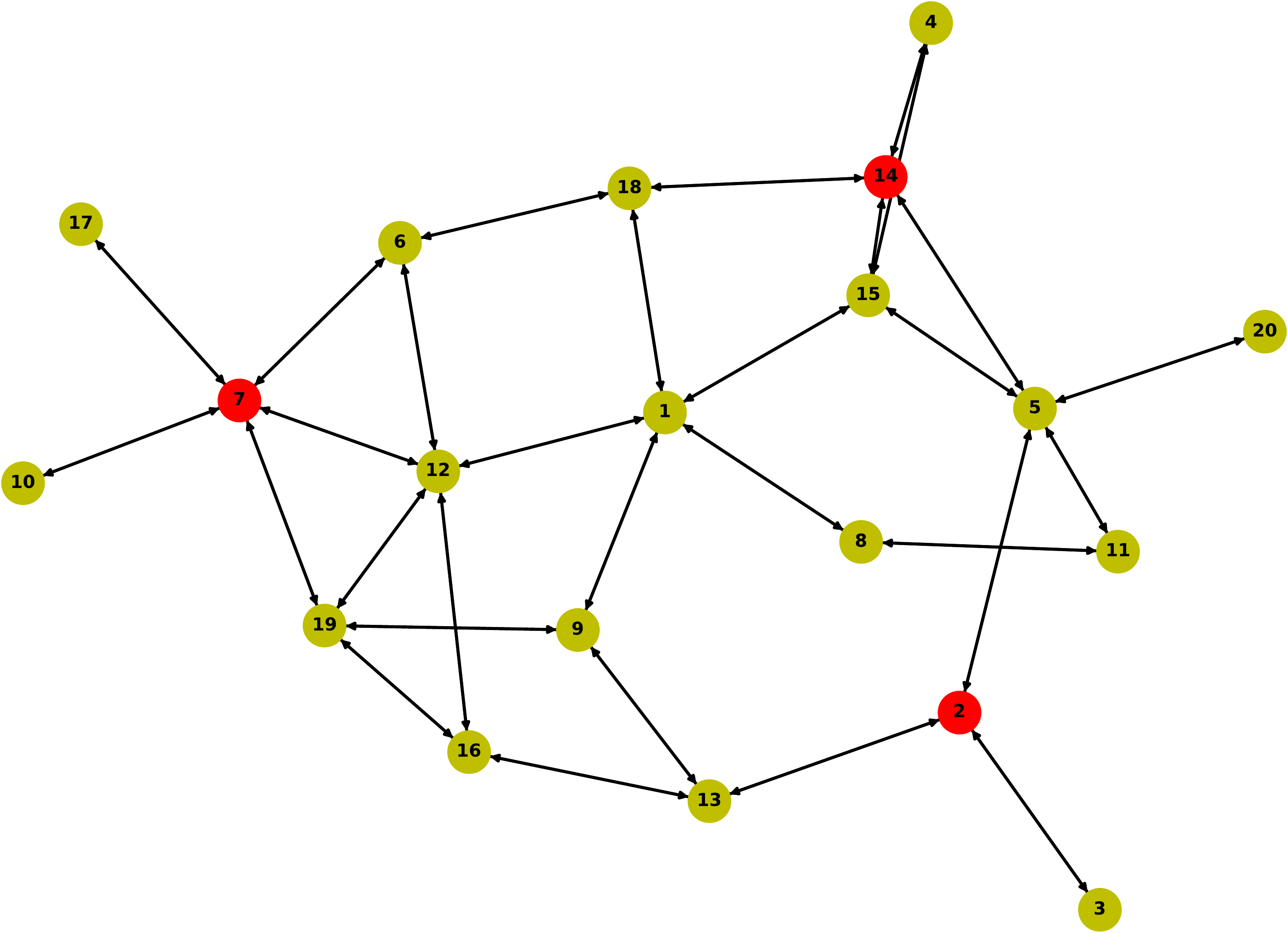}
        \caption{20 nodes 30 edges}
    \end{subfigure}
    \hfill
    \begin{subfigure}[b]{0.30\linewidth}\centering
        \includegraphics[width=\textwidth]{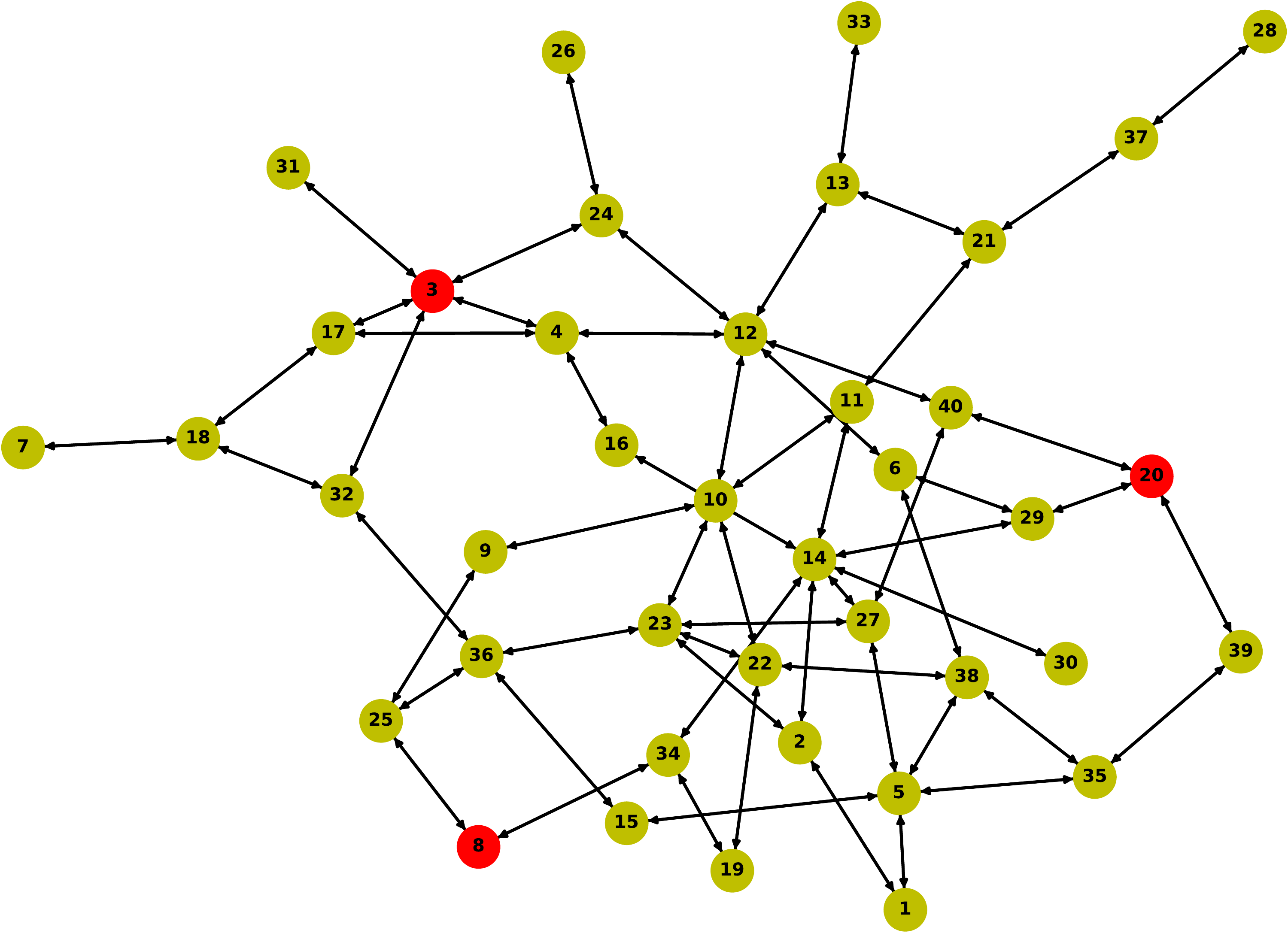}
        \caption{40 nodes 60 edges}
    \end{subfigure}
    \hfill
    \begin{subfigure}[b]{0.30\linewidth}\centering
        \includegraphics[width=\textwidth]{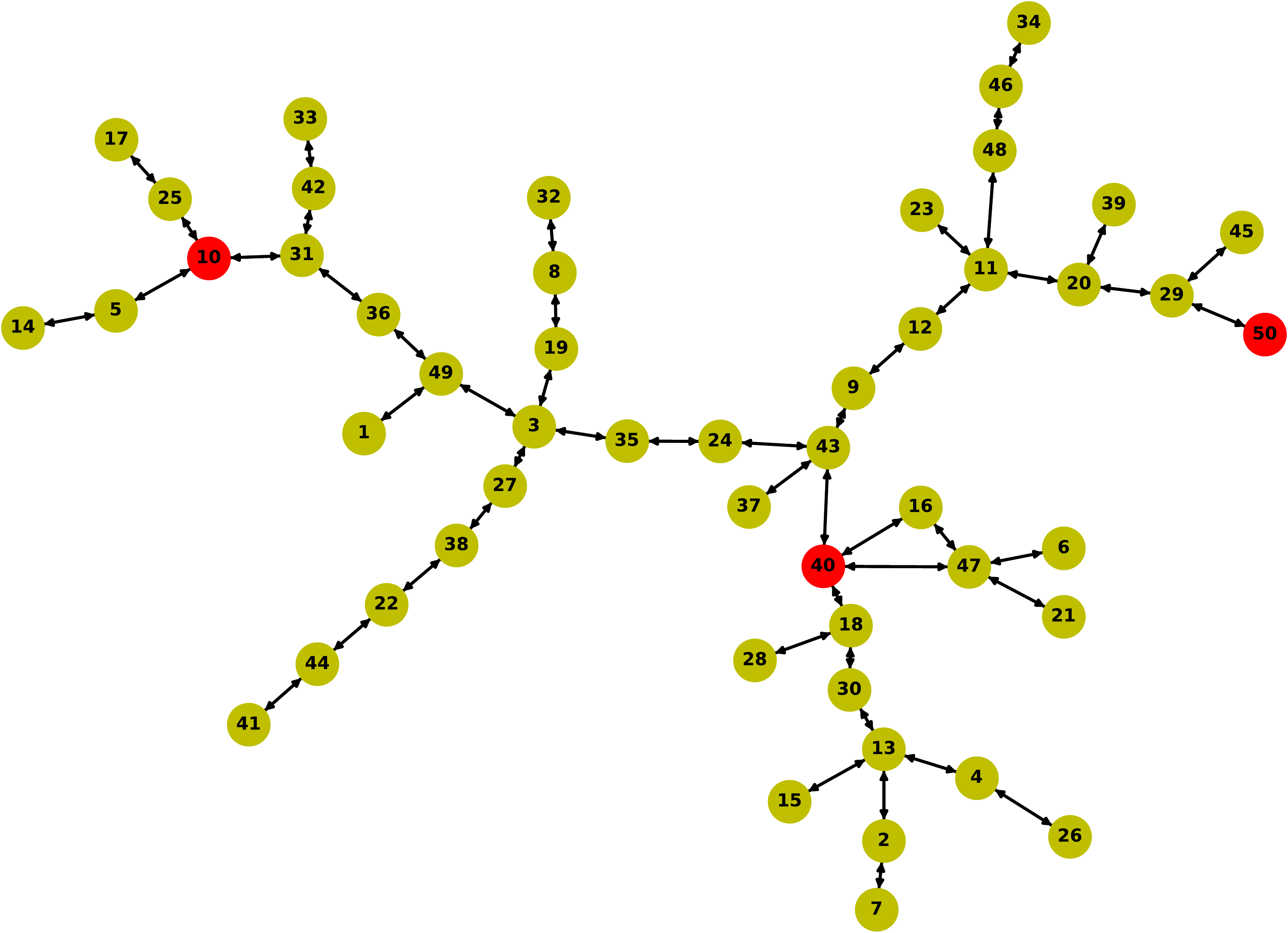}
        \caption{50 nodes 50 edges}
        \label{topo_tree_shape}
    \end{subfigure}
    \begin{subfigure}[b]{0.30\linewidth}\centering
        \includegraphics[width=\textwidth]{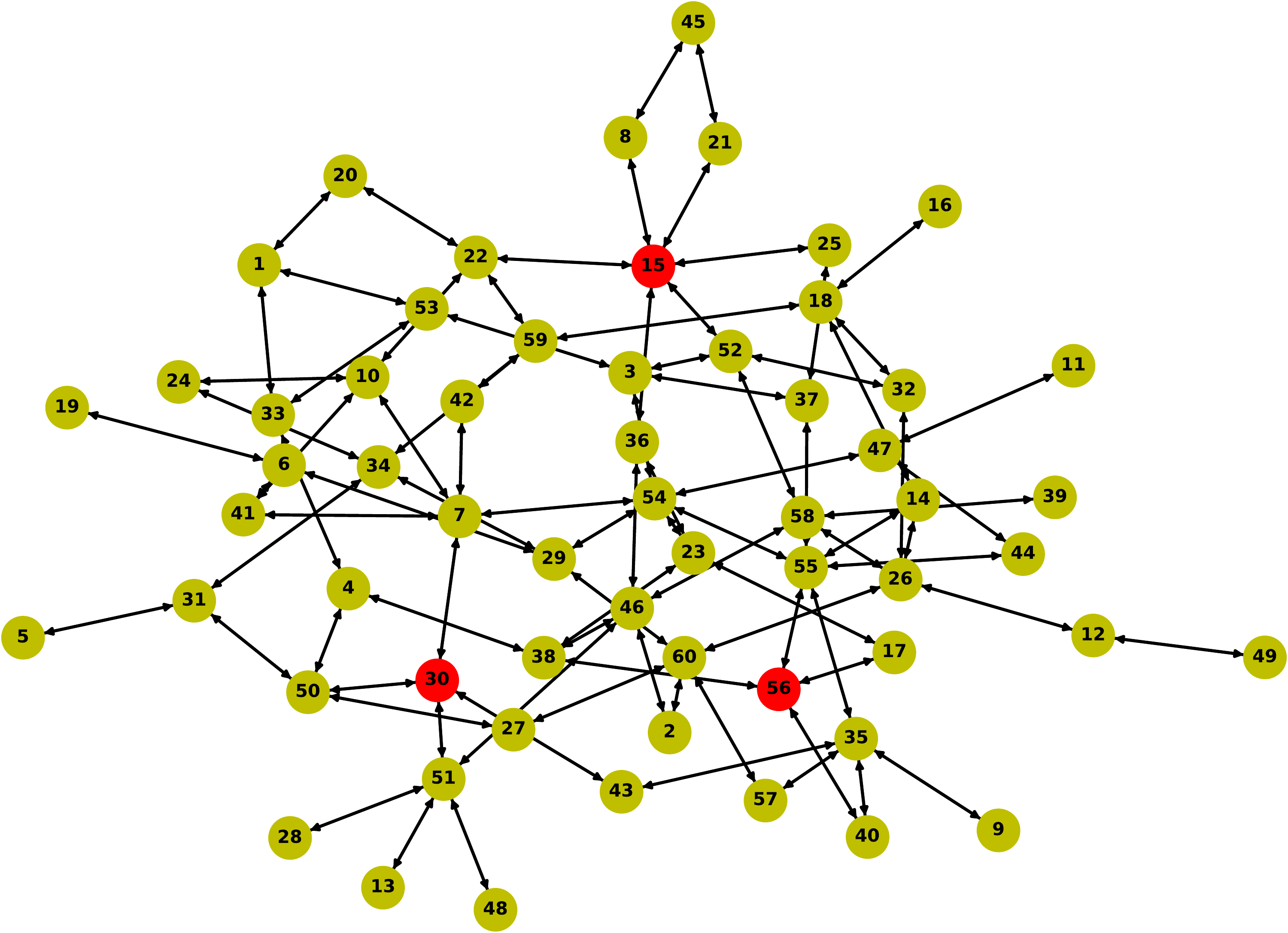}
        \caption{60 nodes 90 edges}
    \end{subfigure}
    \hfill
    \begin{subfigure}[b]{0.30\linewidth}\centering
        \includegraphics[width=\textwidth]{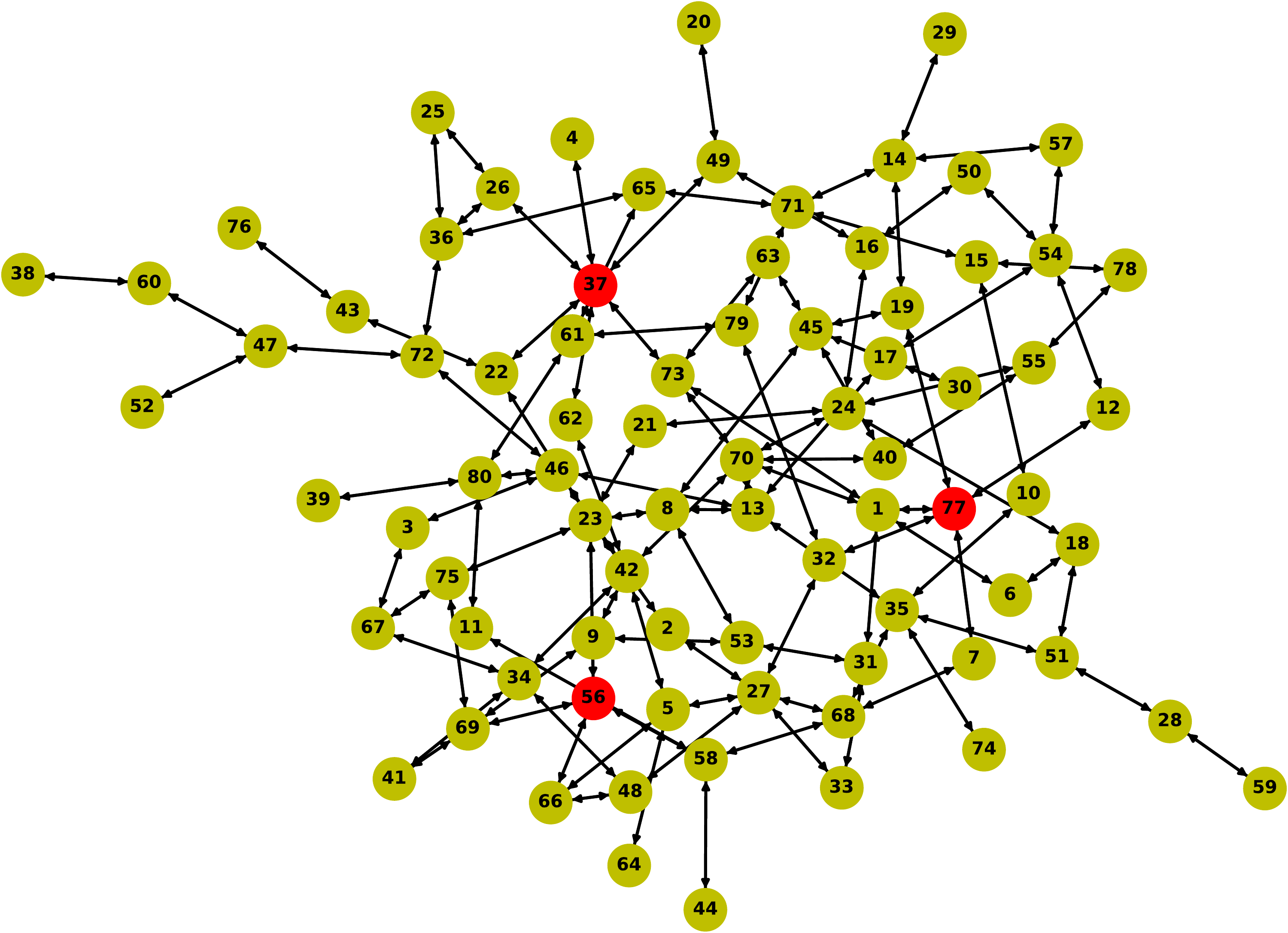}
        \caption{80 nodes 120 edges}
        \label{topo_5e}
    \end{subfigure}
    \hfill
    \begin{subfigure}[b]{0.30\linewidth}\centering
        \includegraphics[width=\textwidth]{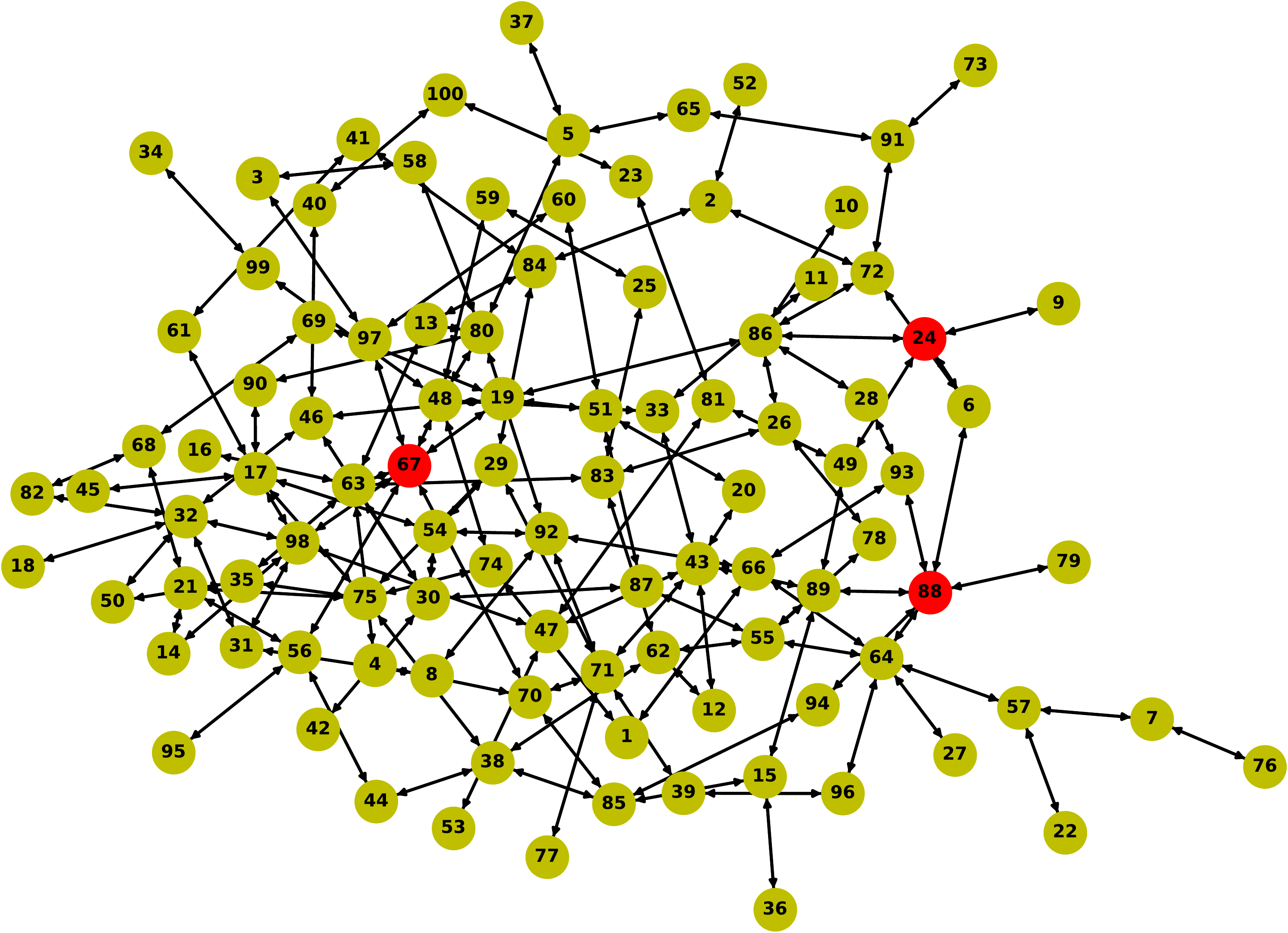}
        \caption{100 nodes 150 edges}
        \label{topo_5f}
    \end{subfigure}
    \caption{Network topologies. Ingress nodes for each graph are colored in red.}
    \label{net_topo}
    \vspace{-6pt}
\end{figure*}

\subsubsection{A real network scenario}\label{subsec_real_topo}
We further considered a real network scenario, with the actual deployment of Base Stations (BSs) collected from the open database OpenCellID\footnote{\url{https://www.opencellid.org}}. Specifically, we considered the ``Citt\`{a} Studi'' area around Politecnico di Milano and selected one mobile operator (Vodafone) with 133 LTE cells falling in such area (see Figure \ref{fig_vodafone}). The BSs deployment shows where the BSs are located but it does not show their interconnection topology nor where the edge clouds are deployed. The reader should note that it is not easy to have access to such piece of information as it is both sensitive for the mobile operator and in continuous evolution. To the best of our knowledge, there is no publicly available true BSs interconnection topology, and for this reason, we decided to infer one as described below.
We performed a clustering on the LTE cells, as illustrated in Figure \ref{fig_vodafone_clusters}, obtaining 30 clusters. 
Finally, we generated the network topology which, as in real mobile scenarios, has a fat tree-like shape with nodes connecting to other nodes. More specifically, starting from the cluster centroids, we connected any two nodes if the distance is lower than a given threshold (800 meters). By doing so, note that some ``leaf'' nodes become connected to more than one \emph{aggregation node} -- i.e., a node that is reached by multiple other nodes -- to increase redundancy and hence reliability of the final topology, as it happens in real networks;
finally, we generated the Minimum Spanning Tree of the geometric graph weighted by the distance and cluster size, while preserving redundant links. The resulting topology is illustrated in Figure \ref{fig_citta_studi_top}; the average node degree resulting from the above procedure is 2.33 and edge clouds can be installed in all nodes (as suggested by 5G specifications).
The structural information for this topology is shown in the last row of Table~\ref{tab_topo}.

\begin{figure*}[!t]
    \centering
    \captionsetup{skip=3pt}
    \captionsetup[subfigure]{skip=3pt}
    \begin{subfigure}[b]{0.327\linewidth}\centering
        \includegraphics[width=\textwidth]{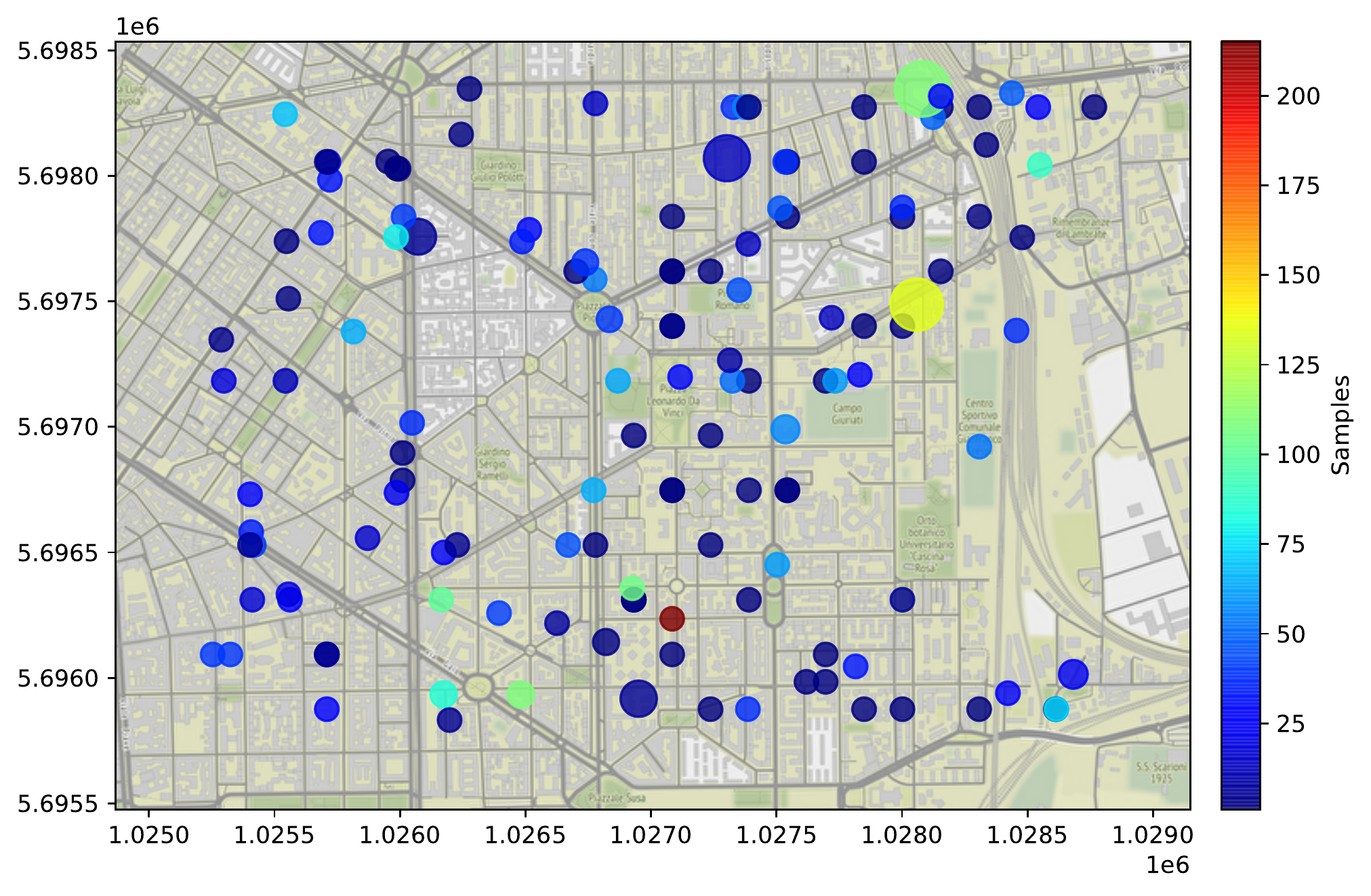}
        \caption{Vodafone LTE cells}
        \label{fig_vodafone}
    \end{subfigure}
    \hfill
    \begin{subfigure}[b]{0.32\linewidth}\centering
        \includegraphics[width=\textwidth]{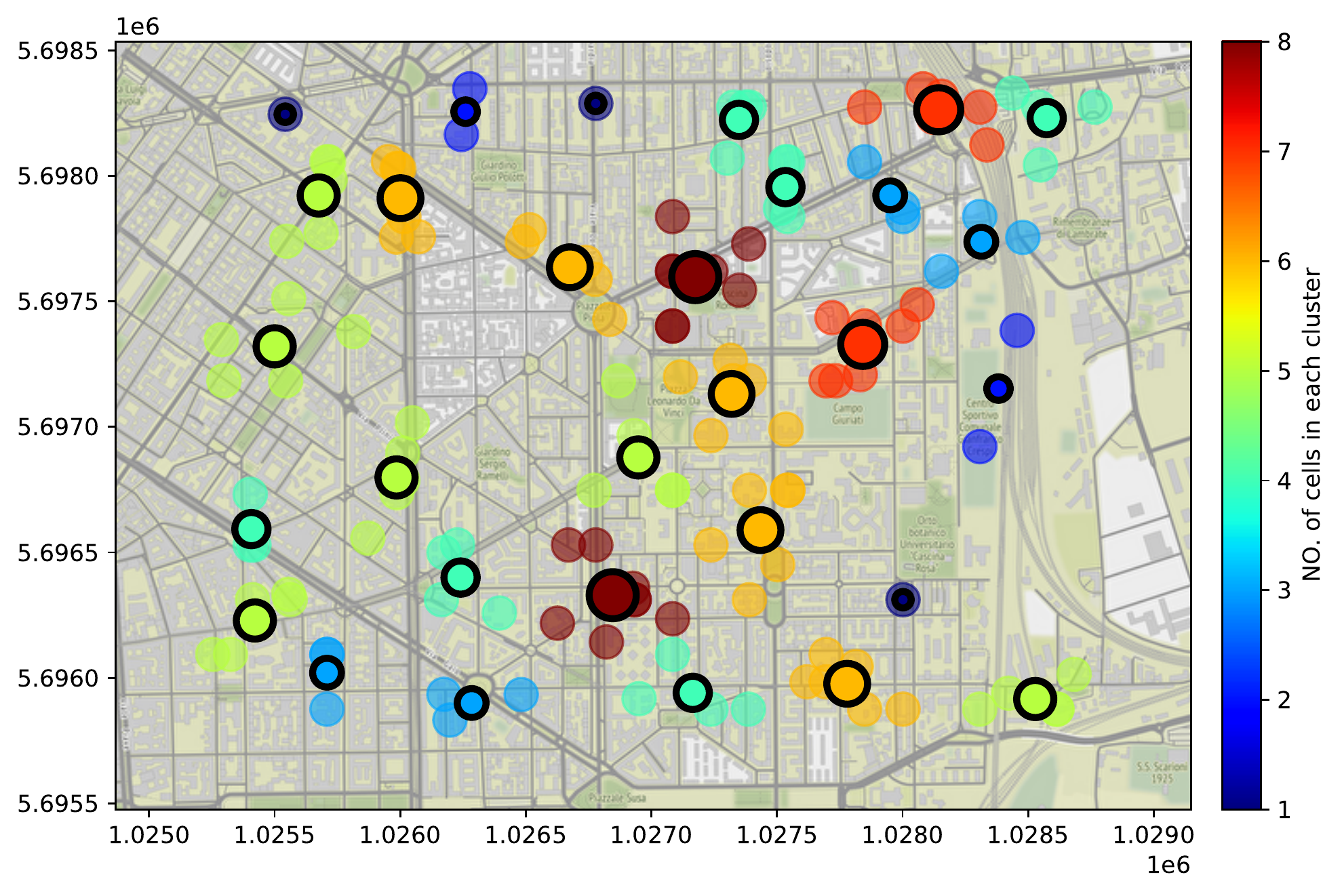}
        \caption{Cell clusters}
        \label{fig_vodafone_clusters}
    \end{subfigure}
    \hfill
    \begin{subfigure}[b]{0.32\linewidth}\centering
        \includegraphics[width=\textwidth]{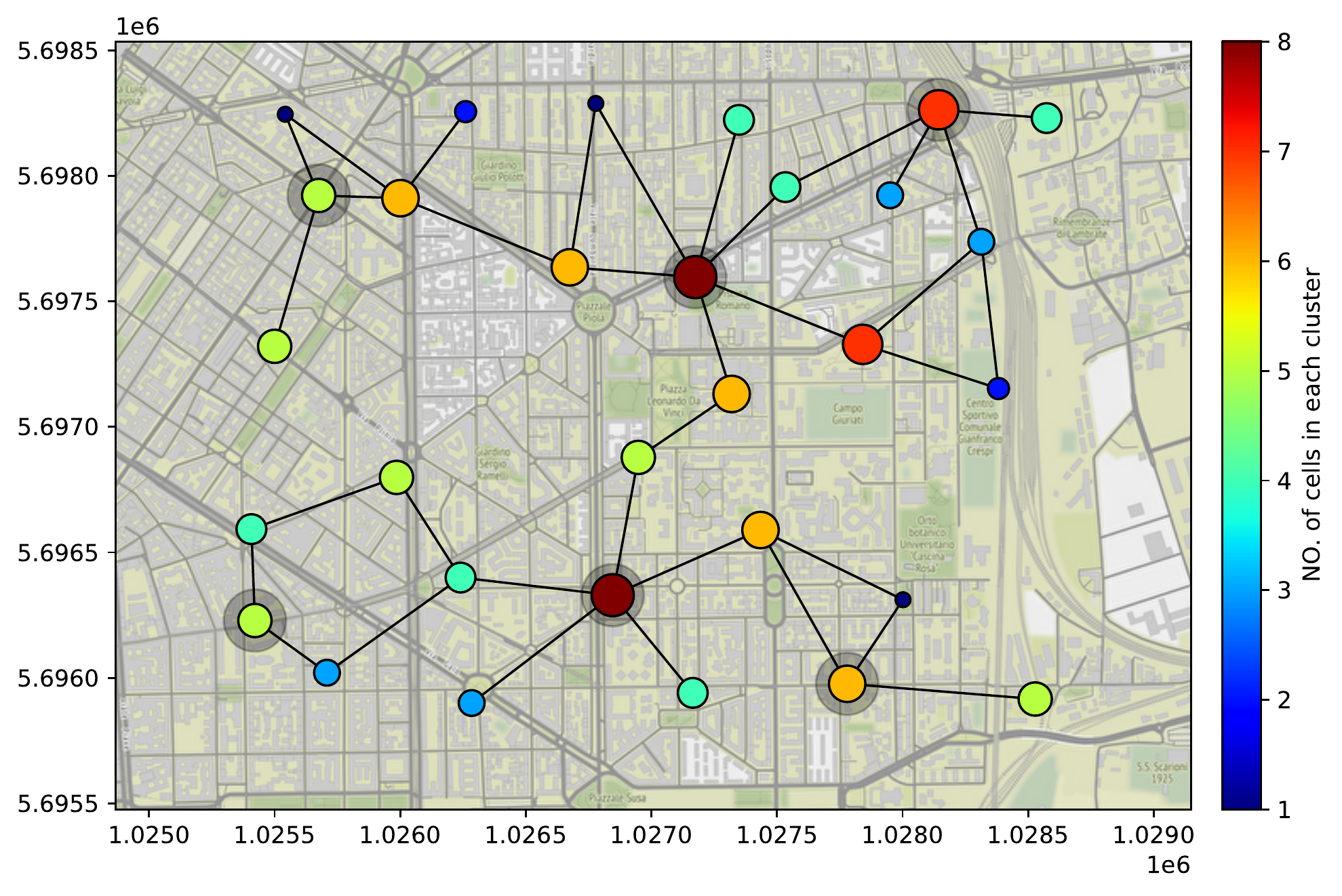}
        \caption{Topology on clusters}
        \label{fig_citta_studi_top}
    \end{subfigure}
    \caption{Citt\`{a} Studi topology with 30 nodes, 35 edges and 6 ingress nodes (marked with gray shadow).}
    \label{fig_citta_studi}
    \vspace{-6pt}
\end{figure*}

\begin{figure}[!ht]
    \captionsetup{skip=3pt}
    \captionsetup[subfigure]{skip=3pt}
    \begin{subfigure}[b]{0.49\linewidth}\centering
        \includegraphics[height=0.687\textwidth]{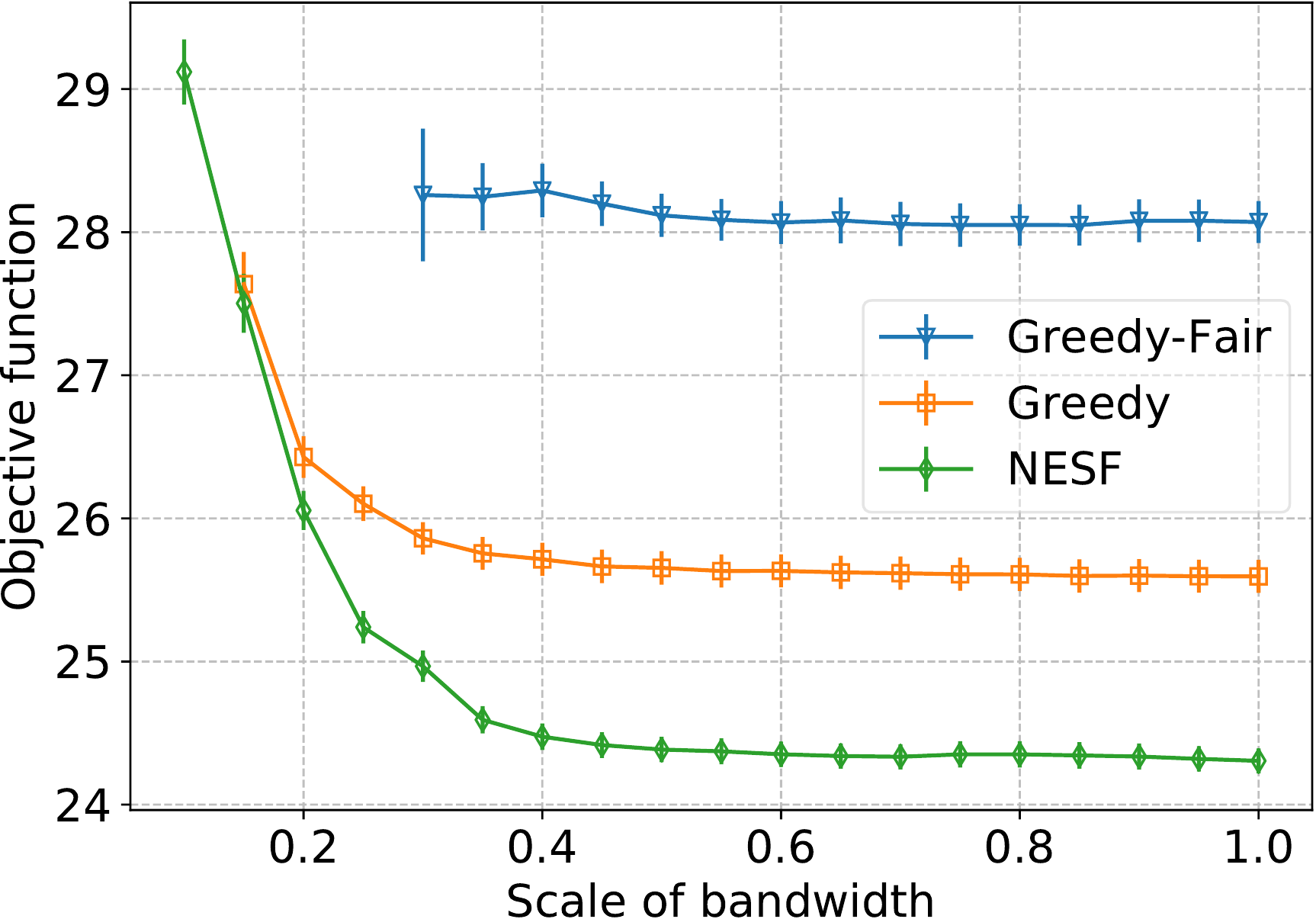}
        \caption{Link bandwidth ($B_l$)}
        \label{fig_B_w3}
    \end{subfigure}
    \hfill
    \begin{subfigure}[b]{0.49\linewidth}\centering
        \includegraphics[height=0.687\textwidth]{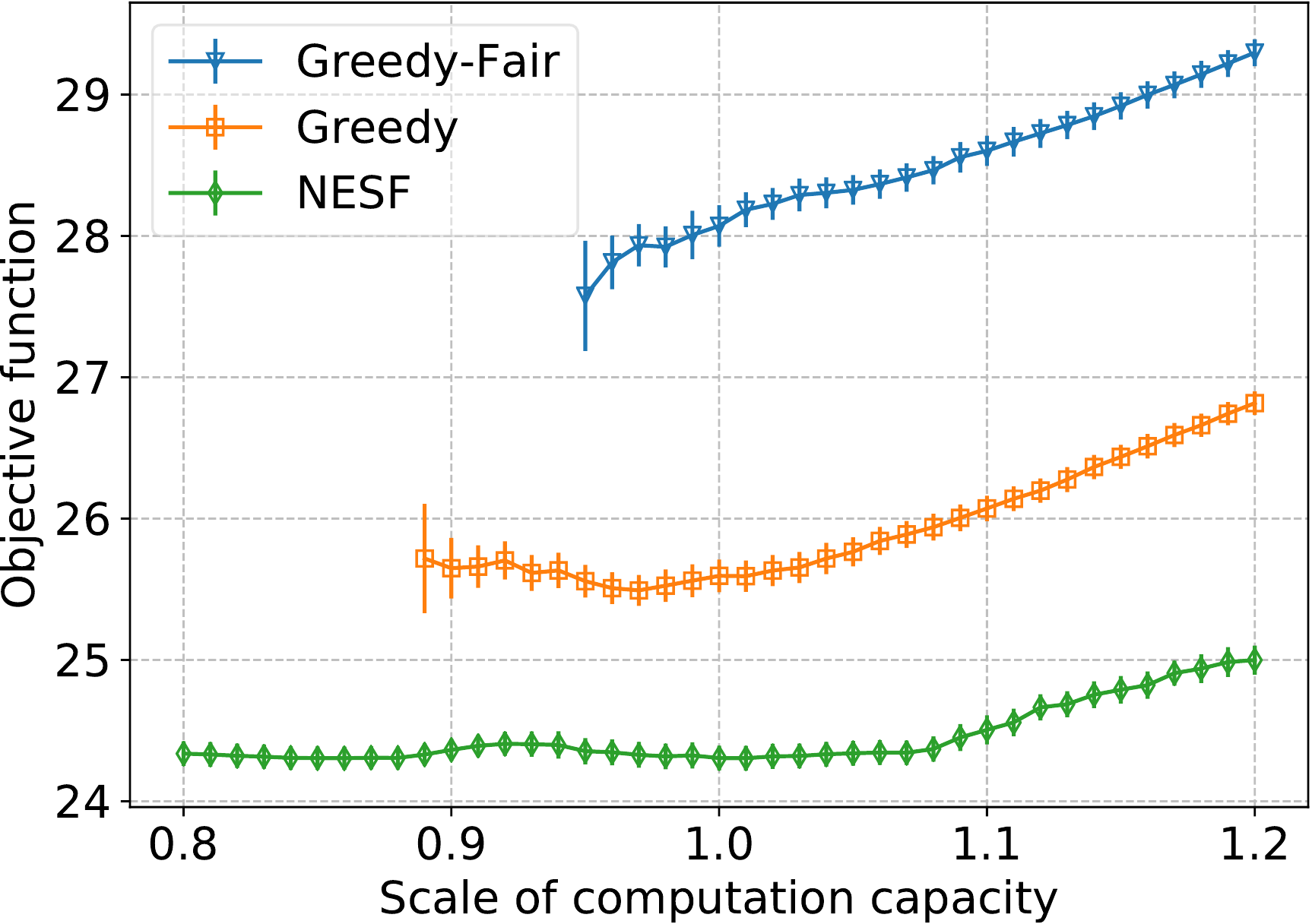}
        \caption{Computation capacity ($D_3$)}
        \label{fig_D3_w3}
    \end{subfigure}
    \caption{Selected numerical results for Citt\`{a} Studi topology.}
    \label{fig_BCD3}
\end{figure}

}

\begin{figure}[!h]
    \captionsetup{skip=3pt}
    \captionsetup[subfigure]{skip=3pt}
    \begin{subfigure}[b]{0.49\linewidth}\centering
        \includegraphics[height=0.680\textwidth]{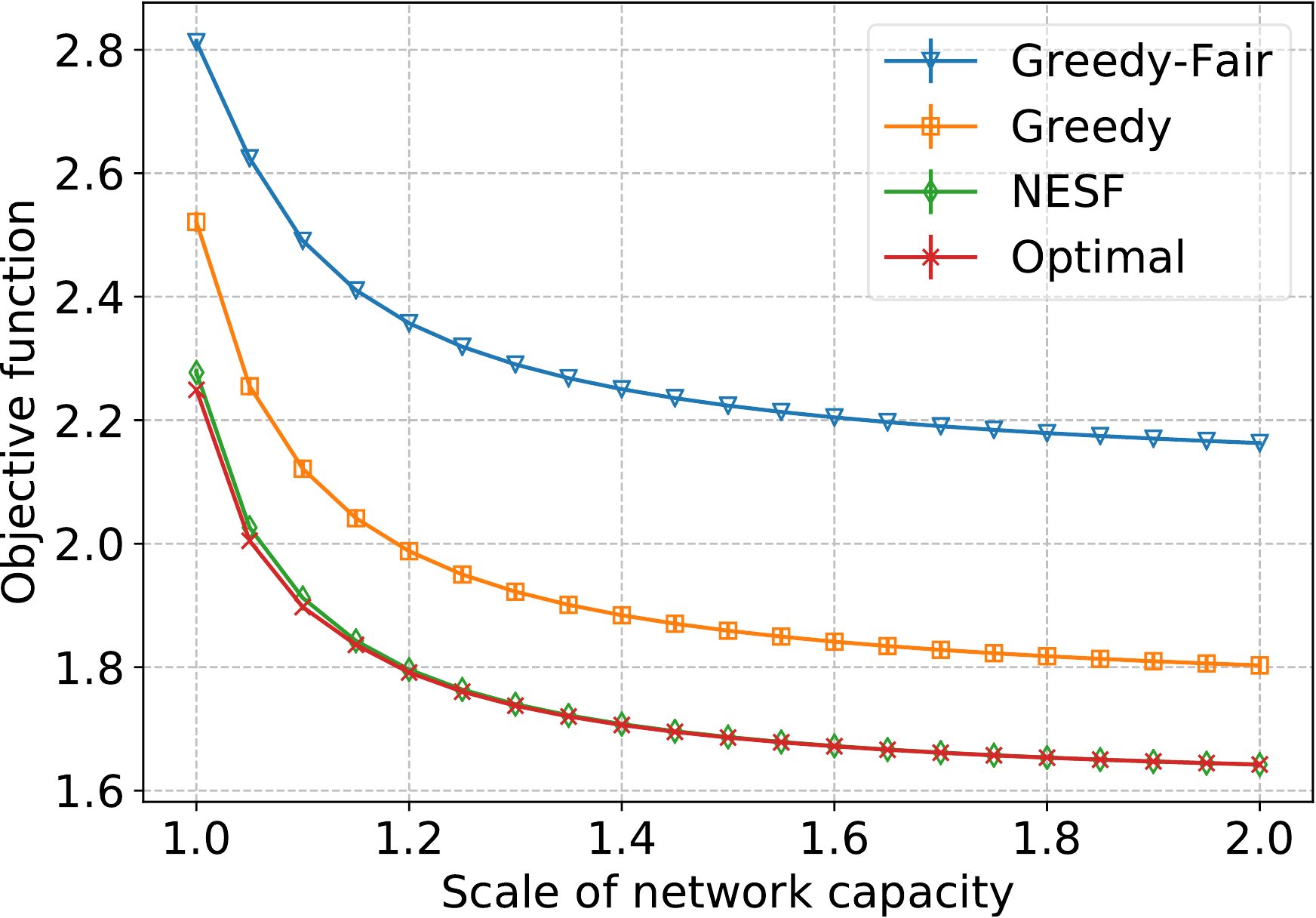}
        \caption{Network capacity ($C_k$)}
        \label{fig_C_opt}
    \end{subfigure}
    \hfill
    \begin{subfigure}[b]{0.49\linewidth}\centering
        \includegraphics[height=0.680\textwidth]{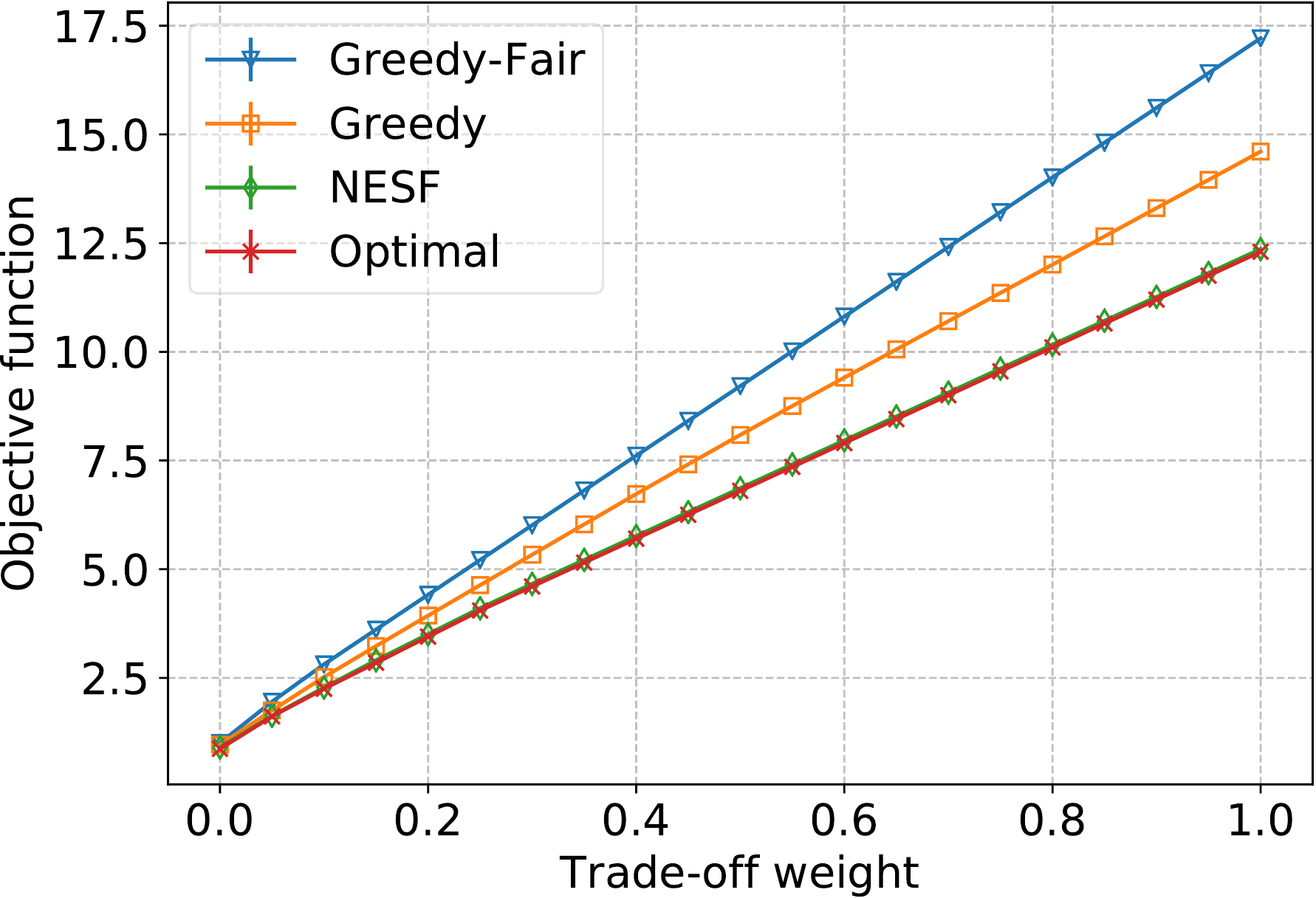}
        \caption{Trade-off weight ($w$)}
        \label{fig_w_opt}
    \end{subfigure}
    \caption{Comparison with the optimum varying two selected parameters ($C_k$ and $w$) in the example network scenario 10N20E of Figure~\ref{fig-toyExample}.}
    \label{fig_opt}
\end{figure}

\begin{figure*}[!t]
    \captionsetup{skip=3pt}
    \captionsetup[subfigure]{skip=0pt}
    \centering
    \begin{subfigure}[b]{0.47\linewidth}\centering
        \includegraphics[width=\textwidth]{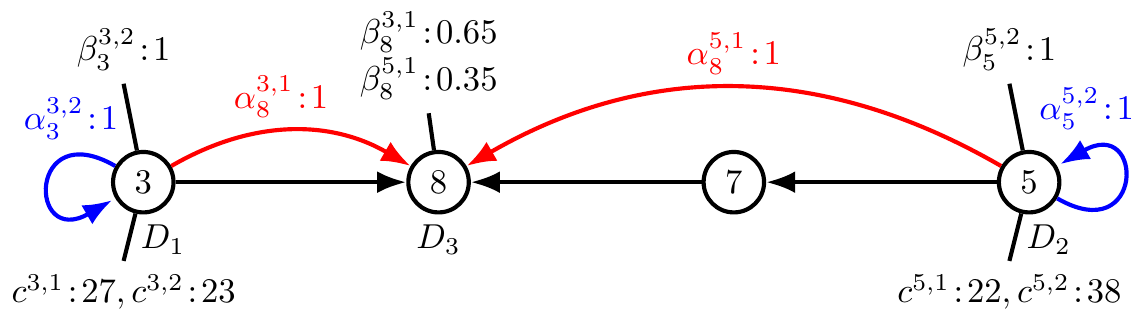}
        \caption{\emph{Optimal}}
        \label{fig_opt_c1}
    \end{subfigure}
    \hfill
    \begin{subfigure}[b]{0.47\linewidth}\centering
        \includegraphics[width=\textwidth]{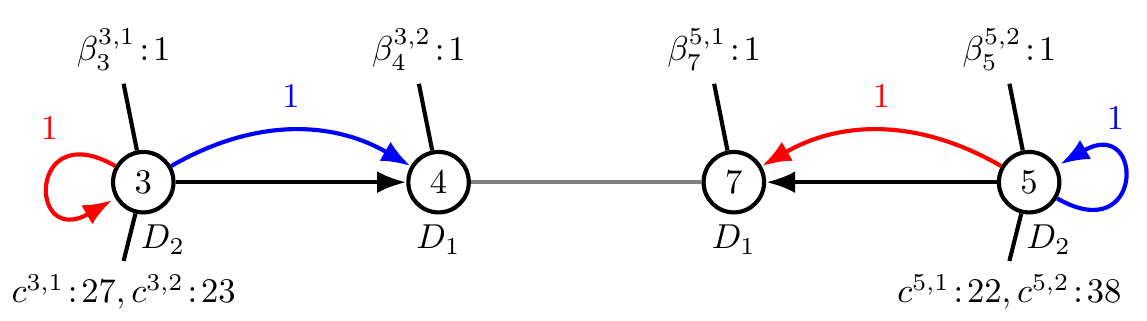}
        \caption{\emph{Greedy}}
        \label{fig_greedy_c1}
    \end{subfigure}\\
    \begin{subfigure}[b]{0.47\linewidth}\centering
        \includegraphics[width=\textwidth]{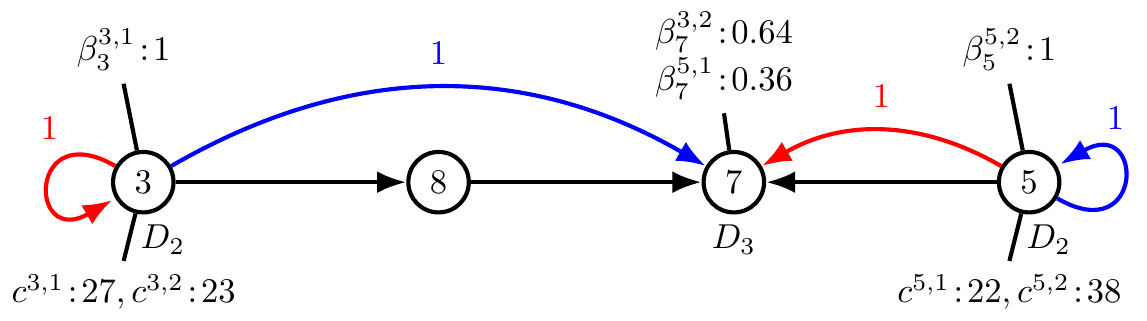}
        \caption{\emph{NESF}}
        \label{fig_sfv6_c1}
    \end{subfigure}
    \hfill
    \begin{subfigure}[b]{0.47\linewidth}\centering
        \includegraphics[width=\textwidth]{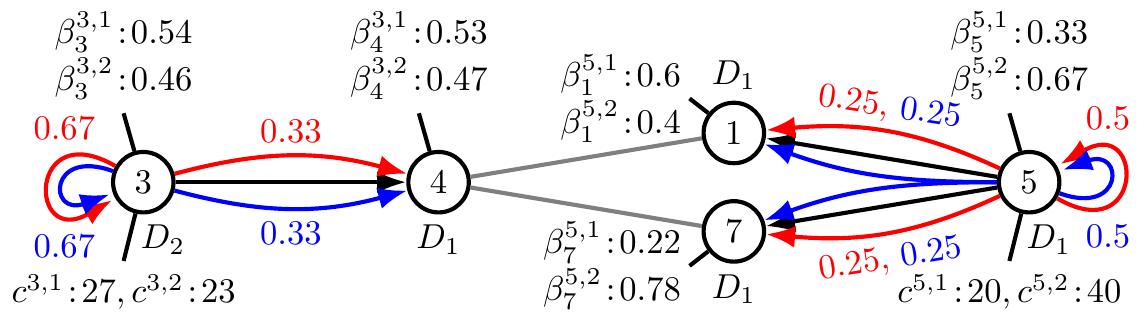}
        \caption{\emph{Greedy-Fair}}
        \label{fig_greedy_f_c1}
    \end{subfigure}
    \caption{Comparison of the solutions achieved by the heuristics and the optimum for the 10N20E topology.}
    \label{fig_comp_opt}
    \vspace{-10pt}
\end{figure*}

\subsection{Experimental Setup}\label{subsection-setup}
We implement our model and heuristics using SCIP (Solving Constraint Integer Programs)\footnote{\url{http://scip.zib.de}}, an open-source framework that solves constraint integer programming problems. All numerical results presented in this section have been obtained on a server equipped with an Intel(R) Xeon(R) E5-2640 v4 CPU @ 2.40GHz and 126 Gbytes of RAM. The parameters of SCIP in our experiments are set to their default values.

\textcolor{black}{
The illustrated results are obtained by averaging over 50 instances with random traffic rates $\lambda^{kn}$ following a Gaussian distribution $N(\mu, \sigma^2)$, where $\mu$ is the value of $\lambda^{kn}$ shown in Table \ref{init_tab} and $\sigma=0.1$ (we recall that the optimization problem is solved under the assumption that the traffic shows only little random variations during the time slot under observation. For this reason, the choice of a Gaussian distribution is appropriate).
We computed 95\% narrow confidence intervals, as shown in the following figures.}

\begin{table}[!t]
    \centering
    \captionsetup{skip=3pt}
    \caption{Parameters setting - Initial (reference) values (for the case of high traffic load with low tolerable latency)}
    \setlength{\tabcolsep}{3.0pt}
    \begin{tabular}{lllr}
        \toprule
        Parameter & Initial value \\
        \midrule
        Link bandwidth $B_l$ \hfill (Gb/s) & $100$ &
            \multicolumn{2}{r}{($l \in \mathcal{L}$)} \\
        Network capacity $C_k$ \hfill (Gb/s) & $60,\; 50,\; 40$ &
            \multicolumn{2}{r}{($k \in \mathcal{K}$)} \\
        Computation level $D_a$ \hfill (Gb/s) & $30,\; 40,\; 50$ &
            \multicolumn{2}{r}{($a\in\mathcal{A}$)} \\
        Computation budget $P$ \hfill (Gb/s) & $300$ & & \\
        Traffic rate $\lambda^{kn}$ \hfill (Gb/s) & \multicolumn{2}{l}{
            $\begin{bmatrix}
                5  & 20 & 7  & 9  & 15 \\
                16 & 4  & 12 & 8  & 6  \\
                7  & 9  & 3  & 12 & 5 
            \end{bmatrix}$} & (${\mathcal{K} \times \mathcal{N}}$) \\
        Tolerable latency $\tau_n$ \hfill (ms) &
            \multicolumn{2}{l}{$1,\; 1.5,\; 2,\; 3,\; 3.5$} & ($n \in \mathcal{N}$) \\
        Weights $\kappa_i,\;w$ & \multicolumn{2}{l}{$0.1,\;0.1$} & ($i \in \mathcal{E}$) \\
       \bottomrule
    \end{tabular}
    \label{init_tab}
\end{table}

In Table \ref{init_tab} we provide a summary of the reference values we define for each parameter \textcolor{black}{for the experiments with the random topologies}. Such values are representative of a scenario with a high traffic load and low tolerable latency relative to the limited communication and computation resources.
Referring to the computation capacity levels and budget in Table~\ref{init_tab}, it is worth noticing that unit ``cycles/s'' is often used for these metrics; for simplicity we transform it into ``Gb/s'' by using the factor ``8bit/1900cycles'', which assumes that processing 1 byte of data needs 1900 CPU cycles in a BBU pool \cite{tang2017system}.

The number of traffic types is set to five. Each traffic type can be dedicated to a specific application case (e.g., video transmission for entertainment, real-time signaling, virtual reality games, audio). Our traffic rates result from the aggregation of traffic generated by multiple users connected at a certain ingress nodes. We select rate values that can be typical in a 5G usage scenario and that almost saturate the wireless network capacity at the ingress nodes that we assume to vary from 40 to 60 Gb/s. The tolerable latency for each traffic type aims at challenging the approach with quite demanding requirements ranging from 1 to 3.5 ms. 
More specifically, the values of traffic rate $\lambda^{kn}$ and tolerable latency $\tau_n$ are designed to cover several different scenarios, i.e., \textit{mice}, \textit{normal} and \textit{elephant} traffic load under \textit{strict}, \textit{normal} and \textit{loose} latency requirements.
For simplicity, in this paper we fix the number of ingress nodes to three.
An in-depth analysis of the impact of the number of ingress nodes on the performance of the optimization algorithm is the subject of our future research.
To make the problem solution manageable, we assume to adopt links of the same bandwidth (100~Gb/s) that are representative of current fiber connections. As in the example of Section~\ref{section-systemoverview}, we assume three possible levels for the computation capacity (30, 40 and 50 Gb/s), under the assumption that, as it happens in typical cloud IaaS, users see a predefined computation service offer. The maximum computation budget is set to 300 Gb/s, which is a relatively low value considering the traffic rates we use in the experiments and the number of available nodes in the considered topologies. Finally, by assigning the same values to weights $\kappa_i,\;w$, we make sure that the two components of the optimization problem, the total latency and the operation cost, have the same importance in the identification of the solution. 

\textcolor{black}{In the network scenario of Section~\ref{subsec_real_topo}, we set the network capacity of each edge $(i,j)$ proportionally to the size of
nodes/clusters to make it scale by a factor $K$ (set according to the specific parameters of our network scenario to 12.5, more precisely using expression $12.5\cdot\max_{n\in\{i,j\}}\{\#\{\text{Node}\}_n\}$) so that, as in real mobile access networks, it can accommodate aggregate traffic coming from edge/leaf nodes to aggregation nodes.
Finally, we select 6 ingress nodes (marked by gray shadow in Figure~\ref{fig_citta_studi_top}), and the traffic rates in Table~\ref{init_tab} are correspondingly duplicated from 3 to 6, while the planning budget is increased to $P=600Gb/s$ for this scenario.}

\begin{figure*}[!t]
    \centering
    \captionsetup{skip=3pt}
    \captionsetup[subfigure]{skip=3pt}
    \begin{subfigure}[b]{0.313\linewidth}\centering
        \includegraphics[width=\textwidth]{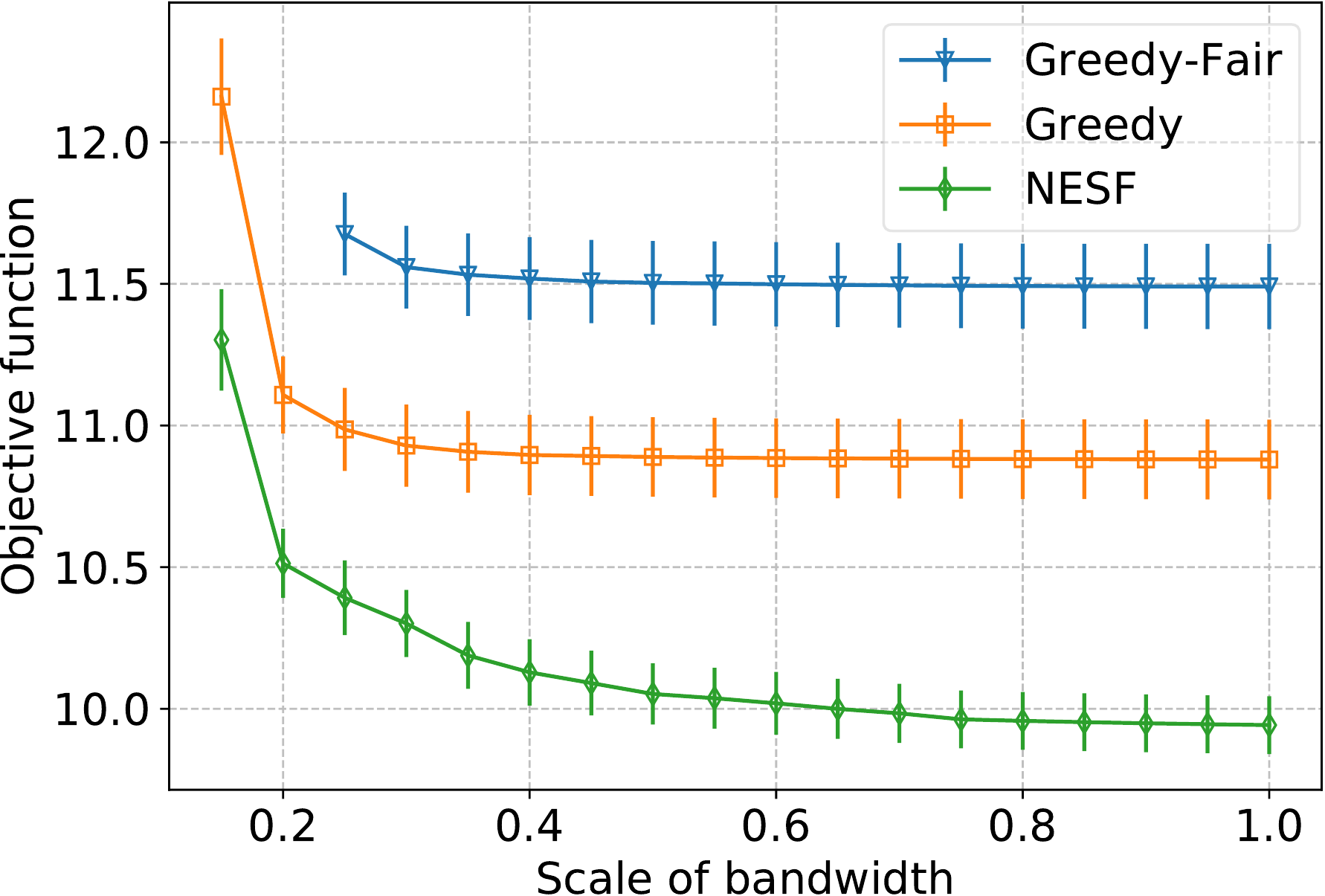}
        \caption{Bandwidth ($B_l$)}
        \label{fig_B}
    \end{subfigure}
    \hfill
    \begin{subfigure}[b]{0.30\linewidth}\centering
        \includegraphics[width=\textwidth]{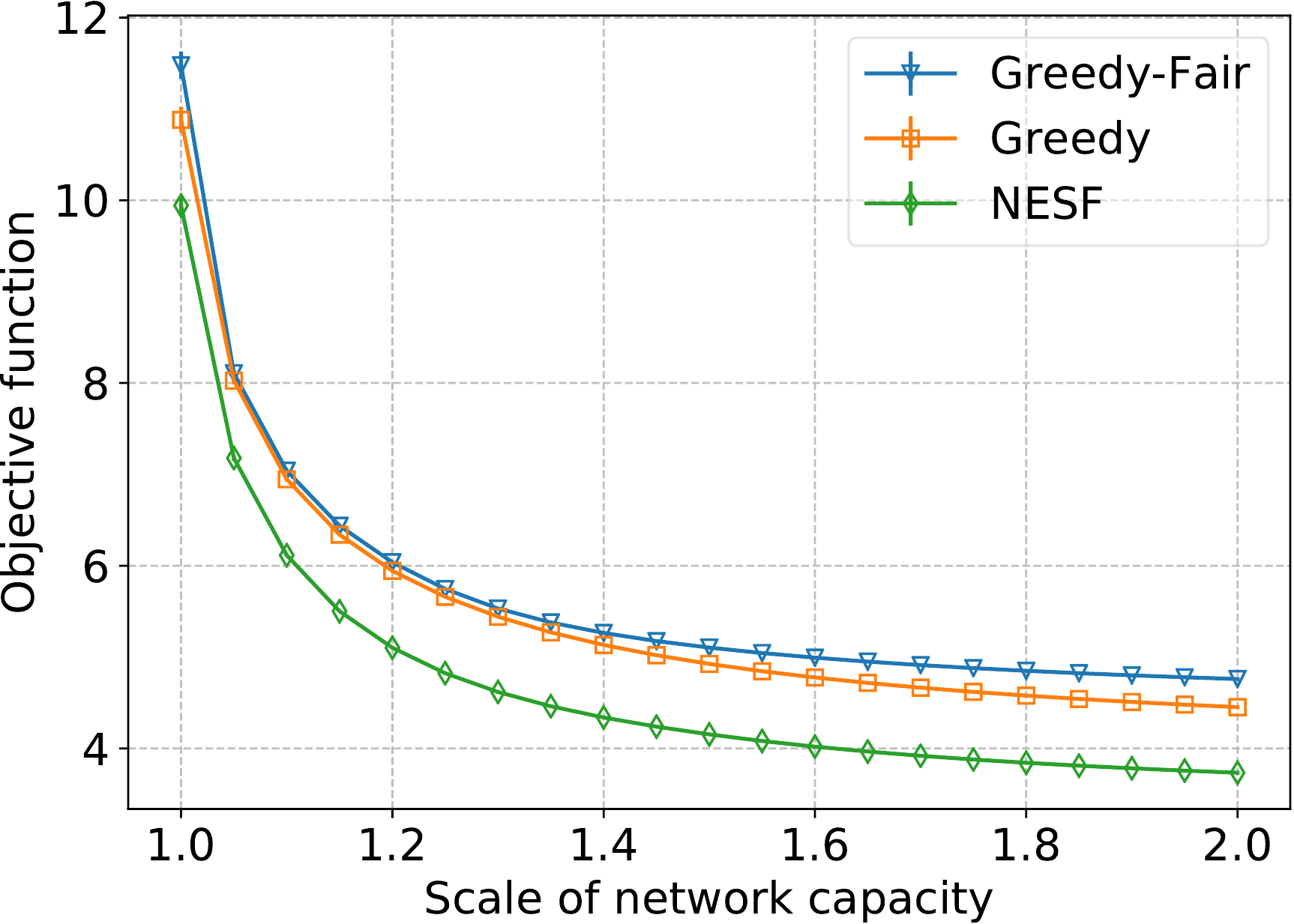}
        \caption{Network capacity ($C_k$)}
        \label{fig_C}
    \end{subfigure}
    \hfill
    \begin{subfigure}[b]{0.31\linewidth}\centering
        \includegraphics[width=\textwidth]{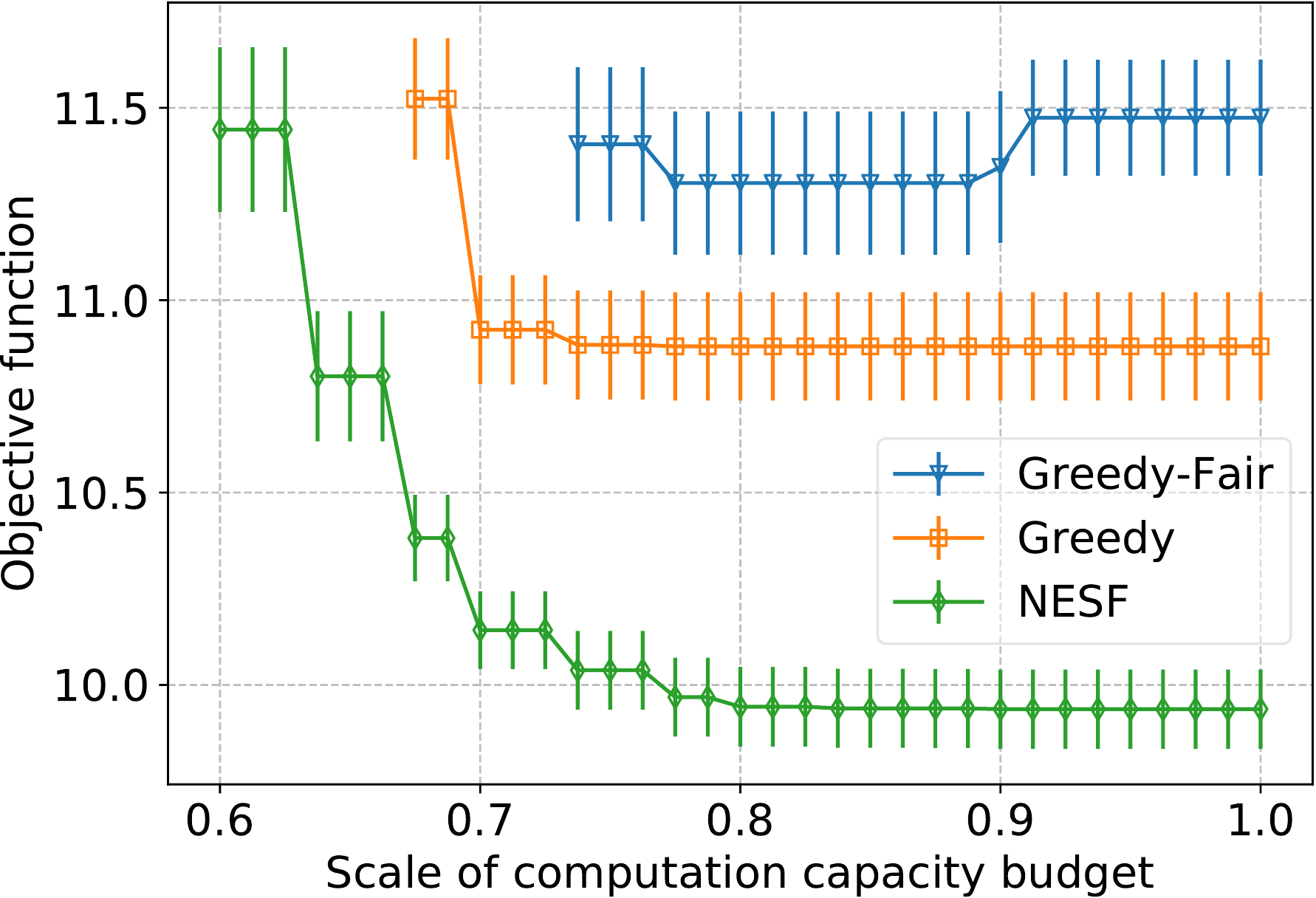}
        \caption{Computation capacity budget ($P$)}
        \label{fig_P}
    \end{subfigure}\\[3pt]
    \begin{subfigure}[b]{0.31\linewidth}\centering
        \includegraphics[width=\textwidth]{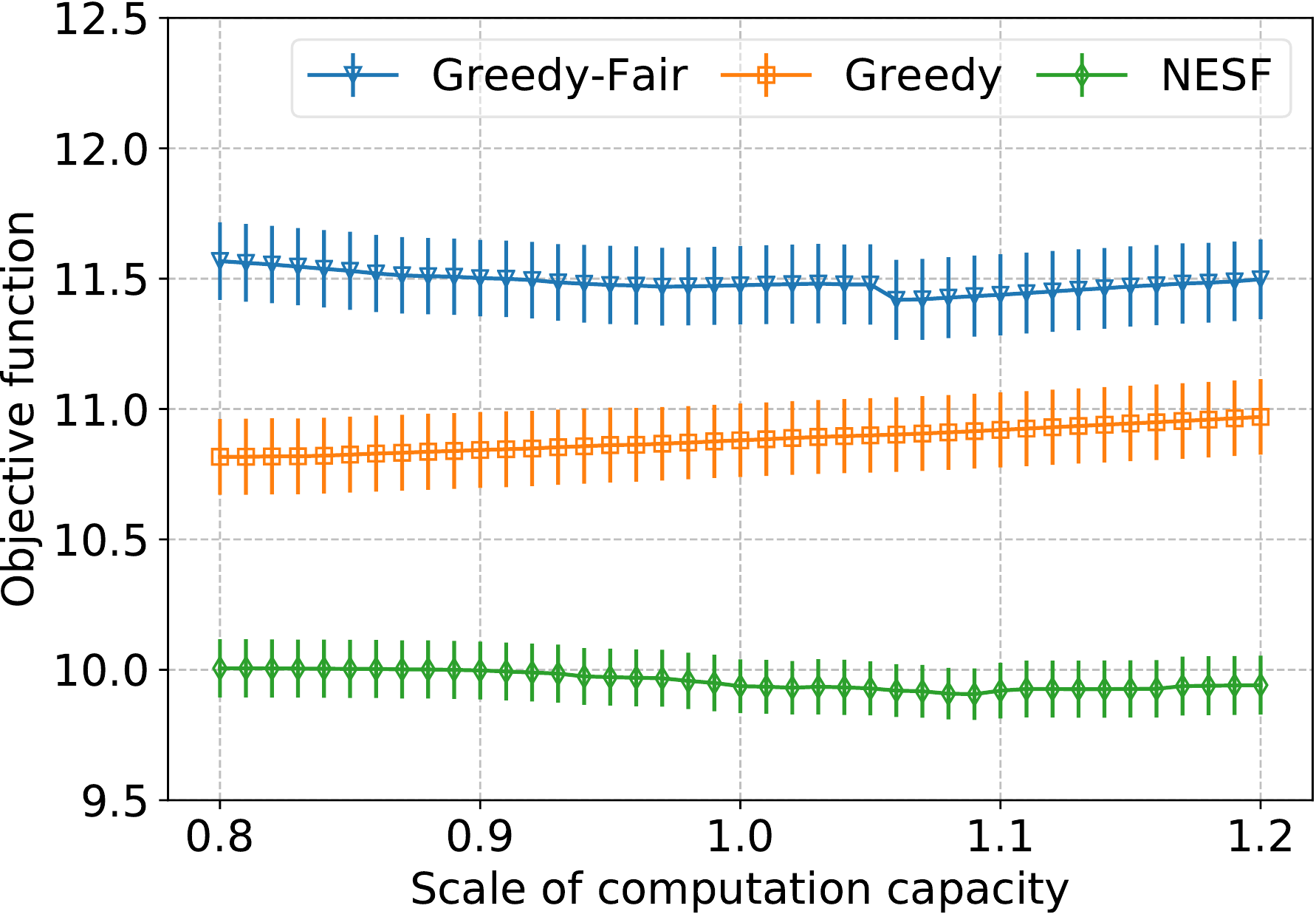}
        \caption{Computation capacity ($D_1$)}
        \label{fig_D1}
    \end{subfigure}
    \hfill
    \begin{subfigure}[b]{0.31\linewidth}\centering
        \includegraphics[width=\textwidth]{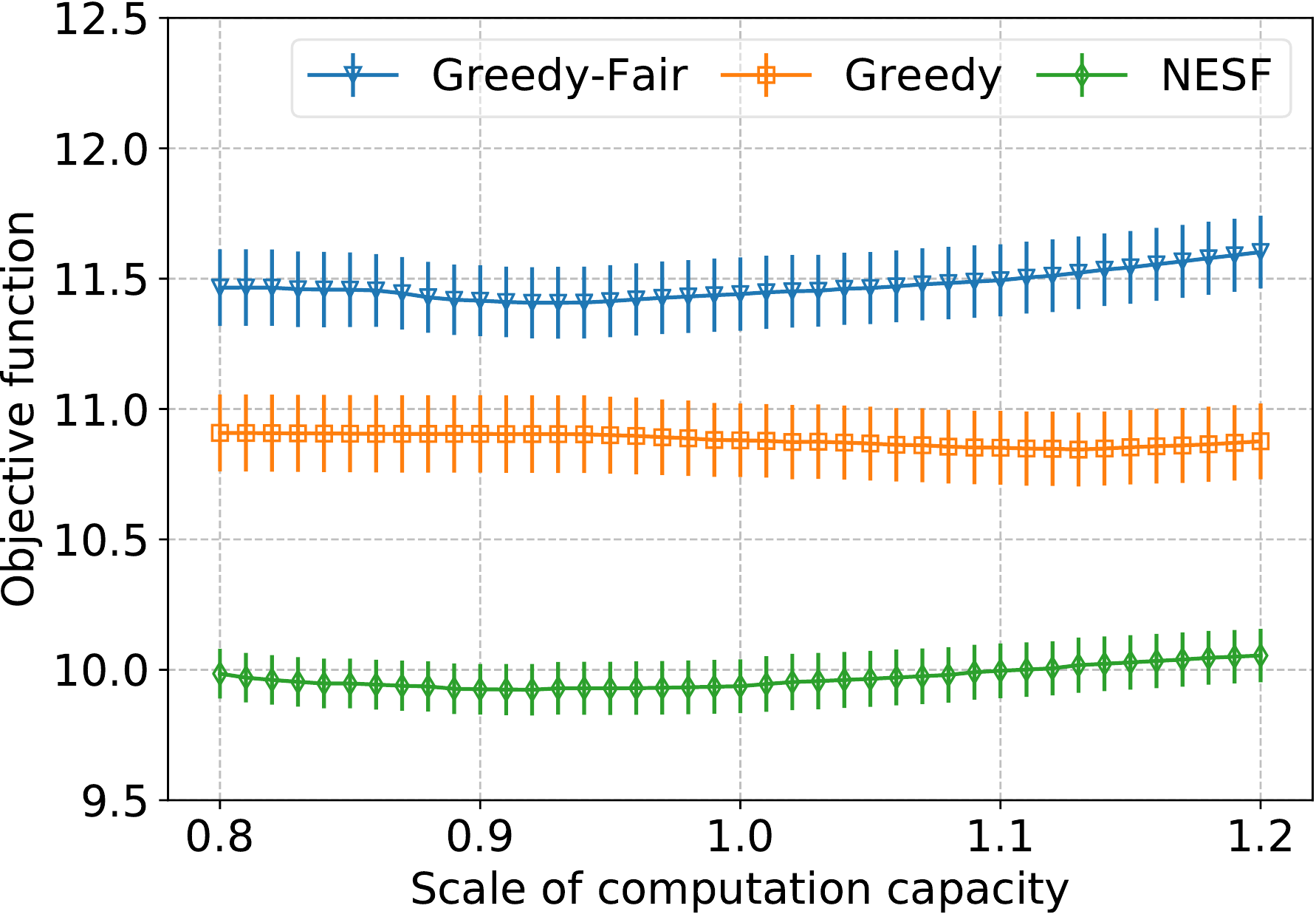}
        \caption{Computation capacity ($D_2$)}
        \label{fig_D2}
    \end{subfigure}
    \hfill
    \begin{subfigure}[b]{0.30\linewidth}\centering
        \includegraphics[width=\textwidth]{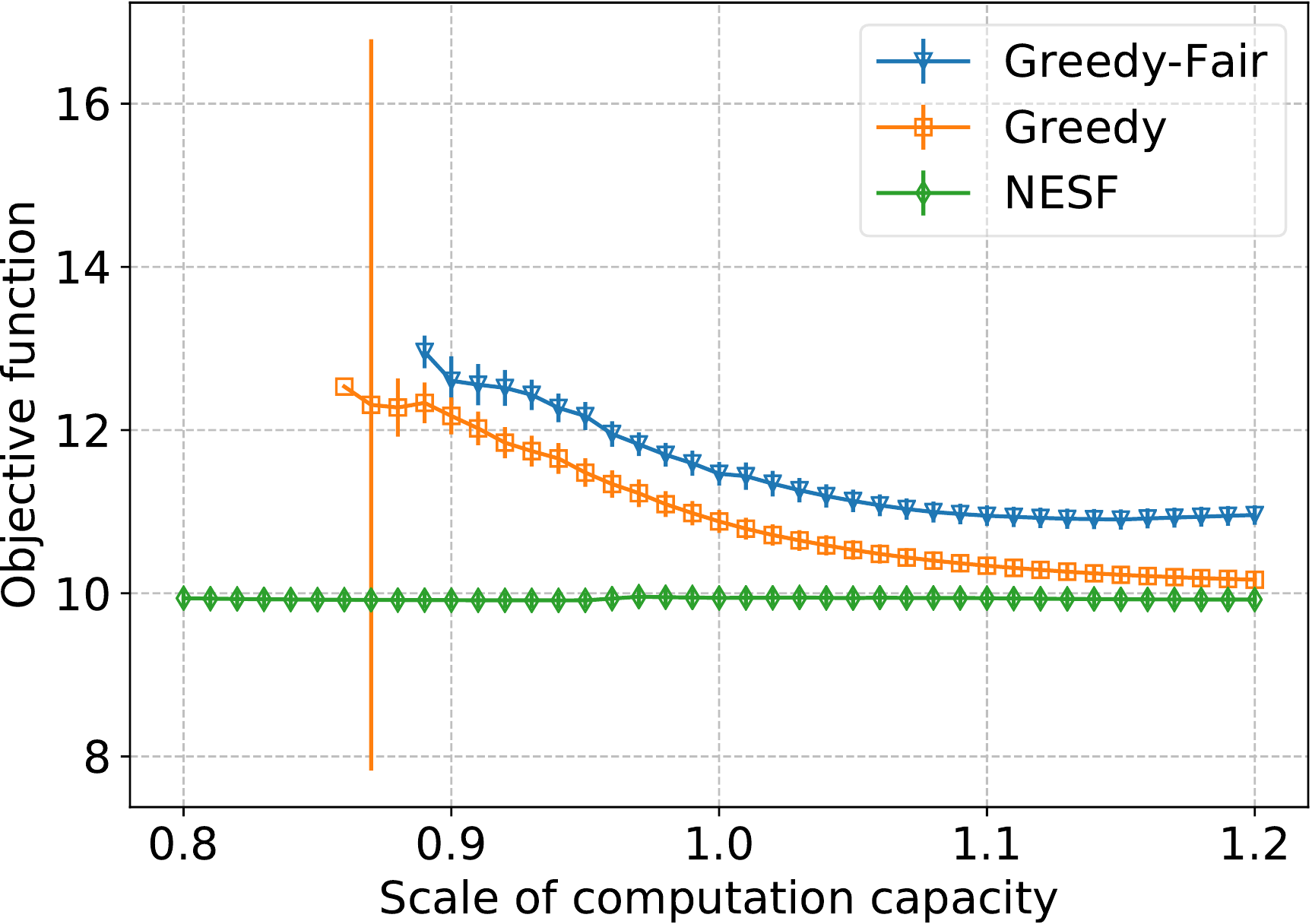}
        \caption{Computation capacity ($D_3$)}
        \label{fig_D3}
    \end{subfigure}\\[3pt]
    \begin{subfigure}[b]{0.30\linewidth}\centering
        \includegraphics[width=\textwidth]{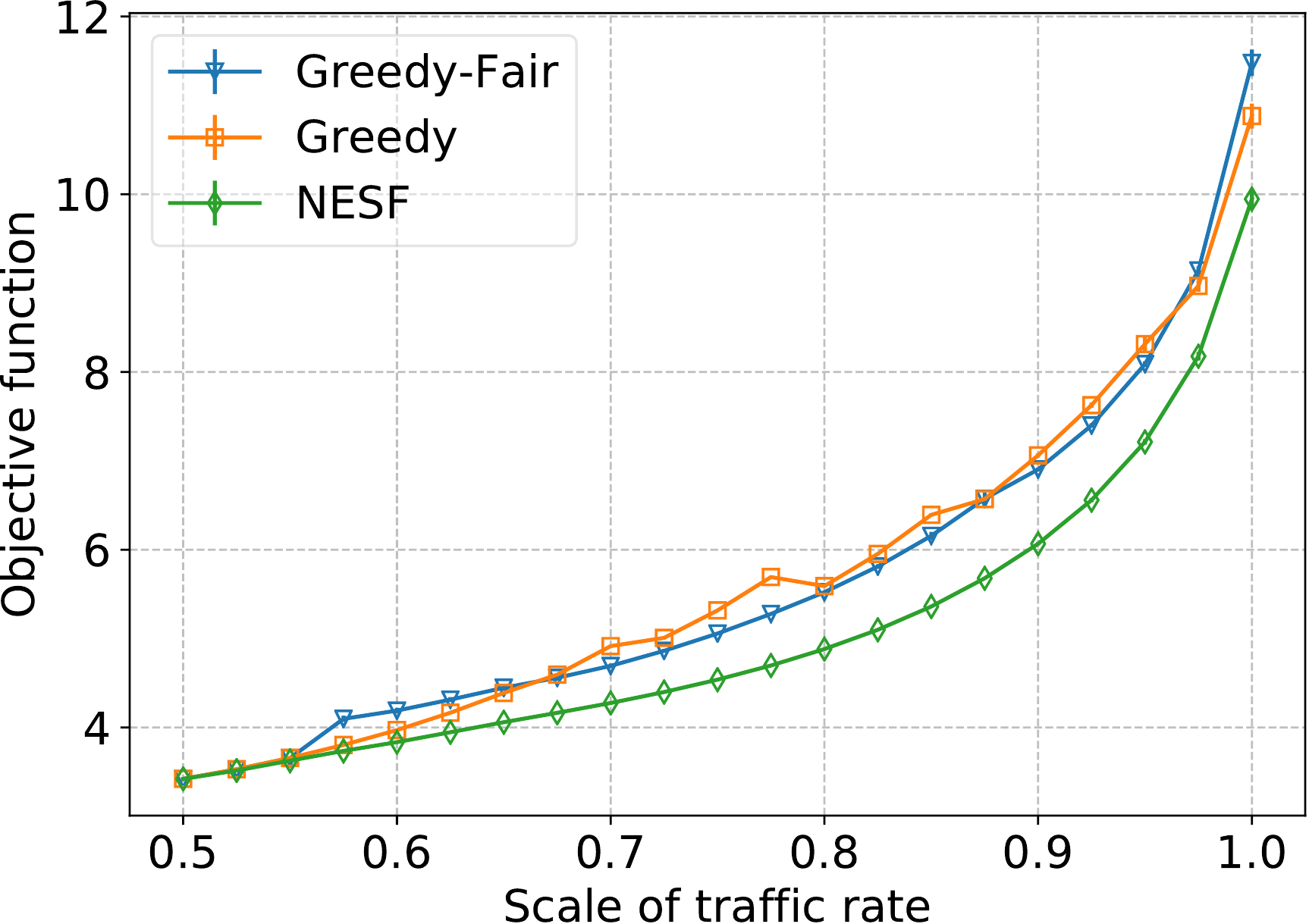}
        \caption{Traffic rate ($\lambda^{kn}$)}
        \label{fig_lambda}
    \end{subfigure}
    \hfill
    \begin{subfigure}[b]{0.31\linewidth}\centering
        \includegraphics[width=\textwidth]{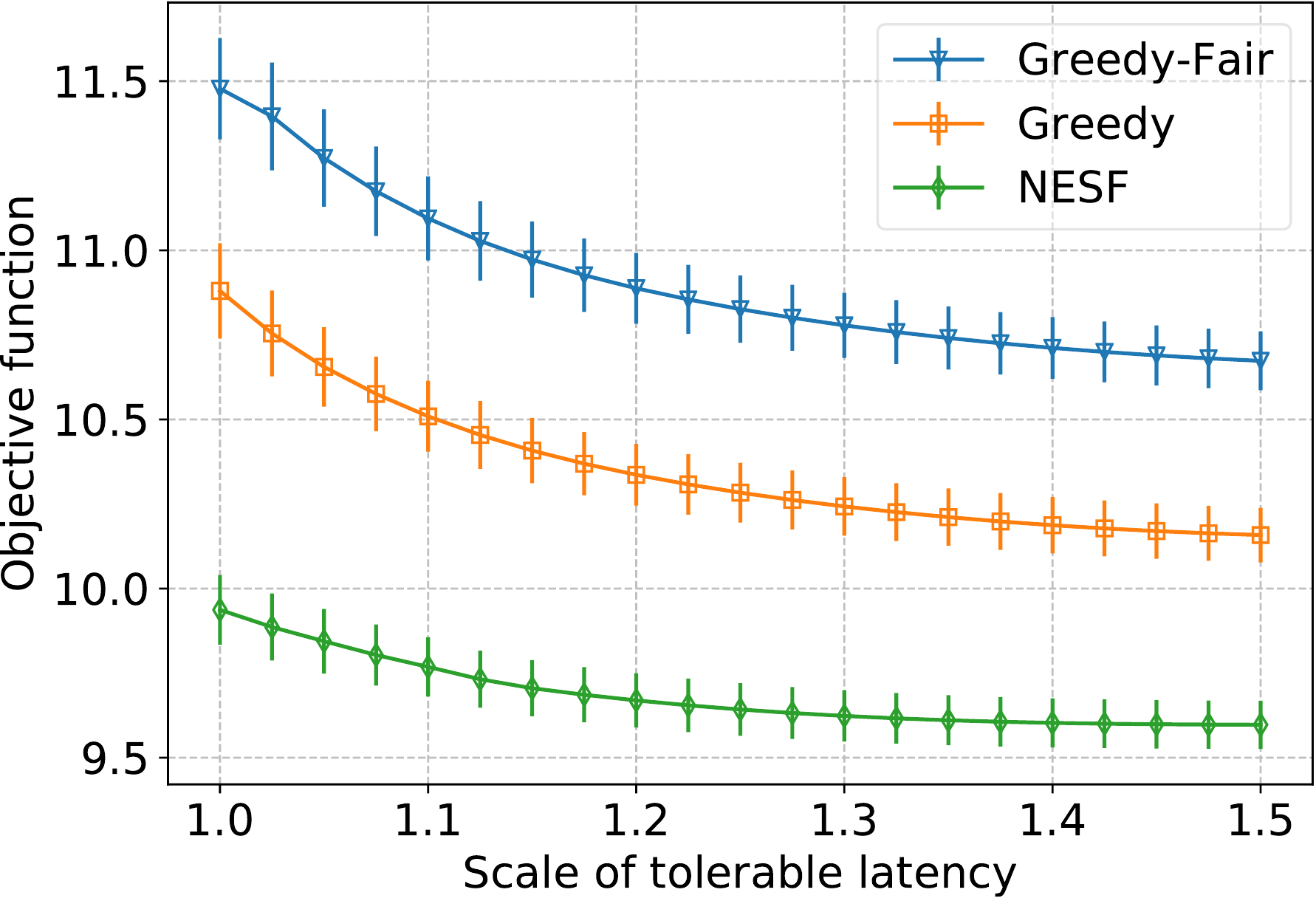}
        \caption{Tolerable latency ($\tau_n$)}
        \label{fig_tau}
    \end{subfigure}
    \hfill
    \begin{subfigure}[b]{0.30\linewidth}\centering
        \includegraphics[width=\textwidth]{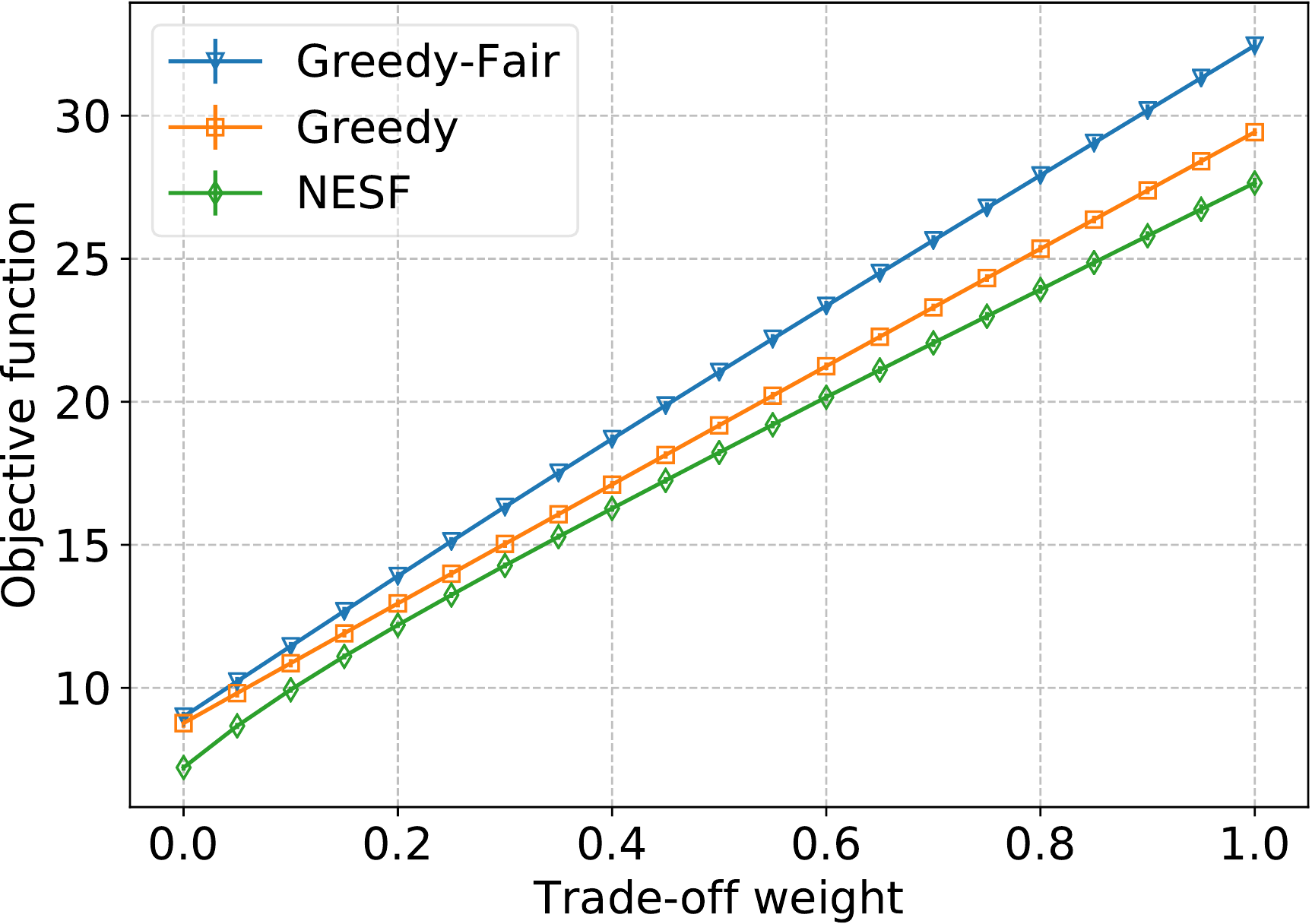}
        \caption{Trade-off weight ($w$)}
        \label{fig_w}
    \end{subfigure}
    \caption{Numerical results for the large-scale network topology \ref{topo_5e}, 80N120E (averaged over 50 instances). }
    \label{fig_results}
    \vspace{-10pt}
\end{figure*}

\textcolor{black}{On such topology, we run the numerical experiments, and the results show very similar trends as those illustrated in Figure~\ref{fig_results}. In Figure~\ref{fig_BCD3} we chose a subset of the results (the objective function value of our optimization model) obtained by scaling the link bandwidth $B_l$ and the computation capacity $D_3$ (those for the network capacity $C_k$ are shown in Fig. \ref{fig_cw}(c)).}

\subsection{Analysis of the optimization results for a small network}
\label{subsection-results-small}
\textcolor{black}{We first compare the results obtained by our proposed heuristic, \emph{NESF}, against the optimum obtained solving model $\mathcal{P}\zcal{1}$ in the simple topology illustrated in Figure \ref{fig-toyExample}, Section~\ref{section-systemoverview}. Note that the original model could be solved only in such a small network scenarios due to a very high computing time.
In Figure~\ref{fig_opt} we show the variation of the objective function (the sum of total latency and operation cost) with respect to two parameters, the network capacity~$C_k$ and the weight~$w$ in the objective function. In these cases, it can be observed that \emph{NESF} obtains near-optimal solutions, practically overlapping with the optimum curve, for the whole range of the parameters, while both \emph{Greedy} and \emph{Greedy Fair} perform worse. The results achieved when the other parameters vary show the same trend. For the sake of space, we do not show them, but they are reported in the supplementary results available here\footref{note_sup}.}

Figure~\ref{fig_comp_opt} shows the configuration of nodes and routing paths for the network (10N20E) with the parameter values defined in Section~\ref{section-systemoverview}. Each sub-figure refers to one of the four considered solutions. Here we highlight the ingress nodes (i.e., $3$ and $5$) and the other nodes which offer computation capacity or support traffic routing. The remaining nodes are not shown for the sake of clarity. The black arrows represent the enabled routing paths. The traffic flow allocation of each solution is marked in red for traffic type 1 and blue for type~2, respectively. The values of all relevant decision variables (see Section~\ref{section-problemformulation}) are shown as well.

Comparing Figures~\ref{fig_opt_c1} and~\ref{fig_sfv6_c1}, we notice that both \emph{Optimal} and \emph{NESF} enable the computation capacity on the ingress nodes and an intermediate node, with one type of traffic kept in the ingress nodes and the other offloaded to the intermediate. The obvious differences between \emph{Optimal} and \emph{NESF} include: i) planning of the computation capacity on ingress node $3$ (i.e., $D_1$ by \emph{Optimal} while $D_2$ by \emph{NESF}), and ii) the intermediate node selected and the consequent routing paths. However, the obtained objective function values (trade-off between the total latency and operation cost) by \emph{Optimal} and \emph{NESF} are respectively $2.25$ and $2.28$, and very close to each other. To further check the reasons behind, we found that the latencies for the traffic of type~1 and 2 are, respectively, $0.49ms$ and $0.55ms$ for \emph{Optimal}, while $0.50ms$ and $0.47ms$ for \emph{NESF}.
Since in this case \emph{NESF} can acquire less total latency at the expense of a little bit higher computation cost, compared with \emph{Optimal}, their corresponding objective function values are close.
Note that the computing time needed to obtain the optimal solution is around 10 hours ($35724$ seconds) while \emph{NESF} is able to compute the approximate solution in only about $1$ second.

The \emph{Greedy} and \emph{Greedy-Fair} approaches tend to enable computation capacity on more nodes. \emph{Greedy-Fair} also splits each type of traffic following multiple paths.
Both aspects result in a higher objective function value.

When increasing the network capacity $C_k$ by the scale factor $1.2$, the resulting solutions remain almost the same, except for the allocation of the wireless network capacity and computation capacity.

\subsection{Analysis of the heuristic results for larger networks}
\label{subsection-results-large}
We investigate the effect of several parameters on the objective function value, with respect to link bandwidth $B_l$, network capacity $C_k$, computation capacity $D_a$ and corresponding total budget $P$, traffic rate~$\lambda^{kn}$, tolerable latency~$\tau_n$ and trade-off weight $w$. We conduct our simulations by scaling one parameter value at a time, starting from the initial values in Table~\ref{init_tab}. Since the goal is to minimize the weighted sum of total latency and operation cost, lower values for the objective function are preferable.

In Figure~\ref{fig_results} we report all results referring to the topology with 80 Nodes and 120 links (Figure~\ref{topo_5e}). All results obtained considering the other topologies in Figure~\ref{net_topo} are available here\footref{note_sup} and show similar trends.

\subsubsection{Effect of the link bandwidth $B_l$}
Figure \ref{fig_B} illustrates the variation of the objective function value (costs w.r.t. latency and computation) versus the link bandwidth $B_l, \forall l \in \mathcal{L}$, the values of which are scaled with respect to its initial ones in Table \ref{init_tab} from $0$ to $1.0$ with a step of $0.05$. In all cases, the problem instance is unfeasible below a certain threshold bandwidth value.
As~$B_l$ increases above the threshold, the cost value achieved by each approach decreases and converges to a smaller value, i.e., around $9.7$ for \emph{NESF} (achieved at $0.9$), $10.84$ for \emph{Greedy} at $0.3$ and $11.48$ for \emph{Greedy-Fair} at $0.4$.
In all cases, \emph{NESF} performs the best among all the approaches, with the following gains: around $11\%$ to \emph{Greedy} and $16\%$ to \emph{Greedy-Fair}.
\emph{Greedy} and \emph{Greedy-Fair} show little flexibility to the variation of link bandwidth.

\subsubsection{Effect of the wireless network capacity $C_k$}
Figures~\ref{fig_C} demonstrates the variation of the objective function value with respect to the wireless network capacity~$C_k, \forall k \in \mathcal{K}$, scaled with respect to the initial values reported in Table~\ref{init_tab} from $1.0$ to $2.0$, which corresponds to the case in which the wireless network shows a capacity comparable to the one of the internal network links.
When $C_k$ increases, the objective function value obtained by each approach decreases quite fast (more than $2$ times) and converges to a specific value.
For \emph{NESF}, the cost decreases from $9.70$ and converges to $3.73$; \emph{Greedy} and \emph{Greedy-Fair} exhibit close performance, i.e., \emph{Greedy} from $10.84$ to $4.45$, \emph{Greedy-Fair} from $11.48$ to $4.76$.
\emph{NESF} still has the best performance among all the approaches, with consistent gaps: around $16\%$ to \emph{Greedy} and up to $22\%$ for \emph{Greedy-Fair}.
This trend reflects the strong effect of the wireless network capacity increase on the minimization of the overall system cost and performance.

\subsubsection{Effect of the computation capacity budget $P$}
Figures \ref{fig_P} shows the trend of the objective function value at the variation of the computation capacity budget $P$, whose value is scaled with respect to the initial one in Table \ref{init_tab} from $0.5$ to $1.0$ with a step of $0.0125$. Clearly, a low power budget challenges the optimization approach that must ensure the available computation capacity is always within this budget. The figure shows that each heuristic has a limit budget value below which it is unable to find a feasible solution ($0.738$ for \emph{Greedy-Fair}, $0.675$ for \emph{Greedy} and $0.60$ for \emph{NESF}). Thus, \emph{NESF} is the most resilient in this case.
As $P$ increases, the cost values obtained by \emph{NESF} and \emph{Greedy} monotonically decrease like staircases, and finally fast converge to specific points, i.e., $9.70$ for \emph{NESF} and $10.84$ for \emph{Greedy}.
The staircase pattern is due to the fact that the optimal solution remains constant when $P$ varies in a small range, and the decreasing trend is also consistent with the real world case.
However, the cost value for \emph{Greedy-Fair} exhibits an opposite trend. This is due to its strategy that tries to use the maximum number of nodes that the budget $P$ can cover, and distribute the traffic load on all of them. This scheme, thus, results in a waste of computation capacity and cost increase in some situations.
Finally, \emph{NESF} still achieves the best performance, with the following gaps: around $11\%$ to \emph{Greedy} and $16\%$ to \emph{Greedy-Fair}.

\subsubsection{Effect of the computation capacity $D_a$}
Figures \ref{fig_D1}, \ref{fig_D2}, and \ref{fig_D3} illustrate the variations of the objective function value with respect to the three levels of computation capacity $D_a$, which are scaled from $0.8$ to $1.2$ w.r.t. the initial values in Table \ref{init_tab} with a step of $0.01$, still keeping the relation $D_1<D_2<D_3$.
\textcolor{black}{In Figures \ref{fig_D1} and \ref{fig_D2}, the objective function values obtained by the three approaches show very small variation when the computation capacity is scaled. In Figure \ref{fig_D3}, there is a clear decreasing trend for the objective function values achieved by both \emph{Greedy} and \emph{Greedy-Fair}. The reason is that many edge nodes are enabled with the $D_3$ computation level, and the increased $D_3$ capacity reduces much of the total latency while not adding much operation cost. The objective function value achieved by \emph{NESF}, on the other hand, almost does not change.}
To summarize, \emph{NESF} could provide better and more stable solutions, compared with the other approaches.

\subsubsection{Effect of the traffic rate $\lambda^{kn}$}
Figure \ref{fig_lambda} shows the objective function value variation versus the traffic rate.
Values $\lambda^{kn}, kn \in \mathcal{K}\times\mathcal{N}$ are scaled from $0.5$ to $1.0$ with respect to the initial value in Table \ref{init_tab}, with a step of~$0.025$.
As traffic $\lambda^{kn}$ increases, the objective function values for all the approaches increase.
We observe that $NESF$ is characterized by a smooth curve, which indicates stability in the solving processing, while both \emph{Greedy} and \emph{Greedy-Fair} exhibit larger fluctuations.
When the scale is $\leqslant 0.55$, i.e., the traffic rate is relatively low, the cost values for all the approaches are the same since the best configuration, i.e., locally computing of the traffic, is easily identified by all of them.
After that point, \emph{NESF} exhibits a better performance with a clear gap (around $14\%$) with respect to the other approaches.

\subsubsection{Effect of the tolerable latency $\tau_n$}
Figure \ref{fig_tau} illustrates the objective function value with respect to the tolerable latency $\tau_n, n\in\mathcal{N}$ scaled from $1.0$ to $1.5$ on the initial value in Table \ref{init_tab}.
When $\tau_n$ increases, the objective function values obtained by all the approaches decrease and converge to specific points, i.e., around $9.48$ for \emph{NESF}, $10.15$ for \emph{Greedy}, and finally $10.64$ for \emph{Greedy-Fair}.
Parameter $\tau_n$ serves in our model as an upper bound (see constraint \eqref{con_tau}), and limits the solution space.
In fact, with a low $\tau_n$ value, the feasible solution set is smaller and the total cost increases, and vice versa.
Finally, \emph{NESF} performs the best, with the following gaps: around $7\%$ with respect to \emph{Greedy}, and $11\%$ to \emph{Greedy-Fair}.

{\color{black}
We further considered more stringent scenarios where we extended the scaling range of the tolerable latency, $\tau_n$, from $0.75$ to $1.5$. The results are shown in Figure \ref{fig_tau_new},
and are related to the \emph{Citt\`{a} Studi} topology (see Figure \ref{fig_citta_studi_top})
and show that, when latency requirements are very stringent (the left part in these figures) the total cost of the network planned to accommodate such stringent requirements sharply increases (see Figure \ref{fig_tau_j}). Please also note that, for some of these extreme values of the scaling parameter, the \emph{Greedy} and \emph{Greedy-Fair} benchmark algorithms were unable to find a feasible solution, while our proposed heuristics (\emph{NESF}) is always able to find a solution.

\begin{figure*}[!htb]
    \centering
    \captionsetup{skip=3pt}
    \captionsetup[subfigure]{skip=3pt}
    \begin{subfigure}[b]{0.31\linewidth}\centering
        \includegraphics[width=\textwidth]{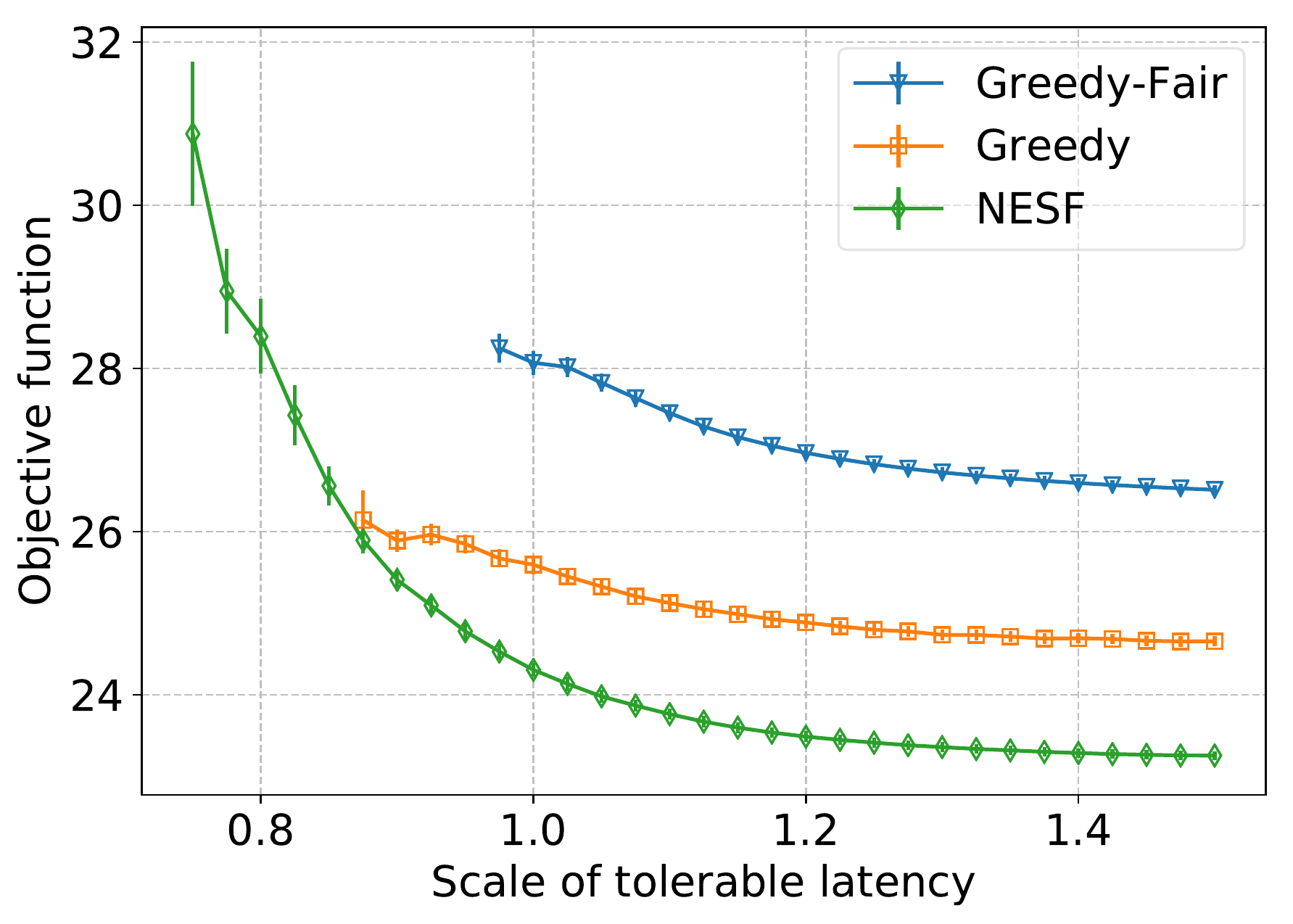}
        \caption{Objective function}
        \label{fig_tau_o}
    \end{subfigure}
    \hfill
    \begin{subfigure}[b]{0.32\linewidth}\centering
        \includegraphics[width=\textwidth]{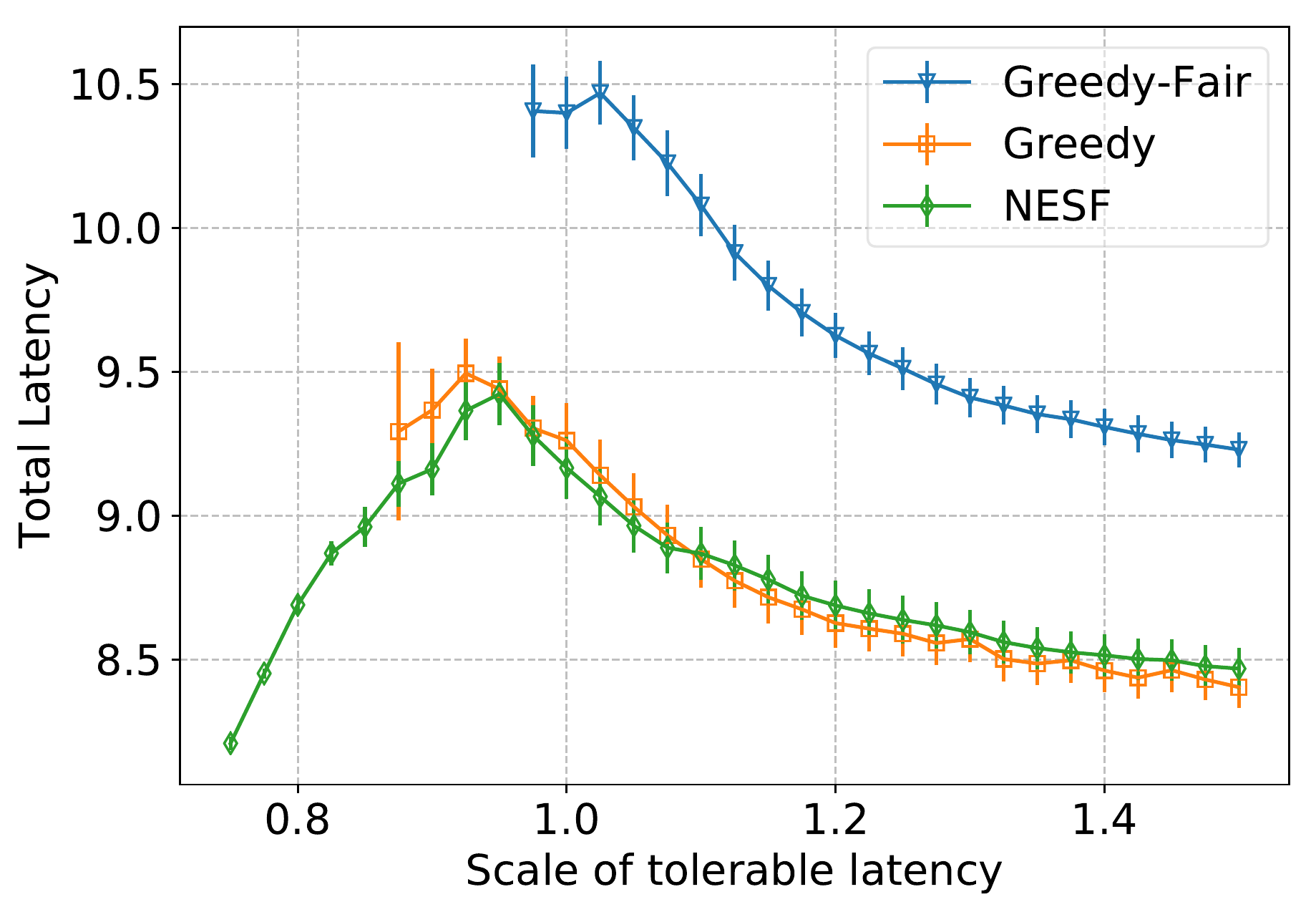}
        \caption{Total latency}
        \label{fig_tau_t}
    \end{subfigure}
    \hfill
    \begin{subfigure}[b]{0.31\linewidth}\centering
        \includegraphics[width=\textwidth]{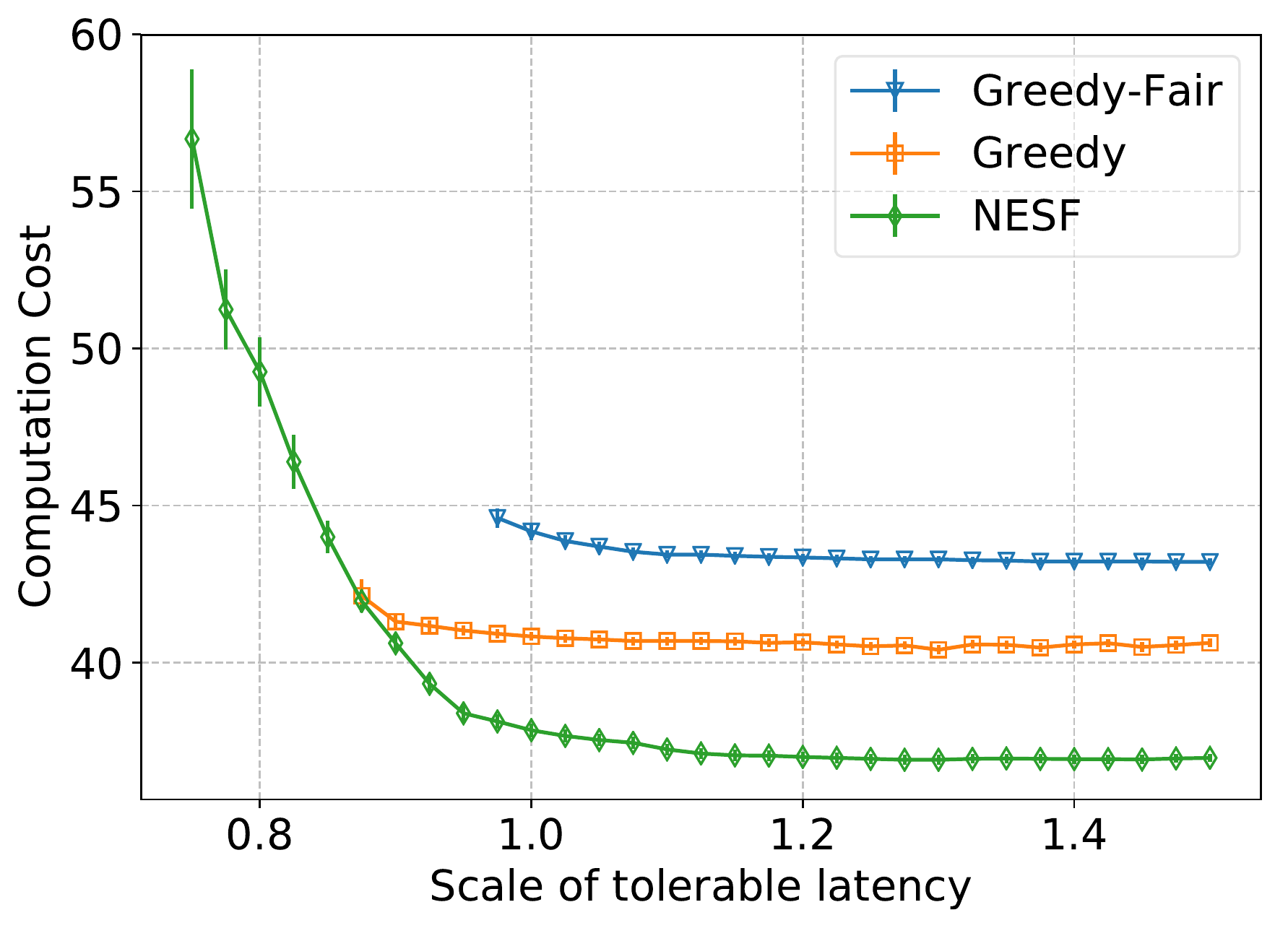}
        \caption{Total network cost}
        \label{fig_tau_j}
    \end{subfigure}
    \caption{Scaling tolerable latency $\tau_n$ from $0.75$ to $1.5$, \emph{Citt\`{a} Studi} topology.}
    \label{fig_tau_new}
    \vspace{-10pt}
\end{figure*}
}

\subsubsection{Effect of the trade-off weight $w$}
This parameter permits to express, in the objective function computation, the relevance of the overall operation cost with respect to the total latency experienced by users. Lower values of $w$ correspond to a lower relevance of the operation cost w.r.t. latency. In 
Figure \ref{fig_w} $w$ is changed from $0$ to $1.0$ with a step of $0.05$.
When $w=0$, the optimization focuses almost exclusively on the total latency. As $w$ increases, the objective function values increase almost linearly for all the approaches.
\textcolor{black}{
The \emph{NESF} algorithm still achieves the best performance, with gaps around $7\%$ with respect to \emph{Greedy} and $16\%$ w.r.t. \emph{Greedy-Fair}.}

{\color{black}

Hereafter we present (Table \ref{tab_w}) numerical results obtained in the ``Citt\`{a} Studi'' topology, to illustrate the impact of the trade-off weight $w$. 
Following the setting of the weight parameter $w$ discussed in Section \ref{subection_OptimizationProblemJPSNC}, which permits to privilege the optimization of the network cost $J$ or the delay $T$, we obtain in this scenario (based on the parameters values), $w_L \approx 0.003$ and $w_U \approx 0.4$.
For simplicity, we select three values for $w$ (viz., 0.003, 0.1, 0.4) to give different priorities to the overall latency and planning cost.

\begin{table}[!t]
    \captionsetup{skip=3pt}
    \caption{Impact of the weight $w$ (solution computed by the \emph{NESF} heuristic).}
    \label{tab_w}
    \centering
    \begin{tabular}{lrrrr}
        \toprule
        & $w$ & $T+wJ$ & $T$ & $J$ \\
        \midrule
        \multicolumn{1}{c|}{\multirow{3}{*}{\tabincell{c}{Scaling \textcolor{black}{link bandwidth} \\ $B_l$ (factor 0.6)}}}
        %                      & $0.003$ & 8.14853386  & 8.00525386 & 47.76 \\
        %\multicolumn{1}{c|}{} & $0.1$   & 12.58340148 & 8.42640148 & 41.57 \\
        %\multicolumn{1}{c|}{} & $0.4$   & 24.35175242 & 9.17575242 & 37.94 \\
                              & $0.003$ & 8.15  & 8.01 & 47.76 \\
        \multicolumn{1}{c|}{} & $0.1$   & 12.58 & 8.43 & 41.57 \\
        \multicolumn{1}{c|}{} & $0.4$   & 24.35 & 9.18 & 37.94 \\
        \midrule
        \multicolumn{1}{c|}{\multirow{3}{*}{\tabincell{c}{Scaling \textcolor{black}{network capacity} \\ $C_k$ (factor 1.5)}}}
        %                      & $0.003$ & 2.21109243 & 2.06985243 & 47.08 \\
        %\multicolumn{1}{c|}{} & $0.1$   & 6.60256042 & 2.85556042 & 37.47 \\
        %\multicolumn{1}{c|}{} & $0.4$   & 17.7240755 & 2.9240755  & 37.00 \\
                              & $0.003$ & 2.21 & 2.07 & 47.08 \\
        \multicolumn{1}{c|}{} & $0.1$   & 6.60 & 2.86 & 37.47 \\
        \multicolumn{1}{c|}{} & $0.4$   & 17.72 & 2.92  & 37.00 \\
        \bottomrule
    \end{tabular}
\end{table}

Let us analyze the results for scaling network capacity~$C_k$ as an example.
If we set $w=0.003$, thus giving priority in the optimization to the minimization of the experienced overall latency $T$, we see that such value is, in average, $2.07$, while the cost of the planned network $J$ is $47.08$. In this case, we tend to plan costlier networks but we can satisfy more stringent latency requirements of users. If on the other hand we set $w=0.4$, thus privileging cost minimization and then reducing latency as second step, we observe that, in average, the latency $T$ is $2.92$ while the average cost of the planned network $J$ is $37.00$. By comparing these two extreme situations we observe that the latency increases of $41\%$, passing from the first scenario to the second, while in parallel the cost reduces of about $21\%$.
Finally, Figure \ref{fig_cw} shows for completeness the whole set of results, that is, the objective function value for the three $w$ settings considered in the previous Table, and for all $C_k$ scaling factors.

\begin{figure*}[!htb]
    \centering
    \captionsetup{skip=3pt}
    \captionsetup[subfigure]{skip=3pt}
    \begin{subfigure}[t]{0.31\linewidth}\centering
        \includegraphics[width=\textwidth]{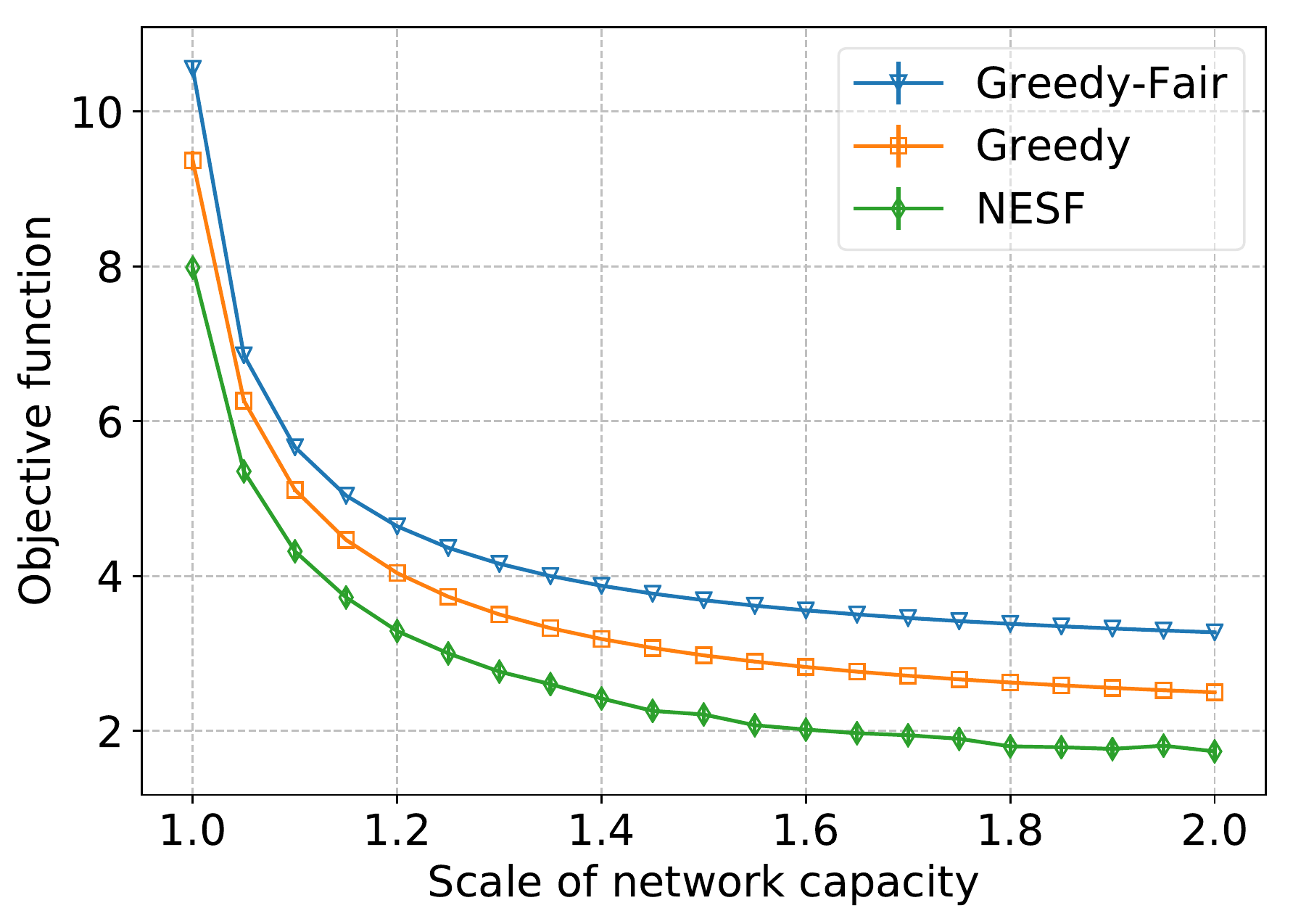}
        \caption{$w=0.003$}
        \label{fig_c_w1}
    \end{subfigure}
    \hfill
    \begin{subfigure}[t]{0.31\linewidth}\centering
        \includegraphics[width=\textwidth]{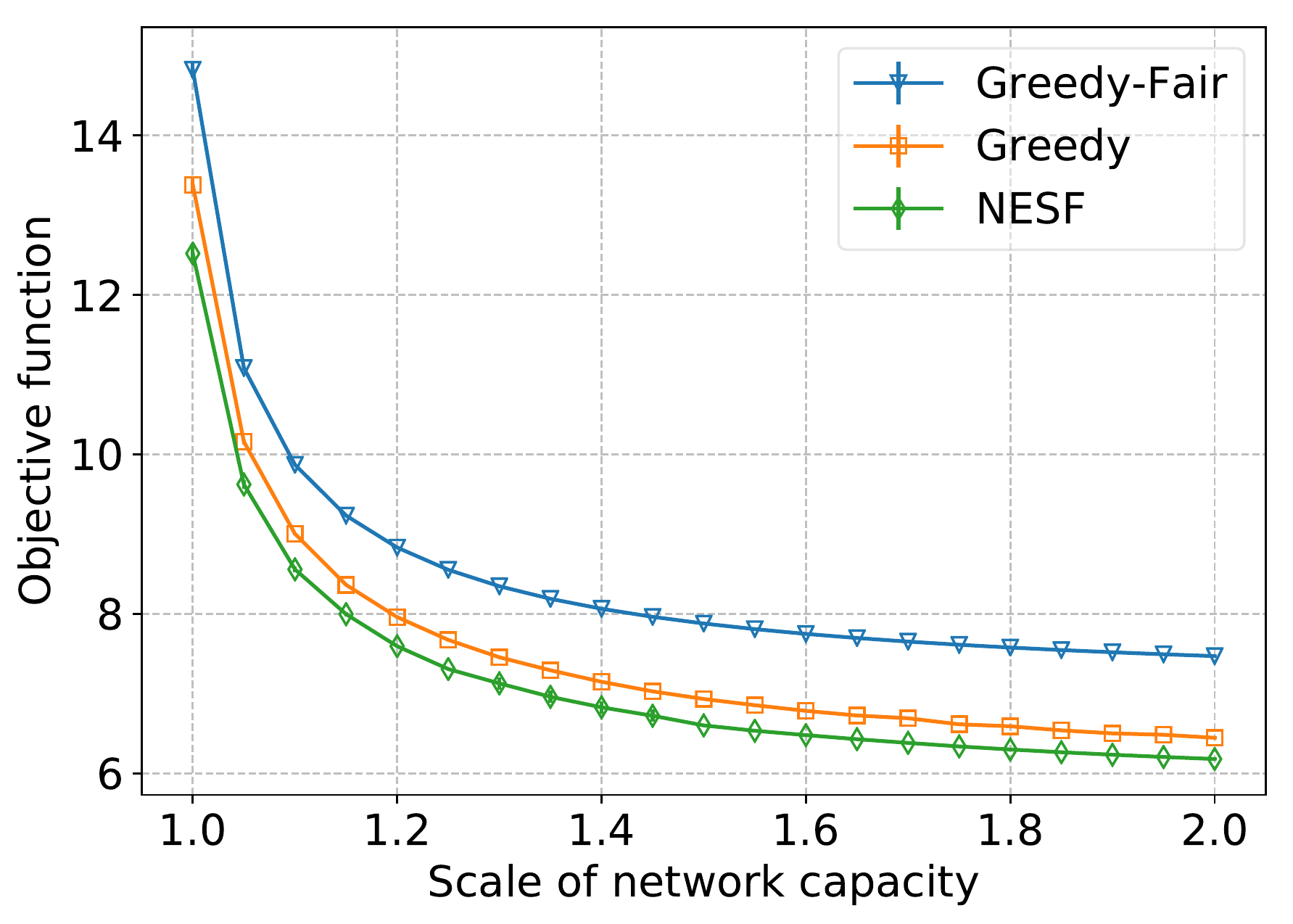}
        \caption{$w=0.1$}
        \label{fig_c_w2}
    \end{subfigure}
    \hfill
    \begin{subfigure}[t]{0.31\linewidth}\centering
        \includegraphics[width=\textwidth]{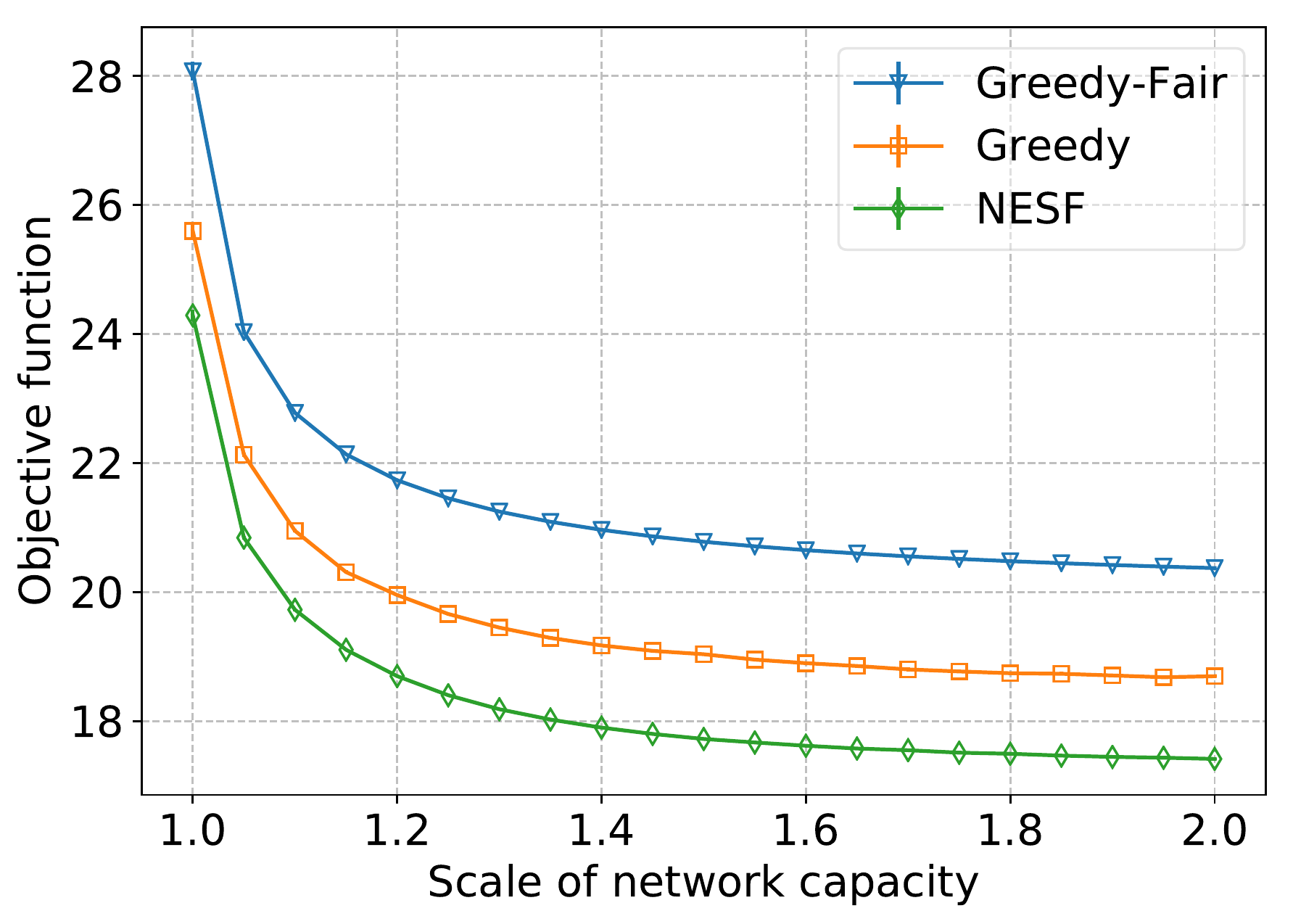}
        \caption{$w=0.4$}
        \label{fig_c_w3bis}
    \end{subfigure}
    \caption{Scaling network capacity $C_k$ under different weight $w$ settings.}
    \label{fig_cw}
    \vspace{-6pt}
\end{figure*}
}

\begin{figure}[!t]
    \centering
    \captionsetup{skip=3pt}
    \includegraphics[width=0.8\linewidth]{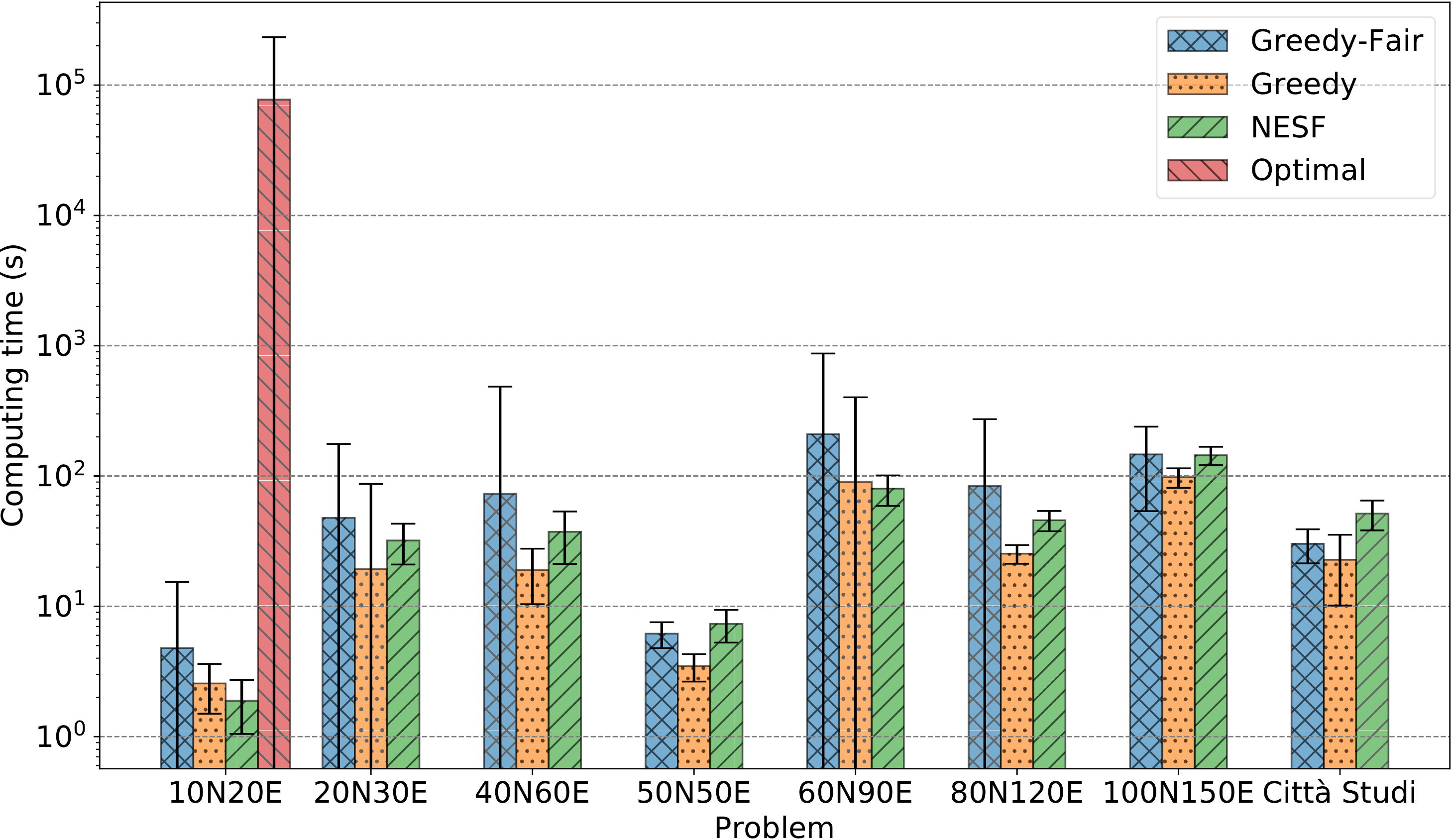}
    \caption{Computing time.}
    \label{fig_time}
    %\vspace{-6pt}
\end{figure}

{\color{black}
\subsubsection{Robustness analysis}

In the same scenario illustrated in Section~\ref{section-systemoverview}, we further quantify the robustness of our proposed model and algorithms. To this aim, we increase the traffic from one ingress node ($\lambda^{n5,t2}$) first from $35$ to $36Gb/s$ and then from $35$ to $40Gb/s$. In both cases, the original scenario, with $\lambda^{n5,t2}=35 Gb/s$, is denoted by the symbol ``$\circ$'' in Tables~\ref{tab_margin_err_n} and \ref{tab_margin_err}, while the changed one is denoted by ``$\ast$''.

We compare the solutions computed by three approaches, where \emph{Optimal} solves the problem optimally, \emph{NESF} is the solution provided by our heuristic, \emph{$\circ$NESF} represents the solution computed for the original instance ($\circ$) by directly applying it to the changed one ($\ast$).
The \emph{Margin} row is computed as \emph{$\circ$NESF} - \emph{NESF}.
We observe that \emph{$\circ$NESF} can directly provide a feasible solution also for the modified scenario with $\lambda^{n5,t2}=36 Gb/s$, very close to the original one in terms of objective function value.

In the second scenario, since traffic increases more consistently (from 35 to 40 Gb/s), we consider a further approach (named \emph{M$\circ$NESF}) to avoid the infeasibility that can be experienced when applying directly, as \emph{$\circ$NESF} does, the solution computed for the original instance ($\circ$) to the changed one ($\ast$). Indeed, all allocation and routing solutions taken for the original problem are still valid (including decisions $c^{kn},\,b^{kn}_i,\,\alpha^{kn}_i,\,\beta^{kn}_i$ and also routing path $\mathcal{R}^{kn}_i$), and we just need to re-optimize planning decisions of computation capacity levels $\delta^a_i$. 
This permits to avoid infeasibility and to obtain very good solutions: in this scenario the objective function of \emph{Optimal} is 2.415, \emph{NESF} 2.479 and \emph{M$\circ$NESF} 2.524, just 1.8\% higher than \emph{NESF}.

\begin{table}[bhtp]
    \setlength{\tabcolsep}{3pt}
    \captionsetup{skip=3pt}
    \caption{\small Robustness analysis for instance \emph{10N20E} ($\circ$: original, $\lambda^{n5,t2}=35Gb/s$; $\ast$: changed scenario with \textbf{$\lambda^{n5,t2}=36Gb/s$)}.}
    \label{tab_margin_err_n}
    \centering
    \begin{tabular}{rr|l|r|l|r|l|r|l}
        \toprule
        & \multicolumn{2}{c|}{$T+wJ$} & \multicolumn{2}{c|}{$T$} & \multicolumn{2}{c|}{$J$} & \multicolumn{2}{c}{Computing time (s)} \\
        \cline{2-3}\cline{4-5}\cline{6-7}\cline{8-9}
        & $\circ$ & $\ast$ & $\circ$ & $\ast$ & $\circ$ & $\ast$ & $\circ$ & $\ast$ \\
        \midrule
        \emph{Optimal} & 2.249 & 2.256 & 1.049 & 1.056 & 12.0 & 12.0 & 38463 & 48521 \\
        \emph{NESF}    & 2.277 & 2.281 & 0.977 & 0.981 & 13.0 & 13.0 & 1.307 & 1.169 \\
        \emph{$\circ$NESF}  & -    & 2.318 & -    & 1.018 & -    & 13.0 & -     & 0.291 \\
        \midrule
        Margin & \multicolumn{2}{r}{0.037} & \multicolumn{2}{r}{0.037} & \multicolumn{2}{r}{0} & \multicolumn{2}{r}{} \\
        \bottomrule
    \end{tabular}
\end{table}

\begin{table}[bhtp]
    \setlength{\tabcolsep}{3pt}
    \captionsetup{skip=3pt}
    \caption{\small Robustness analysis for instance \emph{10N20E} ($\circ$: original, $\lambda^{n5,t2}=35Gb/s$; $\ast$: changed scenario with \textbf{$\lambda^{n5,t2}=40Gb/s$)}.}
    \label{tab_margin_err}
    \centering
    \begin{tabular}{rr|l|r|l|r|l|r|l}
        \toprule
        & \multicolumn{2}{c|}{$T+wJ$} & \multicolumn{2}{c|}{$T$} & \multicolumn{2}{c|}{$J$} & \multicolumn{2}{c}{Computing time (s)} \\
        \cline{2-3}\cline{4-5}\cline{6-7}\cline{8-9}
        & $\circ$ & $\ast$ & $\circ$ & $\ast$ & $\circ$ & $\ast$ & $\circ$ & $\ast$ \\
        \midrule
        \emph{Optimal} & 2.249 & 2.415 & 1.049 & 1.115 & 12.0 & 13.0 & 38463 & 581077 \\
        \emph{NESF}    & 2.277 & 2.479 & 0.977 & 0.979 & 13.0 & 15.0 & 1.307 & 1.273 \\
        \midrule
        \emph{$\circ$NESF} & \multicolumn{8}{c}{directly applying $\circ$ solution to $\ast$: \emph{Infeasible}} \\
        \midrule
        \emph{M$\circ$NESF}  & -    & 2.524 & -    & 1.224 & -    & 13.0 & -     & 0.339 \\
        \midrule
        Margin & \multicolumn{2}{r}{0.045} & \multicolumn{2}{r}{0.245} & \multicolumn{2}{r}{-2} & \multicolumn{2}{r}{} \\
        \bottomrule
    \end{tabular}
\end{table}

}

\subsection{Computing Time}\label{subsection-time}

Figure \ref{fig_time} compares the average computing time of the proposed approaches under all considered network topologies. The computing time for $\mathcal{P}\zcal{1}$ is shown only for the smallest topology and it is already significantly larger than the others.
For the tree-shaped network topology (Figure~\ref{topo_tree_shape}), all approaches are able to obtain the solution very fast, in less than $10s$. This is due to the fact that routing optimization is indeed trivial in such topology.
The computing time is ordered as: \emph{Greedy}$<$\emph{NESF}$<$\emph{Greedy-Fair}.
When considering standard deviation, the order is: \emph{NESF}$<$\emph{Greedy}$<$\emph{Greedy-Fair}, and this shows the stability of our proposed approach in the solving process.
As for the network topology with $100$ nodes and $150$ edges (a general large scale network), \emph{NESF} is able to obtain a good solution in around $100s$, and remains below this value in the other considered cases.
This gives us an indication that the network management component can periodically run \emph{NESF} as a response to changes in the network or in the incoming traffic, and optimize nodes computation capacities and routing paths accordingly. This is a key feature for providing the necessary QoS levels in next-generation mobile network architectures and for updating it dynamically.

\section{Related Work}
\label{section-relatedwork}
Several works have been recently published on the resource management problem in a MEC environment; most of them consider a single mobile edge cloud at the ingress node and do not account for its connection to a larger edge cloud network~\cite{wang2017computation,mao2017stochastic,ma2017cost}.
The following of this section provides a short overview on the various areas that are relevant to the problem we consider. As discussed in the \textbf{Summary} part, ours is the first approach that considers at the same time multiple aspects related to the configuration of an edge cloud network.

\textbf{Network planning:}
The network planning problem in a MEC/Fog/Cloud context tackles the problems concerning nodes placement, traffic routing and computation capacity configuration.
The authors in \cite{ceselli2017mobile} propose a mixed integer linear programming (MILP) model to study cloudlet placement, assignment of access points (APs) to cloudlets and traffic routing problems, by minimizing installation costs of network facilities.
The work in \cite{santoyo2018latency} proposes a MILP model for the problem of fog nodes placement under capacity and latency constraints.
\cite{ma2019cost} presents a model to configure the computation capacity of edge hosts and adjust the cloud tenancy strategy for dynamic requests in cloud-assisted MEC to minimize the overall system cost.

\textbf{Service/content placement:}
\textcolor{black}{The service and content placement problems are considered in several contexts including, among others, micro-clouds, multi-cell MEC etc.}
The work in \cite{wang2016dynamic} studies the dynamic service placement problem in mobile micro-clouds to minimize the average cost over time. The authors first propose an offline algorithm to place services using predicted costs within a specific look-ahead time-window, and then improve it to an online approximation one with polynomial time-complexity.
An integer linear programming (ILP) model is formulated in \cite{he2018s} for serving the maximum number of user requests in edge clouds by jointly considering service placement and request scheduling. The edge clouds are considered as a pool of servers without any topology, which have shareable (storage) and non-shareable (communications, computation) resources. Each user is also limited to use one edge server.
In \cite{farhadi2019service}, the authors extend the work in \cite{he2018s} by separating the time scales of the two decisions: service placement (per frame) and request scheduling (per slot) to reduce the operation cost and system instability.
In \cite{poularakis2019joint}, the authors study the joint service placement and request routing problem in multi-cell MEC networks to minimize the load of the centralized cloud. No topology is considered for the MEC networks. A randomized rounding (RR) based approach is proposed to solve the problem with a provable approximation guarantee for the solution, i.e., the solution returned by RR is at most a factor (more than~3) times worse than the optimum with high probability. However, although it offers an important theoretical result, the guarantee provided by the RR approach is only specific to the formulated optimization problem. %, and relatively far from optimality. %More in general, RR techniques, if employed alone, can exhibit problems of robustness, and may also tend to generate infeasible solutions.
\cite{wang2018service} studies the problem of service entities placement for social virtual reality (VR) applications in the edge computing environment.
\cite{zhang2018optimal} analyzes the mixed-cast packet processing and routing policies for service chains in distributed computing networks to maximize network throughput.

The work in \cite{pu2018online} studies the edge caching problem in a Cloud RAN (C-RAN) scenario, by jointly considering the resource allocation, content placement and request routing problems, aiming at minimizing the system costs over time.
\cite{chen2018joint} formulates a joint caching, computing and bandwidth resources allocation model to minimize the energy consumption and network usage cost. The authors consider three different network topologies (ring, grid and a hypothetical US backbone network, US64), and abstract the fixed routing paths from them using the OSPF routing algorithm.

\textbf{Cloud activation/selection:}
\textcolor{black}{The cloud activation and selection problems are studied as a way to handle the configuration of computation capacity in a MEC environment.}
The authors in~\cite{wang2018cooperative} design an online optimization model for task offloading with a sleep control scheme to minimize the long term energy consumption of mobile edge networks. The authors use a Lyapunov-based approach to convert the long term optimization problem to a per-slot one. No topology is considered for the MEC networks.
\cite{Wang2019icc} proposes a model to dynamically switch on/off edge servers and cooperatively cache services and associate users in mobile edge networks to minimize energy consumption.
\cite{Opadere2019icc} jointly optimizes the active base station set, uplink and downlink beamforming vector selection, and computation capacity allocation to minimize power consumption in mobile edge networks.
\cite{Wu2019icc} proposes a model to minimize a weighted sum of energy consumption and average response time in MEC networks, which jointly considers the cloud selection and routing problems. A population game-based approach is designed to solve the optimization problem.

\textbf{Network slicing:}
The authors in \cite{fossati2019multi} study the resource allocation problem in network slicing where multiple resources have to be shared and allocated to verticals (5G end-to-end services).
\cite{leconte2018resource} formulates a resource allocation problem for network slicing in a cloud-native network architecture, which is based on a utility function under the constraints of network bandwidth and cloud power capacities. For the slice model, the authors consider a simplified scenario where each slice serves network traffic from a single source to a single destination. For the network topology, they consider a 6x6 square grid and a 39-nodes fat-tree.

\textbf{Other perspectives:}
Inter-connected datacenters also share some common research problems with the multi-MEC system.
The work in \cite{Xu2017tcc} studies the joint resource provisioning for Internet datacenters to minimize the total cost, which includes server provisioning, load dispatching for delay sensitive jobs, load shifting for delay-tolerant jobs, and capacity allocation.
\cite{li2016cost} presents a bandwidth allocation model for inter-datacenter traffic to enforce bandwidth guarantees, minimize the network cost, and avoid potential traffic overload on low cost links.

The work in \cite{dinh2017offloading} studies the problem of task offloading from a single device to multiple edge servers to minimize the total execution latency and energy consumption by jointly optimizing task allocation and computational frequency scaling.
In \cite{cheng2018icc}, the authors study task offloading and wireless resource allocation in an environment with multiple MEC servers.
\cite{wang2018dynamic} formulates an optimization model to maximize the profit of a mobile service provider by jointly scheduling network resources in C-RAN and computation resources in MEC.

\textbf{Summary:}
To the best of our knowledge, our paper is the first to propose a complete approach that encompasses both the problem of \textit{planning} cost-efficient edge networks and \textit{allocating resources}, performing optimal routing and minimizing the total traffic latency of transmitting, outsourcing and processing user traffic, under a constraint of user tolerable latency for each class of traffic. We model accurately both link and processing latency, using non-linear functions, and propose both exact models and heuristics that are able to obtain near-optimal solutions also in large-scale network scenarios, that include hundreds of nodes and edges, as well as several traffic flows and classes.

\vspace{-1em}
\section{Conclusion and Future Directions}
\label{section-conclusion}
In this paper, we studied the problem of jointly planning and optimizing the resource management of a mobile edge network infrastructure. We formulated an exact optimization model, which takes into accurate account all the elements that contribute to the overall latency experienced by users, a key performance indicator for these networks, and further provided an effective heuristics that computes near-optimal solutions in a short computing time, as we demonstrated in the detailed numerical evaluation we conducted in a set of representative, large-scale topologies, that include both mesh and tree-like networks, spanning wide and meaningful variations of the parameters' set.

We measured and quantified how each parameter has a distinct impact on the network performance (which we express as a weighted sum of the experienced latency and the total network cost) both in terms of strength and form. Traffic rate and network capacity have the stronger effects, and this is consistent with real network cases. Tolerable latency shows an interesting effect: the lower requirements on latency (or equivalently: the higher value of tolerable latency) the system sets, the lower latency and costs the system will have. This information can be useful for network operators to design the network indicators of services. The computation capacity has relatively smaller effect on the network performance, compared with the other parameters. Another key observation that we draw from our numerical analysis is that as the system capacities (including link bandwidth, network capacity and computation capacity budget) increase, the system performance converges to a plateau, which means that increasing the system capacity over a certain level (which we quantify for each network scenario) will have small effectiveness, and on the contrary, it will increase the total system cost.

\textcolor{black}{Finally, we observe that our models can be extended within the theoretical framework of stochastic optimization, which can be used to guarantee robustness of the solution with respect to the uncertainty in the probabilistic description of traffic demands. Possible extensions of our model could further include explicit modeling of resource scaling across clusters, of VM state and storage synchronization as well as IaaS internal traffic across edge facilities.
}

\section*{Acknowledgment}
{\small
This research was supported by the H2020-MSCA-ITN-2016 SPOTLIGHT under grant agreement No. 722788 and the H2020-ICT-2020-1 PIACERE under grant agreement No. 101000162.\par}

\renewcommand*{\bibfont}{\footnotesize}
\printbibliography
%\vspace{-1em}
%\input{author_biographies}

\clearpage
\appendices
\section{Problem Reformulation}
\label{section-problemreformulation}

Problem $\mathcal{P}\zcal{0}$ formulated in Section \ref{section-problemformulation} cannot be solved directly and efficiently due to the reasons detailed in Section~\ref{subsection-reformulationShort}.

To deal with these problems, we propose in this Appendix an equivalent reformulation of $\mathcal{P}\zcal{0}$, which can be solved very efficiently with the Branch and Bound method. Moreover, the reformulated problem can be further relaxed and, based on that, we propose an heuristic algorithm which can get near-optimal solutions in a short computing time.

To this aim, we first reformulate the processing latency and link latency constraints (viz., constraints~(\ref{eq_tp}) and~(\ref{eq_tl})), and we deal at the same time with the computation planning problem. Then, we handle the difficulties related to variables~$\mathcal{R}^{kn}_i$ and the corresponding routing constraints.

\subsection{Processing Latency} \label{subsec_pl}
In equation \eqref{eq_tp}, the variable $\beta^{kn}_i$ and the function $S_i$ connect the computation capacity allocation and planning problem together, and the processing latency $t^{kn,i}_P$ has therefore a highly nonlinear expression.
To handle this problem, we first introduce an auxiliary variable $p^{kn,a}_i = \beta^{kn}_i\delta^a_i$.
Then, $\beta^{kn}_i S_i$ is replaced by a linearized form $\beta^{kn}_i S_i = \sum_{a\in\mathcal{A}}p^{kn,a}_i D_a$.
\textcolor{black}{
Furthermore, we linearize $p^{kn,a}_i = \beta^{kn}_i\delta^a_i$, which is the product of binary and continuous variables, as follows:
}
\begin{equation}
    \left\{\begin{array}{l}
        0\leqslant p^{kn,a}_i \leqslant \delta^a_i, \\
        0 \leqslant \beta^{kn}_i - p^{kn,a}_i \leqslant 1-\delta^a_i, \label{con_p_beta_delta}
    \end{array}\right.
    \forall k,\forall n,\forall a,\forall i.
\end{equation}

According to the definitions of $\alpha^{kn}_i$ and $b^{kn}_i$, we have the following constraint:
\begin{equation}
    \alpha^{kn}_i\leqslant b^{kn}_i\leqslant M\alpha^{kn}_i,\; \forall k,\forall n,\forall i,\label{con_b_alpha}
\end{equation}
where $M>0$ is a big value; such constraint implies that if $\alpha^{kn}_i=0$, the traffic $kn$ is not processed on node $i$, i.e. $b^{kn}_i=0$.

Based on the above, we can rewrite constraint \eqref{con_4_tp} as:
\begin{equation}
    \left\{\begin{array}{l}
        \alpha^{kn}_i \lambda^{kn}-(1-b^{kn}_i) < \sum_{a\in\mathcal{A}}p^{kn,a}_i D_a,\\ \beta^{kn}_i\leqslant b^{kn}_i,
    \end{array}\right.\!\!
    \forall k,\forall n,\forall i. \label{con_4_tp_new}
\end{equation}

Note that the term $(1-b^{kn}_i)$ permits to implement condition $\alpha^{kn}_i > 0$ in Eq. \eqref{con_4_tp}.

In equation \eqref{eq_tp}, we observe that if $b^{kn}_i=1$, we have:
\[\frac{1}{\beta^{kn}_i S_i - \alpha^{kn}_i \lambda^{kn}} > \frac{1}{S_i}\geqslant\frac{1}{\max_{j\in\mathcal{E}}S_j},\]
otherwise $\beta^{kn}_i S_i - \alpha^{kn}_i \lambda^{kn}=0$ resulting in $t^{kn,i}_P\to\infty$.
To handle this case, we first define a new variable $t^{kn,i}_{P'}$ as follows:
\begin{equation}
    t^{kn,i}_{P'} = \frac{1}{\sum_{a\in\mathcal{A}}p^{kn,a}_i D_a - \alpha^{kn}_i \lambda^{kn}+(1-b^{kn}_i)D_m}, \label{eq_tpp}
\end{equation}
where $D_m$ is the maximum computation capacity that can be installed on a node ($D_m = \max_{a\in\mathcal{A}} D_a$).

From this equation, we have $b^{kn}_i=1\Rightarrow t^{kn,i}_{P'}=t^{kn,i}_P>\frac{1}{D_m}$ and $b^{kn}_i=0\Rightarrow t^{kn,i}_{P'}=\frac{1}{D_m},\;t^{kn,i}_P=0$.
Hereafter, we prove that this reformulation has no influence on the solution of our optimization problem.

The outsourcing latency is defined as the maximum of the processing latency $t^{kn,i}_P$ and link latency $t^{kn,i}_L$ among all nodes. Equation \eqref{eq_t_pl} can be transformed as $t^{kn}_{PL}\geqslant t^{kn,i}_P+t^{kn,i}_L, \forall k,\forall n,\forall i$.
When $b^{kn}_i=0$, $t^{kn,i}_P=t^{kn,i}_L=0$. Thus, based on above, the inequality is equivalent to $t^{kn}_{PL}\geqslant t^{kn,i}_{P'}+t^{kn,i}_L, \forall k,\forall n,\forall i.$

\subsection{Link Latency}
As we stated before, to compute the link latency, we need to determine the routing path $\mathcal{R}^{kn}_i$, and this problem will be specifically handled in the next subsection.
Assuming $\mathcal{R}^{kn}_i$ has been determined, we first introduce a binary variable~$\gamma^{kn,i}_l$ defined as follows:
\begin{equation*}
    \gamma^{kn,i}_l = \left\{
    \begin{array}{cl}
        1, & \text{if } l\in\mathcal{R}^{kn}_i,\\
        0, & \text{otherwise},
    \end{array}\right.
    \forall k,\forall n,\forall i,\forall l.
\end{equation*}
which indicates whether $l$ is used in the routing path $\mathcal{R}^{kn}_i$ or not.
Note that only if traffic $kn$ is processed on node $i$ (i.e., $b^{kn}_i=1$) and $i\neq k$, the corresponding routing path is defined.
Then we have:
\begin{equation}
    \left\{\begin{array}{ll}
        \gamma^{kn,k}_l=0, & \forall k,\forall n,\forall l, \\
        \gamma^{kn,i}_l \leqslant b^{kn}_i, & \forall k,\forall n,\forall i,\forall l.
    \end{array}\right. \label{con_r}
\end{equation}
We now introduce variable $v_l$, defined as follows:
\begin{equation}
    v_l=\frac{1}{B_l - \sum\limits_{k'\in\mathcal{K}}\sum\limits_{n'\in\mathcal{N}} f^{k' n'}_l\lambda^{k' n'}},\; \forall l.\label{eq_vl}
\end{equation}
This permits to transform equation \eqref{eq_tl} as $t^{kn,i}_L = \sum\nolimits_{l\in\mathcal{L}}\gamma^{kn,i}_l v_l$.
We then need to linearize the product of the binary variable $\gamma^{kn,i}_l$ and the continuous variable $v_l$, and to this aim we introduce an auxiliary variable $g^{kn,i}_l=\gamma^{kn,i}_l v_l$, thus also eliminating $t^{kn,i}_L$.
Specifically, we first compute the value range of $v_l$ as follows:
\[B^{-1}_l \leqslant v_l \leqslant V_l = \frac{1}{\max\{B_l-\sum\limits_{k\in\mathcal{K}}\sum\limits_{n\in\mathcal{N}}\lambda^{kn},\epsilon\}},\]
where $\epsilon>0$ is a small value. Based on the above, the linearization is performed by the following constraints.
\begin{equation}
    \left\{\begin{array}{l}
        \gamma^{kn,i}_l B^{-1}_l\leqslant g^{kn,i}_l \leqslant \gamma^{kn,i}_l V_l,\\
        (1-\gamma^{kn,i}_l) B^{-1}_l\leqslant v_l - g^{kn,i}_l \leqslant (1-\gamma^{kn,i}_l)V_l.
    \end{array}\right.\label{con_g_r_v}
\end{equation}
At the same time, the link latency is rewritten as $\sum\nolimits_{l\in\mathcal{L}}g^{kn,i}_l$.

\subsection{Routing Path}
Based on the definitions introduced in the previous subsection, the traffic flow $f^{kn}_l$ can be transformed as:
\begin{equation}
    f^{kn}_l = \sum\nolimits_{i\in\mathcal{E}}\gamma^{kn,i}_l\alpha^{kn}_i.\label{eq_f}
\end{equation}
Due to the product of binary and continuous variables,
$h^{kn,i}_l = \gamma^{kn,i}_l\alpha^{kn}_i$ is introduced for linearization, as follows:
\begin{equation}
    \left\{\begin{array}{l}
        0 \leqslant h^{kn,i}_l \leqslant \gamma^{kn,i}_l, \\
        0 \leqslant \alpha^{kn}_i - h^{kn,i}_l \leqslant 1-\gamma^{kn,i}_l.
    \end{array}\right.\label{con_h_g_a}
\end{equation}

Now we need to simplify the traffic flow conservation constraint (see Eq. \eqref{con_fcc}).
To this aim, and to simplify notation, we first introduce in the network topology a ``dummy'' entry node $0$ which connects to all ingress nodes $k\in\mathcal{K}$. All traffic is coming through this dummy node and going to each ingress node with volume $\lambda^{kn}$, i.e. $f^{kn}_l=1,\forall k,\forall n,\forall l\in \mathcal{F}$, where $\mathcal{F}$ is the dummy link set defined as $\mathcal{F}=\{(0,k)\,|\,k\in\mathcal{K}\}$. Then, we extend the definition of $\mathcal{I}_i$ to $\mathcal{I}_i=\{j\in\mathcal{E}\,|\,(j,i)\in\mathcal{L}\cup\mathcal{F}\}$. Equation \eqref{con_fcc} is hence transformed as:
\begin{equation}
    \sum\limits_{j\in\mathcal{I}_i} f^{kn}_{ji} - \sum\limits_{j\in\mathcal{O}_i} f^{kn}_{ij}=\alpha^{kn}_i,\; \forall k,\forall n,\forall i.\label{con_fcc_new}
\end{equation}

Correspondingly, we add the following constraints to the set $\mathcal{F}$ of dummy links:
\begin{equation}
    \left\{\begin{array}{ll}
        \gamma^{kn,i}_{0k}=b^{kn}_i, & \forall k,\forall n,\forall i,\\
        \gamma^{kn,i}_{0k'}=0, & \forall k,\forall n,\forall i,\forall k'\neq k.
    \end{array}\right. \label{con_r_b}
\end{equation}

\textcolor{black}{
The final stage of our procedure is the definition of the constraints that guarantee all desirable properties that a routing path must respect: the fact that a \textit{single path} (traffic is unsplittable) is used, the flow conservation constraints that provide \textit{continuity} to the chosen path, and finally the absence of \textit{cycles} in the routing path $\mathcal{R}^{kn}_i$.
We would like to highlight that the traffic $kn$ can be only split at ingress node $k$, and each proportion of such traffic is destined to an edge node $i$; this is why we have multiple routing paths $\mathcal{R}^{kn}_i,i\in\{1,2,\cdots\}$.
}

To this aim, we introduce the following conditions, and prove that satisfying them along with the constraints illustrated before can guarantee that such properties are respected:
\begin{itemize}
    \item For an arbitrary node $i$, the number of ingress links used by a path $\mathcal{R}^{kn}_{i'}$ is one, and thus variables $\gamma^{kn,i'}_{ji}$ should satisfy the following condition:
        \begin{equation}
            \sum\nolimits_{j\in\mathcal{I}_i} \gamma^{kn,i'}_{ji} \leqslant 1,\;\forall k,\forall n,\forall i,i'. \label{cond1}
        \end{equation}
    \item The flow conservation constraint (see Eq. \eqref{con_fcc_new}) implements the continuity of a traffic flow.
    \item Every routing path should have an end or a destination to avoid loops. This can be ensured by the following equation:
        \begin{equation}
            \gamma^{kn,i}_{ij}=0,\;\forall k,\forall n,\forall (i,j)\in\mathcal{L}. \label{cond3}
        \end{equation}
\end{itemize}

\noindent\textcolor{black}{The proof is as follows:}

a) Substitute Eq. \eqref{eq_f} into \eqref{con_fcc_new} and make the transformation:
\begin{align*}
    &\sum\limits_{j\in\mathcal{I}_i}\sum\limits_{i'\in\mathcal{E}} \gamma^{kn,i'}_{ji}\alpha^{kn}_{i'} - \sum\limits_{j\in\mathcal{O}_i}\sum\limits_{i'\in\mathcal{E}} \gamma^{kn,i'}_{ij} \alpha^{kn}_{i'}\\
    = & \sum\limits_{i'\in\mathcal{E}}\alpha^{kn}_{i'}\sum\limits_{j\in\mathcal{I}_i} \gamma^{kn,i'}_{ji} - \sum\limits_{i'\in\mathcal{E}}\alpha^{kn}_{i'}\sum\limits_{j\in\mathcal{O}_i} \gamma^{kn,i'}_{ij}\\
    = & \sum\limits_{i'\in\mathcal{E}}\alpha^{kn}_{i'}(\sum\limits_{j\in\mathcal{I}_i} \gamma^{kn,i'}_{ji} - \sum\limits_{j\in\mathcal{O}_i} \gamma^{kn,i'}_{ij})=\alpha^{kn}_i
\end{align*}

b) Based on constraints \eqref{con_r} and \eqref{con_r_b}, we have:
\[\text{if }\alpha^{kn}_{i'}=0, \text{ then } \sum\limits_{j\in\mathcal{I}_i} \gamma^{kn,i'}_{ji} - \sum\limits_{j\in\mathcal{O}_i} \gamma^{kn,i'}_{ij} = 0.\]

c) From a) and b), we have:
\begin{empheq}[left=\empheqlbrace]{alignat*=4}
    &\sum\limits_{j\in\mathcal{I}_i}\gamma^{kn,i}_{ji}&&-\sum\limits_{j\in\mathcal{O}_i}\gamma^{kn,i}_{ij}&&=1,&&\;\forall k,\forall n,\forall i\;|\;\alpha^{kn}_i>0,\\
    &\sum\limits_{j\in\mathcal{I}_i}\gamma^{kn,i'}_{ji}&&- \sum\limits_{j\in\mathcal{O}_i}\gamma^{kn,i'}_{ij}&&=0,&&\;\forall k,\forall n,\forall i,\forall i'\neq i.
\end{empheq}

d) Based on c), constraint \eqref{con_r_b}, conditions \eqref{cond1} and \eqref{cond3} can be written as:
\begin{empheq}[left=\empheqlbrace]{alignat=2}
    &\sum\limits_{j\in\mathcal{I}_k}\gamma^{kn,i}_{jk}  &&=1,\;\forall k,\forall n,\forall i\;|\;\alpha^{kn}_i>0,\label{con_d1}\\
    &\sum\limits_{j\in\mathcal{I}_i}\gamma^{kn,i}_{ji}  &&=1,\;\forall k,\forall n,\forall i\;|\;\alpha^{kn}_i>0,\label{con_d2}\\
    &\sum\limits_{j\in\mathcal{I}_i}\gamma^{kn,i'}_{ji} &&= \sum\limits_{j\in\mathcal{O}_i}\gamma^{kn,i'}_{ij}\leqslant 1,\;\forall k,\forall n,\forall i,\forall i'\neq i.\label{con_d3}
\end{empheq}

Their practical meaning is explained as follows:

\begin{itemize}
    \item \eqref{con_d1} ensures \textcolor{black}{$(0,k)$ to be the first link in any routing path $\mathcal{R}^{kn}_i$ if $\alpha^{kn}_i>0$},
    \item \eqref{con_d2} ensures \textcolor{black}{$i$ to be the end node of the last link in any routing path $\mathcal{R}^{kn}_i$ if $\alpha^{kn}_i>0$},
    \item \eqref{con_d3} ensures that \textcolor{black}{if $i\in\mathcal{E}\backslash\{i'\}$ is an intermediate node in a routing path $\mathcal{R}^{kn}_{i'}$, $i$ should have only one input link and one output link. It also indicates the continuity of a traffic flow}.
\end{itemize}

e) \textcolor{black}{Given a non-empty routing path $\mathcal{R}^{kn}_{i'}$} ($\alpha^{kn}_{i'}>0$), check the validity by using the following conditions:

\begin{itemize}
    \item Let $i=k$ in \eqref{con_d3}, then based on \eqref{con_d1}, $\sum\limits_{j\in\mathcal{O}_k} \gamma^{kn,i'}_{kj}=1$;
    \item Assume $(k,j')$ is a link of $\mathcal{R}^{kn}_{i'}$, then $\gamma^{kn,i'}_{kj'}=1$.
    \item If $j'=i'$, then the path is found, otherwise, continue with the following steps:
    \item Let $i=j'$ in \eqref{con_d3}, due to $\gamma^{kn,i'}_{kj'}=1$, $\sum\limits_{j\in\mathcal{O}_{j'}} \gamma^{kn,i'}_{j'j} = 1$;
    \item Assume $(j',j'')$ is a link of $\mathcal{R}^{kn}_{i'}$, then $\gamma^{kn,i'}_{j'j''}=1$.
    \item Check $j''=i'$ in the same way as the above steps, the whole path $k\to i'$ must be found.
\end{itemize}

Thus, if all the conditions are satisfied, $\mathcal{R}^{kn}_{i'}$ must be a valid routing path having the three properties (unsplittability, traffic continuity, absence of cycles).

\subsection{Final Reformulated Problem}
Based on the reformulation of routing and the demonstrations in the above subsections, the flow conservation constraints can be further improved and the flow variable $f^{kn}_{ij}$ can be eliminated as follows:
\begin{empheq}[left=\empheqlbrace]{alignat=3}
    &\sum\limits_{j\in\mathcal{I}_i} \gamma^{kn,i}_{ji} && = b^{kn}_i,\;&&\forall k,\forall n,\forall i,\label{con_sr_b}\\
    &\sum\limits_{j\in\mathcal{I}_i} \gamma^{kn,i'}_{ji}&& = \sum\limits_{j\in\mathcal{O}_i} \gamma^{kn,i'}_{ij},\;&&\forall k,\forall n,\forall i,\forall i'\neq i.\label{con_r_r}
\end{empheq}

Equation \eqref{eq_T} contains a maximization form, to get rid of which we use a standard technique by introducing variable $T_n=\max_{k\in\mathcal{K}}\{t^{kn}_W + t^{kn}_{PL}\}$ and linearize it as $T_n\geqslant t^{kn}_W + t^{kn}_{PL}, \forall k,\forall n$ (in Section \ref{subsec_pl}, a similar transformation has been performed on $t^{kn}_{PL}$ (see Eq. \eqref{eq_t_pl})).
Since the arguments of the two maximizations are independent, based on the reformulation of processing latency, equation \eqref{con_tau} can be transformed as:
\begin{equation}
    t^{kn}_W + t^{kn,i}_{P'} + \sum\nolimits_{l\in\mathcal{L}}g^{kn,i}_l \leqslant T_n \leqslant \tau_n,\; \forall k,\forall n,\forall i.\label{con_T}
\end{equation}

Finally, the equivalent reformulation of $\mathcal{P}\zcal{0}$ can be written as:

\begin{alignat}{1}
    \mathcal{P}\zcal{1}: \min_{\substack{c^{kn},b^{kn}_i,\alpha^{kn}_i,\\\beta^{kn}_i,\delta^a_i,\gamma^{kn,i}_l}} & \; \sum_{n\in\mathcal{N}}T_n + w \sum_{i \in \mathcal{E}} \kappa_i S_i, \notag \\
    \text{s.t.} \quad\;\;
        & \; \eqref{eq_si}-\eqref{con_P},\,\eqref{eq_tw}-\eqref{con_cn_n},\,\eqref{con_p_beta_delta}-\eqref{con_g_r_v},\notag \\
        & \; \eqref{con_h_g_a},\,\eqref{con_r_b}-\eqref{cond3},\,\eqref{con_sr_b}-\eqref{con_T}. \notag
\end{alignat}

In problem $\mathcal{P}\zcal{1}$, $c^{kn},b^{kn}_i,\alpha^{kn}_i,\beta^{kn}_i,\delta^a_i$ and $\gamma^{kn,i}_l$ are the main decision variables, while other auxiliary variables like $T_n, S_i, h^{kn,i}_l, v_l$, etc. are not shown here for simplicity.
All the variables are bounded.
Since constraints \eqref{eq_tw}, \eqref{eq_tpp} and \eqref{eq_vl} are quadratic while the others are linear, $\mathcal{P}\zcal{1}$ is a mixed-integer quadratically constrained programming (MIQCP) problem, for which commercial and freely available solvers can be used, as we discussed in the numerical evaluation section.

\end{document}